\documentclass[fleqn,10pt]{article}
\usepackage{latexsym, graphicx, epsfig, amsmath, amssymb,amsfonts}
\usepackage{natbib,amsthm,version}
\usepackage{amsbsy,bm,multirow,enumerate}
\usepackage[titletoc,page]{appendix}
\usepackage[mathscr]{eucal}
\usepackage{mathtools}
\usepackage{color}
\usepackage{subfigure}
\usepackage[utf8]{inputenc}
\usepackage[english]{babel}
\usepackage{float}
\usepackage{caption}
\usepackage{systeme}
\usepackage[affil-it]{authblk}

\begin{document}

\title{ 
	A consistent and conservative model and its scheme for $N$-phase-$M$-component incompressible flows 
} 

\author[1]{
	Ziyang Huang%
	\thanks{Email: \texttt{huan1020@purdue.edu}}}

\author[1,2]{
	Guang Lin%
	\thanks{Email: \texttt{guanglin@purdue.edu}; Corresponding author}}

\author[1]{
	Arezoo M. Ardekani%
	\thanks{Email: \texttt{ardekani@purdue.edu}; Corresponding author}}

\affil[1]{
	School of Mechanical Engineering, Purdue University, West Lafayette, IN 47907, USA}
\affil[2]{
	Department of Mathematics, Purdue University, West Lafayette, IN 47907, USA}

\date{}

\maketitle


\begin{abstract}
In the present work, we propose a consistent and conservative model for multiphase and multicomponent incompressible flows, where there can be arbitrary numbers of phases and components. Each phase has a background fluid called the pure phase, each pair of phases is immiscible, and components are dissolvable in some specific phases. The model is developed based on the multiphase Phase-Field model including the contact angle boundary condition, the diffuse domain approach, and the analyses on the proposed consistency conditions for multiphase and multicomponent flows. The model conserves the mass of individual pure phases, the amount of each component in its dissolvable region, and thus the mass of the fluid mixture, and the momentum of the flow. It ensures that no fictitious phases or components can be generated and that the summation of the volume fractions from the Phase-Field model is unity everywhere so that there is no local void or overfilling. It satisfies a physical energy law and it is Galilean invariant. 
A corresponding numerical scheme is developed for the proposed model, whose formal accuracy is 2nd-order in both time and space. It is shown to be consistent and conservative and its solution is demonstrated to preserve the Galilean invariance and energy law. Numerical tests indicate that the proposed model and scheme are effective and robust to study various challenging multiphase and multicomponent flows.
\end{abstract}

\vspace{0.05cm}

\section{Introduction}\label{Sec Introduction}
Multiphase and multicomponent flow problems are ubiquitous and have various applications. The problems include strong interactions among phases and components, leading to challenges in developing analytical or numerical solutions.
Before we summarize the related studies, we first clarify the terminology in the present work, since we notice that ``multiphase'' and ``multicomponent'' are commutable in some literature. In the present work, we consider that phases are immiscible with each other and the domain is occupied by at least one of the phases, and that components are materials dissolvable in some specific phases and their presence can increase the local density and viscosity based on their amounts (or concentrations). 

The multiphase problems are more difficult to be modeled numerically since a successful numerical model should at least be able to overcome the numerical diffusion and dispersion due to the discontinuity at the moving and deforming phase interfaces. Many efforts have been focused on the two-phase problems in recent decades and the one-fluid formulation \citep{Tryggvasonetal2011,ProsperettiTryggvason2007} is the most popular framework where the motions of different phases are modeled by a single momentum equation. Under this framework, the locations of the two phases can be specified by various methods, e.g., the Front-Tracking method \citep{UnverdiTryggvason1992,Tryggvasonetal2001}, the Level-Set method \citep{OsherSethian1988,Sussmanetal1994,SethianSmereka2003,Gibouetal2018,Gibouetal2019}, the conservative Level-Set method \citep{OlssonKreiss2005,Olssonetal2007,ChiodiDesjardins2017}, the Volume-of-Fluid (VOF) method \citep{HirtNichols1981,ScardovelliZaleski1999,Popinet2009,OwkesDesjardins2017}, the "THINC" method \citep{Xiaoetal2005,Iietal2012,XieXiao2017,Qianetal2018}, and the Phase-Field (or Diffuse-Interface) method \citep{Andersonetal1998,Jacqmin1999,Shen2011,Huangetal2020}. The effect of the surface tension can be modeled by, e.g., the continuous surface force model (CSF) \citep{Brackbilletal1992} and the ghost fluid method (GFM) \citep{Fedkiwetal1999,Lalanneetal2015}, and they are integrated into the flow dynamics by the balanced-force algorithm \citep{Francoisetal2006,Popinet2018}. 
Some recent studies start considering the problems including more than two phases and extending the two-phase models to the three- or multi-phase cases with, e.g., the Volume-of-Fluid (VOF) and Moment-of-Fluid (MOF) methods \citep{Schofieldetal2009,Schofieldetal2010,Francois2015,PathakRaessi2016}, and the Level-Set method \citep{Smithetal2002,Losassoetal2006,Starinshaketal2002}. Nevertheless, a growing number of studies use a Phase-Field model due to its simplicity and effectiveness for three-phase problems, e.g., \citep{BoyerLapuerta2006,Boyeretal2010,KimLowengrub2005,Kim2007,ZhangWang2016,Zhangetal2016}, and for $N$-phase problems, e.g., \citep{BoyerMinjeaud2014,Kim2009,LeeKim2015,Kim2012,WuXu2017,KimLee2017}. Some encouraging progresses on Phase-Field modeling for $N$-phase flows have been done by Dong \citep{Dong2014,Dong2015,Dong2017,Dong2018}, and both the Phase-Field equation and its boundary condition are studied. The latest model in \citep{Dong2018} is the first Phase-Field model that is fully reduction consistent. Based on the Phase-Field model in \citep{Dong2018} and the consistency analysis \citep{Huangetal2020}, Huang et al. developed a consistent and conservative scheme for multiphase incompressible flows \citep{Huangetal2020N}. More recently, Huang et al. develop a consistent and conservative volume distribution algorithm for multiphase flows and apply it to both the Cahn-Hilliard and conservative Allen-Cahn models \citep{Huangetal2020B}, so that the models and their numerical solutions are consistent, conservative, and bounded.   

Despite the active studies on the modeling and simulation of two-/multi-phase flows, the studies including multiple components in a multiphase system are relatively rare, probably because the multiphase problem itself is adequately complicated, and a general and physically plausible multiphase model, e.g., the one in \citep{Dong2018,Huangetal2020N}, is developed very recently. When there are multiple phases and components in a system, a component can dissolve in several different phases, and, at the same time, there can be different components present in a single phase. It is challenging to incorporate all these relationships between phases and components in a general model. In addition, the phases are moving, deforming, and even experiencing topological changes, and the components inside the phases need to respond to these motions appropriately, which casts another challenge in modeling. The appearance of the components may change the local density, which influences the mass transport and the consistency of the model.
We notice that some existing works can be categorized into the two-phase-one-component problems, e.g., the problems including a surfactant \citep{Teigenetal2011,Shietal2019,Soligoetal2019surfactant,Zhuetal2020}, where the concentration of the surfactant changes the surface tension and, as a result, introduces the Marangoni effect, and the problems including phase change \citep{Giussanietal2020,Scapinetal2020}, where the vapor generated from the liquid is dissolvable in the condensible gas. 
However, including these additional physics in the problems having multiple phases and components is out of the scope of the present study. 
We confine our work to incompressible flows without phase change and assume that the Marangoni effect due to the presence of the components is negligible.  

In the present work, we develop a consistent and conservative model for multiphase and multicomponent incompressible flows. The model allows arbitrary numbers of phases and components appearing simultaneously. 
Each component can exist in different phases. In each phase, there is a background fluid called the pure phase, and multiple components can be dissolved in this pure phase. Therefore, each phase is a ``solution'' of its pure phase (or background fluid) as the ``solvent'' and the components dissolved in this phase as the ``solutes''.
Individual pure phases and components have their own densities and viscosities. Each pair of phases has a surface tension and each component has a diffusivity in a given phase. At a wall boundary, each pair of phases has a contact angle. 
The model is based on the multiphase Phase-Field model including the contact angle boundary condition \citep{Dong2017,Dong2018,Huangetal2020N}, the diffuse domain approach \citep{Lietal2009}, and the consistency analysis \citep{Huangetal2020}.
The Phase-Field model introduces a set of order-parameters to indicate the location of each phase. The interfaces between phases are replaced by interfacial regions whose thickness is preserved by the balance of the thermodynamical compression and diffusion.
The diffuse domain approach is applied to replace the convection-diffusion equation of a component defined in a specific phase with its mathematical equivalent equation defined in the whole domain. 
The consistency analysis is performed based on the consistency conditions for multiphase and multicomponent flows, proposed in the present work. It ensures that the flow dynamics is correctly modeled for different phases and components, and that no fictitious phases or components are allowed to be generated.
The resulting model conserves the mass of individual pure phases, the amount of each component in its dissolvable region, and thus the mass of the fluid mixture, and the momentum of the multiphase and multicomponent flows. It ensures that the summation of the volume fractions from the Phase-Field model is unity everywhere so that there is no local void or overfilling. It satisfies a physical energy law and it is Galilean invariant. It satisfies all the consistency conditions, which are the consistency of reduction, the consistency of volume fraction conservation, the consistency of mass conservation, and the consistency of mass and momentum transport. It should be noted that the consistency conditions play a critical role in avoiding fictitious phases and components, in deriving the energy law, and in proving the Galilean invariance of the model. 
The model is flexible to allow a component to cross a phase interface with either zero flux or continuous flux based on the property of the phase interface if the component is dissolvable in both sides of the interface. 
In addition to study the dynamics of the multiphase and multicomponent flows, the present model is applicable to study some multiphase flows, where the miscibility of each two phases can be different. 

The corresponding numerical scheme is developed for the proposed model, which is an extension of our previous schemes for multiphase flows \citep{Huangetal2020N,Huangetal2020B}. 
The scheme is formally 2nd-order accurate in both time and space and it preserves the properties of the model. The scheme conserves the mass of individual pure phases, the amount of each component in its dissolvable region, and therefore the mass of the fluid mixture. The momentum is exactly conserved by the scheme with the conservative method for the interfacial force, while it is essentially conserved with the balanced-force method. 
All the consistency conditions are shown to be satisfied in the discrete level by the scheme. This is very important for the problems including large density ratios \citep{Huangetal2020} and to ensure that no fictitious phases or components can be generated by the scheme.
The numeral solution appears to preserve the Galilean invariance and satisfy the energy law.
The properties and capabilities of the proposed model and scheme are demonstrated numerically.

The rest of the paper is presented as follows.
In Section \ref{Sec Model}, the problem of interest is defined in detail, and the consistent and conservative $N$-phase-$M$-component model is introduced, followed by the consistency analysis, derivation of the energy law, and the proof of Galilean invariance.
In section \ref{Sec Numerical method}, the numerical scheme for the proposed model is introduced, followed by the analysis and discussion about its accuracy, consistency, and conservation properties.
In Section \ref{Sec Results}, multiple numerical cases are performed to demonstrate the properties of the proposed model and scheme, and to show their capability for handling challenging problems.
Finally, we conclude the present study in Section \ref{Sec Conclusion}.

\section{The consistent and conservative $N$-phase-$M$-component model}\label{Sec Model}
In this section, we first define the problem of interest. Then, the multiphase Phase-Field model, the diffuse domain approach for individual components, the material properties, and the momentum equation are introduced. The consistency conditions for the $N$-phase-$M$-component system are proposed and the consistency analysis is performed. Thanks to satisfying the consistency conditions, the proposed model satisfies a physical energy law and is Galilean invariant. At the end of the section, we summarize the properties of the $N$-phase-$M$-component model, and provide an example of a multiphase problem where each pair of phases can be either miscible or immiscible. We then illustrate the applicability of the model in different circumstances of cross-interface transports of a component that is dissolvable in both sides of the interface. 

\subsection{Definitions}\label{Sec Definitions}
The problem considered in the present work includes multiple phases and components, and we use $N$ ($N \geqslant 1$) to denote the number of phases and $M$ ($M \geqslant 0$) for the number of components. 

The $N$ phases are specified by defining a set of order parameters $\{\phi_p \}_{p=1}^N \in [-1,1]$, which are the volume fraction contrasts of individual phases. At the location of $\phi_p=1$, there is only Phase $p$, while $\phi_p=-1$ represents the absence of Phase $p$. The volume fractions of individual phases are related to  $\{\phi_p \}_{p=1}^N$ by 
\begin{equation} \label{Eq Volume fraction}
\chi_p=\frac{1+\phi_p}{2}, 1 \leqslant p \leqslant N,
\end{equation} 
so that $\{\chi_p\}_{p=1}^N \in [0,1]$. Every location should be at least occupied by one of the phases while at most occupied by all of them. As a result, the volume fractions are not independent of each other and should satisfy the summation constraint 
\begin{equation} \label{Eq Summation constraint chi}
\sum_{p=1}^N \chi_p=1,
\end{equation}
or equivalently
\begin{equation}\label{Eq Summation constraint phi}
\sum_{p=1}^N \phi_p=2-N,
\end{equation}
everywhere. Each phase has a background fluid called the pure phase, and $\rho_p^{\phi}$ and $\mu_p^{\phi}$ denote the density and viscosity, respectively, of pure Phase $p$. Every pair of phases is immiscible, and therefore there is a pairwise surface tension between them, denoted by  $\sigma_{p,q}$ for Phases $p$ and $q$. The pairwise surface tensions build a $N \times N$ matrix which is symmetric and has zero diagonal. 
At a wall boundary, there is a pairwise contact angle $\theta_{p,q}^W$ of Phases $p$ and $q$, measured on the side of Phase $p$.
Again, the pairwise contact angles build a $N \times N$ matrix whose entries satisfy $\theta_{p,q}^W+\theta_{q,p}^W=\pi$.
As already mentioned, the Marangoni effect of the components is assumed to be negligible, and consequently, the pairwise surface tensions and contact angles do not change  due to the appearance of the components.

The $M$ components are represented by a set of concentrations $\{C_p\}_{p=1}^M$. Individual components have their densities $\{ \rho_p^C \}_{p=1}^M$ and viscosities $\{ \mu_p^C \}_{p=1}^M$. Each component can be dissolved inside different phases. On the other hand, there can be multiple components inside a specific phase. To indicate the dissolvabilities among the components and the phases, the dissolvability matrix with dimension $M \times N$ is defined as
\begin{equation}\label{Eq Dissolvability matrix}
I^M_{p,q}
=
\left\{
\begin{array}{l}
1, \textrm{if Component $p$ is dissolvable in Phase $q$}, \\
0, \textrm{else}.
\end{array}
\right.
\end{equation}
The diffusion coefficient of Component $p$ in Phase $q$ is denoted by $D_{p,q}$.
Noted that values of $\{C_p\}_{p=1}^M$ are meaningful only inside the corresponding phases they dissolve in. To obtain a meaningful value in the entire domain of interest, one needs to use, e.g., $I^M_{p,q} \chi_q C_p$, to represent the amount of Component $p$ in Phase $q$.
As pointed out in \cite{SzulczewskiJuanes2013}, there are two commonly used ways to interpret concentration. The first case is to quantify the amount of a dissolved solute of a solution such as the ``molar concentration'' of salt in a salt water solution. The second case is to quantify the composition of miscible fluids such as the ``volume fraction'' of ethanol in a water-ethanol system.
Here, the components act like ``solutes'' while the pure phases behave as ``solvents''. Therefore, the concentrations of the components are interpreted as the first case, e.g., the ``molar concentration''. We further assume that inside each phase the ``solution'' compound of the pure phase (as the “solvent”) and the components (as the “solutes”) dissolved in that phase is dilute. In other words, in a specific phase, the volume fractions of the components dissolved in it are negligibly small compared to the volume fraction of the pure phase or the background fluid of that phase. Therefore, the volume of each phase will not be changed by either the appearance or cross-interface transport of the components. This assumption of diluteness is also consistent with neglecting the Marangoni effect of the components. 
It should be noted that the concentrations in the proposed model can also be interpreted as the second case, e.g., the ``volume fraction'', in a subset of problems where each phase has a single diffusion coefficient shared by all the components dissolved in it and the cross-interface transport of those components is prohibited. Although the assumption of diluteness can be removed in these problems, one has to be cautious about the assumption of neglecting the Marangoni effect of the components when applying the proposed model to these problems. 
In the present work, we loosely call $\{ \rho_p^C \}_{p=1}^M$ and $\{ \mu_p^C \}_{p=1}^M$ densities and viscosities, respectively, regardless of which interpretation of the concentrations is employed. The actual meaning of $\{ \rho_p^C \}_{p=1}^M$ depends on the concrete definition of the concentrations. For example, $\rho_p^C$ is the molar mass of Component $p$ when the concentrations are ``molar concentration'', while it is the density of pure Component $p$ minus the density of the pure phase the component dissolves in when the concentrations mean ``volume fraction''. The same works for $\{ \mu_p^C \}_{p=1}^M$.

The densities of the pure phases and components are all constant, and phase changes are not considered in the present work, in addition to the aforementioned assumptions. As a result, the velocity of the flow $\mathbf{u}$ is divergence-free, i.e.,
\begin{equation}\label{Eq Divergence-free}
\nabla \cdot \mathbf{u}=0.
\end{equation}

\subsection{Governing equations}\label{Sec Governing equations}
The governing equations of the $N$-phase-$M$-component model are described in detail, which include the multiphase Phase-Field model, the diffuse domain approach for the components, the material properties, and the momentum equation.

\subsubsection{The Phase-Field equation}\label{Sec Phase-Field}
The dynamics of the volume fraction contrasts $\{\phi_p\}_{p=1}^N$ is governed by the following multiphase Phase-Field model \citep{Dong2018,Huangetal2020N} defined in the whole domain of interest $\Omega$,
\begin{equation}\label{Eq Phase-Field}
\frac{\partial \phi_p}{\partial t}
+
\nabla \cdot (\mathbf{u} \phi_p)
=
\nabla \cdot \left(\sum_{q=1}^N M_{p,q}^{\phi} \nabla \xi_q \right), 
\quad 1 \leqslant p \leqslant N,
\end{equation}
where $M_{p,q}^{\phi}$ is the mobility, and $\xi_p$ is the chemical potential of Phase $p$. 
The homogeneous Neumann boundary condition is employed to the chemical potentials at the boundary of $\Omega$ unless otherwise specified.
The definitions of $M_{p,q}^{\phi}$ and $\xi_p$ are
\begin{equation}\label{Eq Mobility}
M_{p,q}^{\phi}=M^{\phi}(\phi_p,\phi_q)=
\left\{
\begin{array}{l}
-M_0^{\phi} (1+\phi_p)(1+\phi_q), p \neq q, \\
M_0^{\phi} (1+\phi_p)(1-\phi_q), p=q,
\end{array}
\right.
\quad 1 \leqslant p,q \leqslant N,
\end{equation}
and 
\begin{equation}\label{Eq Chemical Potential}
\xi_p=\sum_{q=1}^N \lambda_{p,q} \left( \frac{1}{\eta^2} \left( g'_1(\phi_p)-g'_2(\phi_p+\phi_q) \right)+\nabla^2 \phi_q \right), \quad 1 \leqslant p \leqslant N,
\end{equation}
along with the mixing energy density
\begin{equation}\label{Eq mixing energy density}
\lambda_{p,q}=\frac{3}{2\sqrt{2}} \eta \sigma_{p,q}, \quad 1 \leqslant p,q \leqslant N,
\end{equation}
and with the two potential functions
\begin{equation}\label{Eq Potential functions}
g_1(\phi)=\frac{1}{4} (1-\phi^2)^2, \quad g_2(\phi)=\frac{1}{4} \phi^2 (\phi+2)^2,
\end{equation}
where $M_0^{\phi}$ is a positive constant, $\eta$ represents the thickness of the interfaces, and $g_1'(\phi)$ and $g_2'(\phi)$ are the derivatives of $g_1(\phi)$ and $g_2(\phi)$ with respect to $\phi$, respectively. 

Equivalently, the Phase-Field equation, Eq.(\ref{Eq Phase-Field}), can be written as
\begin{equation}\label{Eq Phase-Field Conservation}
\frac{\partial \phi_p}{\partial t}+\nabla \cdot \mathbf{m}_{\phi_p}=0, \quad 1 \leqslant p \leqslant N,
\end{equation}
where 
\begin{equation}\label{Eq Phase-Field flux}
\mathbf{m}_{\phi_p}
=
\mathbf{u}\phi_p-\sum_{q=1}^N M_{p,q}^{\phi} \nabla \xi_q, \quad 1 \leqslant p \leqslant N,
\end{equation}
is the Phase-Field flux. 

At a wall boundary, the contact angle boundary condition in \citep{Dong2017,Huangetal2020N}, i.e.,
\begin{equation}\label{Eq Contact angle bc}
\mathbf{n} \cdot \nabla \phi_p =\sum_{q=1}^N \zeta_{p,q} \frac{1+\phi_p}{2} \frac{1+\phi_q}{2},
\quad 1 \leqslant p \leqslant N,
\end{equation}
where
\begin{equation}\label{Eq Contact angle coefficient}
\zeta_{p,q}=\frac{2\sqrt{2}}{\eta} \cos(\theta_{p,q}^W),
\quad 1 \leqslant p,q \leqslant N,
\end{equation}
is implemented, 
and the contact angles $\{\theta_{p,q}^W\}_{p,q=1}^N$ are set to be $\frac{\pi}{2}$ unless otherwise specified in the present study.

The Phase-Field model Eq.(\ref{Eq Phase-Field}) is first developed by Dong \citep{Dong2018} in terms of volume fractions and is later reformulated in its conservative form in terms of volume fraction contrasts by Huang et al. in \citep{Huangetal2020N}. One important property of the Phase-Field model is that it satisfies the consistency of reduction, which is defined in Section \ref{Sec Consistency analysis}. This property grantees that no fictitious phase is allowed by the model to appear. 
Each $\phi_p$ is governed by a conservative law, and thus, with a proper boundary condition, e.g., that the normal velocity vanishes at the domain boundary, $\frac{d}{dt} \int_{\Omega} \phi_p d\Omega=0$ is true for all $p$. This is equivalent to $\frac{d}{dt} \int_{\Omega} \chi_p d\Omega=0$ for all $p$ from Eq.(\ref{Eq Volume fraction}), which represents the conservation of the phase volumes, and further to $\frac{d}{dt} \int_{\Omega} \rho_p^\phi \chi_p d\Omega=0$ for all $p$, which is the mass conservation of the pure phases. Moreover, it can be shown that the summation constraint, i.e., Eq.(\ref{Eq Summation constraint phi}), is implied by Eq.(\ref{Eq Phase-Field}). 
Summing Eq.(\ref{Eq Phase-Field}) over $p$, both sides of the summed equation become zero, thanks to the definition of $M_{p,q}^{\phi}$ in Eq.(\ref{Eq Mobility}), if Eq.(\ref{Eq Summation constraint phi}) is true. In other words, Eq.(\ref{Eq Summation constraint phi}) is the solution to the summed equation. More details about the Phase-Field model are referred to \cite{Dong2018}. 
Both studies in \citep{Dong2018,Huangetal2020N} indicate that the Phase-Field model is effective to model multiphase flows when it is appropriately combined with the momentum equation. An alternative Phase-Field model is the conservative Allen-Cahn model for multiphase flows developed in \citep{Huangetal2020B}, which shares the same properties as the model introduced in the present work. However, it doesn't include the effect of the contact angles.

\subsubsection{The concentration equation}\label{Sec Component}
Suppose Component $p$ is dissolvable in Phase $q$, i.e., $I_{p,q}^M=1$, the dynamics of Component $p$ is modeled by the convection-diffusion equation, i.e.,
\begin{equation}\label{Eq Component p, Omega_q}
\frac{\partial{C_p}}{\partial t}
+
\nabla \cdot (\mathbf{u}C_p)
=
\nabla \cdot (D_{p,q} \nabla C_p),
\end{equation}
defined in $\Omega_q$, which is the part of the domain occupied by Phase $q$. The boundary condition of Component $p$ at $\Gamma_{q,r}$, i.e., the boundary between $\Omega_q$ and $\Omega_r$, is
\begin{equation}\label{Eq Boundary condition Component p,qr}
\mathbf{n}_{q,r} \cdot ( D_{p,q} \nabla C_p )
=
\left\{
\begin{array}{l}
0, \textrm{if $I_{p,r}^M=0$} \\
-\mathbf{n}_{r,q} \cdot ( D_{p,r} \nabla C_p ), \textrm{if $I_{p,r}^M=1$},
\end{array}
\right.,
\quad q \neq r,
\end{equation}
where $\mathbf{n}_{q,r}$ is the normal vector pointing from $\Omega_q$ to $\Omega_r$, and it is obvious that $\mathbf{n}_{q,r}=-\mathbf{n}_{r,q}$. The boundary condition Eq.(\ref{Eq Boundary condition Component p,qr}) tells that if Component $p$ is unable to dissolve in $\Omega_r$, i.e., $I_{p,r}^M=0$, no diffusive flux is allowed across $\Gamma_{q,r}$. On the other hand, if Component $p$ is also dissolvable in $\Omega_r$, i.e., $I_{p,r}^M=1$, the diffusive flux is continuous at $\Gamma_{q,r}$. The convective flux is zero at $\Gamma_{q,r}$ by considering that the velocities of the flow and of $\Gamma_{q,r}$ are the same. 
As already mentioned in Section \ref{Sec Definitions}, the effect of cross-interface transports of components, i.e., Eq.(\ref{Eq Boundary condition Component p,qr}), on changing phase volumes is not counted, thanks to the assumption of diluteness. In addition, by an appropriate setup, the proposed model allows to remove the cross-interface transports of components without modifying Eq.(\ref{Eq Boundary condition Component p,qr}). Related details are in Section \ref{Sec Summary model}.

Since Phase $q$ can be moving, deforming, or even experiencing topological change, $\Omega_q$ is time-dependent and its boundary can be extremely complex, which casts great challenges on solving Eq.(\ref{Eq Component p, Omega_q}) defined in it and on assigning Eq.(\ref{Eq Boundary condition Component p,qr}) at its boundary. It is desirable to extend the definition of Eq.(\ref{Eq Component p, Omega_q}) from in $\Omega_q$ to in the whole domain of interest $\Omega$, which is fixed, and to applied the boundary condition Eq.(\ref{Eq Boundary condition Component p,qr}) implicitly like modeling the surface tension with the continuum surface force.  
The mathematical equivalence of different differential operators defined in a time-dependent and complex domain to those defined in a larger and fixed domain is proposed by \citep{Lietal2009}, where some additional terms appear to take the boundary condition into account.
Consequently, Eq.(\ref{Eq Component p, Omega_q}) defined in $\Omega_q$, along with Eq.(\ref{Eq Boundary condition Component p,qr}) at $\{\Gamma_{q,r}\}_{r=1,r\neq q}^N$, has the following mathematically equivalent form,
\begin{equation}\label{Eq Component p, Omega}
\frac{\partial{(H_q C_p)}}{\partial t}
+
\nabla \cdot (\mathbf{u} H_q C_p)
=
\nabla \cdot (H_q D_{p,q} \nabla C_p)
+ 
\sum_{r=1,r \neq q}^N \mathbf{n}_{q,r} \cdot (D_{p,q} \nabla C_p) \delta_{q,r},
\end{equation}
defined in the whole domain of interest $\Omega$,
where $H_q$ and $\delta_{q,r}$ are the indicator functions of $\Omega_q$ and $\Gamma_{q,r}$ such that, for a smooth function $f$,
\begin{equation}\label{Eq H_q, delta_{q,r} }
\int_{\Omega_q} f d\Omega = \int_{\Omega} H_q f d\Omega,
\quad
\int_{\Gamma_{q,r}} f ds = \int_{\Omega} \delta_{q,r} f d\Omega.
\end{equation}
We again have used the equality of the velocity of the flow to that of $\Gamma_{q,r}$. The last term in Eq.(\ref{Eq Component p, Omega}) represents the contribution from the boundary condition at $\{\Gamma_{q,r}\}_{r=1,r\neq q}^N$.
A complete derivation from Eq.(\ref{Eq Component p, Omega_q}) to Eq.(\ref{Eq Component p, Omega}) is available in \cite{Lietal2009} and interested readers should refer to that. 
Next, we sum Eq.(\ref{Eq Component p, Omega}) over all the $q$ that $I_{p,q}^M=1$, and the resultant equation for Component $p$ is
\begin{equation}\label{Eq Component H}
\frac{\partial ( H_p^M C_p ) }{\partial t}
+
\nabla \cdot (\mathbf{u} H_p^M C_p)
=
\nabla \cdot ( D_p \nabla C_p),
\end{equation}
where
\begin{equation}\label{Eq H_p^M}
H_p^M=\sum_{q=1,I_{p,q}^M=1}^N H_q=\sum_{q=1}^N I_{p,q}^M H_q,
\end{equation}
\begin{equation}\label{Eq D_p H}
D_p=\sum_{q=1, I_{p,q}^M=1}^N H_q D_{p,q}=\sum_{q=1}^N I_{p,q}^M H_q D_{p,q}.
\end{equation}
Due to the boundary condition Eq.(\ref{Eq Boundary condition Component p,qr}), the summation of the last term in Eq.(\ref{Eq Component p, Omega}) over all the components is zero, given that $\delta_{q,r}=\delta_{r,q}$.
$H_p^M$ indicates the dissolvable region of Component $p$. $D_p$ can be understood as an effective diffusion coefficient of Component $p$ in the whole domain $\Omega$, whose value is zero in $\Omega_r|_{I_{p,r}^M=0}$ and $D_{p,q}$ in $\Omega_q|_{I_{p,q}^M=1}$.

As shown in \citep{Lietal2009}, the exact indicator function $H$ can be approximated by a smoothed indicator function $H^{\varepsilon}$, which has a transition from $0$ to $1$ with a thickness $\varepsilon$, and the resulting equation using the smoothed indicator function asymptotically converges to the original equation and boundary condition, e.g., Eq.(\ref{Eq Component p, Omega_q}) and Eq.(\ref{Eq Boundary condition Component p,qr}) in the present case, as $\varepsilon$ tends to zero. Such a method is called the diffuse domain approach, which is developed by Li et al. \citep{Lietal2009} for solving a partial differential equation (PDE) in a complex domain. As a result, the volume fractions $\{\chi_p\}_{p=1}^N$ from the Phase-Field model is a direct candidate of $\{H_p^\varepsilon\}_{p=1}^N$ and $\varepsilon$ is the same as $\eta$ in this case.

Before replacing $\{H_p\}_{p=1}^N$ by $\{\chi_p\}_{p=1}^N$, we notice that Eq.(\ref{Eq Component H}) implies the dynamic equations for $\{H_p\}_{p=1}^N$. If $C_p$ is homogeneous, Eq.(\ref{Eq Component H}) along with Eq.(\ref{Eq H_p^M}) becomes
\begin{equation}\label{Eq H_p}
\frac{\partial{H_p}}{\partial{t}}
+
\nabla \cdot (\mathbf{u} H_p)
=
0,
\quad 1 \leqslant p \leqslant N.
\end{equation}
If $H_p$ is replaced by $\chi_p$, Eq.(\ref{Eq H_p}) does not hold, since, from Eq.(\ref{Eq Volume fraction}), Eq.(\ref{Eq Phase-Field Conservation}) and Eq.(\ref{Eq Phase-Field flux}), the dynamics of the volume fractions is governed by
\begin{equation}\label{Eq chi_p}
\frac{\partial \chi_p}{\partial t}+\nabla \cdot \mathbf{m}_{\chi_p}=0, \quad 1 \leqslant p \leqslant N,
\end{equation}
where
\begin{equation} \label{Eq Volume fraction flux}
\mathbf{m}_{\chi_p}=\frac{1}{2}(\mathbf{u}+\mathbf{m}_{\phi_p}), \quad 1 \leqslant p \leqslant N,
\end{equation}
is the volumetric fluxes. As a result, directly replacing $\{H_p\}_{p=1}^N$ by $\{\chi_p\}_{p=1}^N$ leads to inconsistency of Eq.(\ref{Eq H_p}) with Eq.(\ref{Eq chi_p}). In order to assure the consistency, the dynamic equations for $\{\chi_p\}_{p=1}^N$, derived from the component equations, should be the same as Eq.(\ref{Eq chi_p}) instead of Eq.(\ref{Eq H_p}), when $\{\chi_p\}_{p=1}^N$ from the Phase-Field model is used to approximate $\{H_p\}_{p=1}^N$. We call such a consistency the consistency of volume fraction conservation, which is important for the energy law and the Galilean invariance of the model, as shown in Sections \ref{Sec Energy law} and \ref{Sec Galilean invariance}, and has not been noticed and discussed in the original diffuse domain approach \citep{Lietal2009}.

Consequently, the consistent component equation is
\begin{equation}\label{Eq Component}
\frac{\partial (\chi_p^M C_p)}{\partial t}
+
\nabla \cdot (\mathbf{m}_{\chi_p^M} C_p)
=
\nabla \cdot (D_p \nabla C_p),
\quad 1 \leqslant p \leqslant M,
\end{equation}
where
\begin{equation}\label{Eq chi_p^M}
\chi_p^M=\sum_{q=1}^N I_{p,q}^M \chi_q,  \quad 1 \leqslant p \leqslant M,
\end{equation}
\begin{equation}\label{Eq chi_p^M flux}
\mathbf{m}_{\chi_p^M}=\sum_{q=1}^N I_{p,q}^M \mathbf{m}_{\chi_q}, \quad 1 \leqslant p \leqslant M,
\end{equation}
\begin{equation}\label{Eq D_p}
D_p=\sum_{q=1}^N I_{p,q}^M \chi_q D_{p,q},  \quad 1 \leqslant p \leqslant M.
\end{equation}
Unless otherwise specified, the homogeneous Neumann boundary condition is applied to Eq.(\ref{Eq Component}) at the boundary of domain $\Omega$.
It is obvious that Eq.(\ref{Eq chi_p}) is recovered by Eq.(\ref{Eq Component}) when $C_p$ is homogeneous. From Eq.(\ref{Eq Component}), the total amount of Component $p$ inside its dissolvable region $\chi_p^M$ is conserved, i.e., $\frac{d}{dt} \int_{\Omega} (\chi_p^M C_p) d\Omega=0$. After defining the component flux as
\begin{equation}\label{Eq Component flux}
\mathbf{m}_{C_p}=\mathbf{m}_{\chi_p^M} C_p-D_p \nabla C_p,\quad 1 \leqslant p \leqslant M,
\end{equation}
Eq.(\ref{Eq Component}) can be written as
\begin{equation}\label{Eq Component Conservation}
\frac{\partial (\chi_p^M C_p)}{\partial t}
+
\nabla \cdot \mathbf{m}_{C_p}
=
0,
\quad 1 \leqslant p \leqslant M.
\end{equation}

\subsubsection{Density and viscosity}\label{Sec Density and viscosity}
Phase $p$ is composed of its pure phase and the components dissolved in it, both of which contribute to the mass in $\Omega_p$. As a result, the density of Phase $p$ in $\Omega_p$, denoted by $\rho_p$, is 
\begin{equation}\label{Eq Density p}
\rho_p
=
\rho_p^\phi + \sum_{q=1,I_{q,p}^M=1}^M \rho_q^C C_q
=
\rho_p^\phi + \sum_{q=1}^M I_{q,p}^M \rho_q^C C_q.
\end{equation}
The density of the fluid mixture in $\Omega$, denoted by $\rho$, is the volume average of $\{\rho_p\}_{p=1}^N$, i.e., 
\begin{equation}\label{Eq Density}
\rho
=
\sum_{p=1}^N \rho_p \chi_p 
=
\sum_{p=1}^N \rho_p^\phi \chi_p  + \sum_{p=1}^M \rho_p^C \chi_p^M C_p,
\end{equation}
where the first term in the rightmost is contributed from the pure phases (or the background fluids) and the second term appears due to the components.
Similarly, the viscosity of the fluid mixture in $\Omega$ is
\begin{equation}\label{Eq Viscosity}
\mu
=
\sum_{p=1}^N \mu_p^\phi \chi_p  + \sum_{p=1}^M \mu_p^C \chi_p^M C_p. 
\end{equation}

\subsubsection{The momentum equation}\label{Sec Momentum}
The velocity of the flow is governed by the momentum equation
\begin{equation}\label{Eq Momentum}
\frac{\partial (\rho \mathbf{u})}{\partial t}
+
\nabla \cdot (\mathbf{m}\otimes\mathbf{u})
=
-\nabla P
+
\nabla \cdot [\mu(\nabla \mathbf{u}+\nabla \mathbf{u}^T)]
+
\rho \mathbf{g}
+
\mathbf{f}_s,
\end{equation}
where $\mathbf{m}$ is the consistent mass flux, $P$ is the pressure that enforces the divergence-free constraint Eq.(\ref{Eq Divergence-free}), $\mathbf{g}$ is the gravity, and $\mathbf{f}_s$ is the surface force that models the multiphase interfacial tensions. 

The consistent mass flux reads
\begin{equation}\label{Eq Mass flux}
\mathbf{m}
=
\sum_{p=1}^N \rho_p^\phi \mathbf{m}_{\chi_p}  + \sum_{p=1}^M \rho_p^C \mathbf{m}_{C_p}. 
\end{equation}
Applying the consistent mass flux in the inertial term of the momentum equation is critical to obtain the energy law and the Galilean invariance.

From \citep{Huangetal2020N}, the surface force reads
\begin{equation}\label{Eq Surface force}
\mathbf{f_s}=\frac{1}{2} \sum_{p=1}^N \xi_p \nabla \phi_p.
\end{equation}
It is shown in \citep{Huangetal2020N} that Eq.(\ref{Eq Surface force}) is equivalent to $\frac{1}{2} \sum_{p,q=1}^N \nabla \cdot (\lambda_{p,q} \nabla \phi_p \otimes \nabla \phi_q)$. Thus, the momentum of the fluid mixture is conserved by the momentum equation Eq.(\ref{Eq Momentum}) if the gravity is neglected. 

\subsection{Consistency analysis}\label{Sec Consistency analysis}
The consistency analysis is of great importance for building up a physical multiphase model, as shown in our previous studies of Phase-Field modeling for two- and multi-phase flows \citep{Huangetal2020, Huangetal2020N, Huangetal2020CAC}. Specifically, the consistency analysis in the present study is based on the following consistency conditions for $N$-Phase-$M$-Component flows. The consistency conditions are
\begin{itemize}
	\item \textit{Consistency of reduction:} 
	A $N$-phase-$M$-component system should be able to recover the corresponding $N'$-phase-$M'$-component system ($1 \leqslant N' \leqslant N-1$, $ 0 \leqslant M' \leqslant M-1$), when ($N-N'$) phases and $(M-M')$ components are absent. 
	
	\item \textit{Consistency of volume fraction conservation:} 
	The conservation equation of the volume fractions derived from the component equation should be consistent with the one derived from the Phase-Field equation when the component is homogeneous. 
	
	\item \textit{Consistency of mass conservation:}
	The mass conservation equation should be consistent with the one derived from the Phase-Field equation, the component equation, and the density of the fluid mixture. The mass flux $\mathbf{m}$ in the mass conservation equation should lead to a zero mass source. 
	
	\item \textit{Consistency of mass and momentum transport:}
	The momentum flux in the momentum equation should be computed as a tensor product between the mass flux and the velocity, where the mass flux should be identical to the one in the mass conservation equation.
\end{itemize}

The consistency of reduction is enriched from the $N$-phase one by adding the component part. It is important because it assures that no fictitious phase or component is generated by the model. Any phases or components that are absent at $t=0$ will not appear at $\forall t>0$, without considering any sources or injection, if the model satisfies the consistency of reduction. The consistency of reduction of the Phase-Field model, i.e., Eq.(\ref{Eq Phase-Field}), and the terms in the density, i.e., Eq.(\ref{Eq Density}), the viscosity, i.e., Eq.(\ref{Eq Viscosity}), and the mass flux, i.e., Eq.(\ref{Eq Mass flux}), related to $\{\phi_p\}_{p=1}^N$, has been shown in \citep{Dong2018, Huangetal2020N}. The momentum equation is reduction consistent as long as the density $\rho$, the viscosity $\mu$, and the mass flux $\mathbf{m}$ are consistent as well. Thus, our attention is paid on the part related to the components. For a specific $p$ and from Eq.(\ref{Eq Component}), we can obtain $C_p \equiv 0, \forall t>0$, if $C_p=0$ initially. As a result, the contribution of $C_p$ to the density Eq.(\ref{Eq Density}), the viscosity Eq.(\ref{Eq Viscosity}), and the mass flux Eq.(\ref{Eq Mass flux}) disappears. Consequently, the $M$-component system with the absence of Component $p$ reduces to the corresponding $(M-1)$ system. We can repeat the procedure $(M-M')$ times so that the $M$-component system reduces to the corresponding $M'$-component system ($0 \leqslant M' \leqslant M-1$). Combining with the analysis in \citep{Huangetal2020N}, the proposed $N$-phase-$M$-component model satisfies the consistency of reduction.

The consistency of volume fraction conservation is newly proposed in the present work for the diffuse domain approach, and has been considered in Section \ref{Sec Component} when building up the component equation Eq.(\ref{Eq Component}). In the original diffuse domain approach, it presumes that the smoothed indicator function has the same sharp-interface dynamics as the exact one Eq.(\ref{Eq H_p}), i.e., the phase interface is advected by the flow velocity. However, this is not the case when the smoothed indicator function is chosen from the Phase-Field model, where the sharp interface is replaced by the interfacial region whose thickness is small but finite. There are allowable thermodynamical compression and diffusion, in addition to the flow advection, inside the interfacial region, and this casts a discrepancy between the Phase-Field model and the original diffuse domain approach. The consistency of volume fraction conservation aims to remove the discrepancy by incorporating the thermodynamical effects in the Phase-Field model to the component equation. This consistency condition is essential for the model satisfying the energy law and the Galilean invariance, see Sections \ref{Sec Energy law} and \ref{Sec Galilean invariance}.

The consistency of mass conservation and the consistency of mass and momentum transport are identical to those for two- and multi-phase flows \citep{Huangetal2020,Huangetal2020N, Huangetal2020CAC}. Based on the definitions of the density of the fluid mixture Eq.(\ref{Eq Density}) and the consistent mass flux Eq.({\ref{Eq Mass flux}}), it can be shown that the density is governed by 
\begin{equation}\label{Eq Mass}
\frac{\partial \rho}{\partial t}
+
\nabla \cdot \mathbf{m}
=
\sum_{p=1}^N \rho_p^\phi
\underbrace{\left(
\frac{\partial \chi_p}{\partial t}  
+
\nabla \cdot \mathbf{m}_{\chi_p}
\right)}_{0}
+ 
\sum_{p=1}^M \rho_p^C
\underbrace{\left(
\frac{\partial (\chi_p^M C_p)}{\partial t}
+ 
\nabla \cdot \mathbf{m}_{C_p}
\right)}_{0}
=0,
\end{equation}
thanks to Eq.(\ref{Eq chi_p}) from the Phase-Field equation Eq.(\ref{Eq Phase-Field}) and the component equation Eq.(\ref{Eq Component Conservation}). Consequently, the consistency of mass conservation is true with the consistent mass flux defined in Eq.(\ref{Eq Mass flux}). Since the consistent mass flux Eq.(\ref{Eq Mass flux}) has been applied to the inertial term of the momentum equation Eq.(\ref{Eq Momentum}), the consistency of mass and momentum transport is satisfied as well. These two consistency conditions are critical for problems with large density ratios, as indicated in our previous studies \citep{Huangetal2020,Huangetal2020N, Huangetal2020CAC}, since they honor the physical coupling between the mass and momentum. Moreover, they are essential for the model satisfying the energy law and the Galilean invariance, see Sections \ref{Sec Energy law} and \ref{Sec Galilean invariance}.

\subsection{Energy law}\label{Sec Energy law}
The proposed $N$-phase-$M$-component model satisfies the following physical energy law. We define the free energy density, component energy density, and kinetic energy density to be
\begin{equation}\label{Eq free energy density}
e_F=\sum_{p,q=1}^N \frac{\lambda_{p,q}}{2} \left(
\frac{1}{\eta^2} \left( g_1(\phi_p)+g_1(\phi_q)-g_2(\phi_p+\phi_q) \right)-\nabla \phi_p \cdot \nabla \phi_q \right),
\end{equation}
\begin{equation}\label{Eq component energy density}
e_C=\sum_{p=1}^M \frac{1}{2} \gamma_{C_p} \chi_p^M C_p^2,
\end{equation}
\begin{equation}\label{Eq Kinetic energy density}
e_K=\frac{1}{2} \rho \mathbf{u} \cdot \mathbf{u},
\end{equation}
respectively, where $\{\gamma_{C_p}\}_{p=1}^M$ are positive dimensional constants so that $e_C$ has a unit $[\mathrm{J/m^3}]$ and we choose them to be unity without loss of generality. The corresponding energies are the integrals of the energy densities over the domain. It should be noted that the chemical potential of Phase $p$, i.e., Eq.(\ref{Eq Chemical Potential}), is the functional derivative of the free energy with respect to $\phi_p$, i.e., $\xi_p=\frac{\delta E_F}{\delta \phi_p}=\frac{\delta}{\delta \phi_p} \int_\Omega e_F d\Omega$.

After multiplying the Phase-Field equation Eq.(\ref{Eq Phase-Field}) with the chemical potential Eq.(\ref{Eq Chemical Potential}) and then summing over all the phases $p$, we obtain the governing equation for the free energy density, i.e.,
\begin{eqnarray}\label{Eq Free energy}
\frac{1}{2} \frac{\partial e_F}{\partial t}
+
\sum_{p.q=1}^N \frac{\lambda_{p,q}}{4} \nabla \cdot \left( 
\frac{\partial \phi_p}{\partial t} \nabla \phi_q
+
\frac{\partial \phi_q}{\partial t} \nabla \phi_p \right)
+
\mathbf{u} \cdot \frac{1}{2} \sum_{p=1}^N \xi_p \nabla \phi_p \\
\nonumber
=
\frac{1}{2} \sum_{p,q=1}^N \nabla \cdot \left( \xi_p M_{p,q}^\phi \nabla \xi_q \right)
-
\frac{1}{2} \sum_{p,q=1}^N M_{p,q}^\phi \nabla \xi_p \cdot \nabla \xi_q.
\end{eqnarray}

After multiplying the component equation Eq.(\ref{Eq Component}) with $\gamma_{C_p} C_p$ and then summing over all the components $p$, we obtain the governing equation for the component energy density, i.e.,
\begin{equation}\label{Eq Component energy}
\frac{\partial e_C }{\partial t}
+
\sum_{p=1}^M \nabla \cdot \left( \mathbf{m}_{\chi_p^M}\frac{1}{2} \gamma_{C_p} C_p^2 \right)
=
\sum_{p=1}^M \nabla \cdot ( \gamma_{C_p} D_p C_p \nabla C_p)
-
\sum_{p=1}^M \gamma_{C_p} D_p \nabla C_p \cdot \nabla C_p.
\end{equation}
Eq.(\ref{Eq chi_p}) has been used in the derivation.

After performing the dot product between $\mathbf{u}$ and the momentum equation Eq.(\ref{Eq Momentum}), we obtain the governing equation for the kinetic energy density, i.e.,
\begin{eqnarray}\label{Eq Kinetic energy}
\frac{\partial e_K}{\partial t}
+
\nabla \cdot \left( \mathbf{m} \frac{\mathbf{u} \cdot \mathbf{u}}{2} \right)
=
-\nabla \cdot (\mathbf{u}P)
+
\nabla \cdot \left( \mu(\nabla \mathbf{u}+\nabla \mathbf{u}^T)\cdot \mathbf{u} \right)
-
\frac{1}{2} \mu (\nabla \mathbf{u}+\nabla \mathbf{u}^T):(\nabla \mathbf{u}+\nabla \mathbf{u}^T)
+
\mathbf{u}\cdot \mathbf{f}_s.
\end{eqnarray}
We have neglected the gravity and used Eq.(\ref{Eq Mass}) in the derivation.

When we combine Eq.(\ref{Eq Free energy}), Eq.(\ref{Eq Component energy}) and Eq.(\ref{Eq Kinetic energy}) together, and assume that all the fluxes vanish at the domain boundary, we obtain the energy law for the $N$-phase-$M$-component system, i.e.,
\begin{eqnarray}\label{Eq Total energy}
\frac{dE_T}{dt}
=
\frac{d}{dt} \left( \frac{1}{2} E_F + E_C + E_K \right)
=
\frac{d}{d t} \int_{\Omega} \left( \frac{1}{2} e_F+e_C+e_K \right) d\Omega\\
\nonumber
=
-
\int_\Omega \frac{1}{2} \sum_{p,q=1}^N M_{p,q}^{\phi} \nabla \xi_p \cdot \nabla \xi_q d\Omega
-
\int_{\Omega} \sum_{p=1}^M \gamma_{C_p} D_p \nabla C_p \cdot \nabla C_p d\Omega
-
\int_{\Omega} \frac{1}{2} \mu (\nabla \mathbf{u}+\nabla \mathbf{u}^T):(\nabla \mathbf{u}+\nabla \mathbf{u}^T) d\Omega.
\end{eqnarray}
Eq.(\ref{Eq Total energy}) states that the total energy of the $N$-phase-$M$-component system, which includes the free energy, the component energy, and the kinetic energy, is not increasing with time without any external input. The right-hand side (RHS) of Eq.(\ref{Eq Total energy}) shows three factors to reduce the total energy of the multiphase and multicomponent system. The first one comes from the thermodynamical non-equilibrium, contributed by the Phase-Field model. It should be noted that it is effective only in the interfacial region since the mobility Eq.(\ref{Eq Mobility}) is zero inside the bulk-phase region. The second one results from the inhomogeneity of the components in their dissolvable regions, contributed by the component equation. It should be noted that $D_p$ is zero where $C_p$ is not dissolvable so there is no contribution from those regions to the total energy. The last one is due to the viscosity of the fluids, contributed from the momentum equation.  
To include the effect of the contact angles, i.e., the contact angles are other than $\frac{\pi}{2}$, a wall energy functional has to be introduced. However, so far, it is still an open question whether the contact angle boundary condition Eq.(\ref{Eq Contact angle bc}) from \citep{Dong2017} corresponds to a multiphase wall functional. 

At the end of this section, we need to emphasize the significance of the consistency conditions in deriving the energy law. The left-hand side (LHS) of Eq.(\ref{Eq Component energy}) is derived from
\begin{eqnarray}\label{Eq Component energy Left}
C_p \left( 
\frac{\partial (\chi_p^M C_p)}{\partial t} 
+
\nabla \cdot (\mathbf{m}_{\chi_p^M} C_p) 
\right)
=
C_p \left( 
\chi_p^M \frac{\partial C_p}{\partial t} 
+
\mathbf{m}_{\chi_p^M} \cdot \nabla  C_p \right)
+ 
C_p^2  \underbrace{\left(  \frac{\partial \chi_p^M }{\partial t} 
+
\nabla \cdot \mathbf{m}_{\chi_p^M}   \right) }_{0}\\
\nonumber
=
\chi_p^M \frac{\partial \frac{1}{2}C_p^2}{\partial t} 
+
\mathbf{m}_{\chi_p^M} \cdot \nabla  \frac{1}{2} C_p^2
+ 
\frac{C_p^2}{2} \underbrace{ \left(  \frac{\partial \chi_p^M }{\partial t} 
+
\nabla \cdot \mathbf{m}_{\chi_p^M}   \right) }_{0}\\
\nonumber
=
\frac{\partial (\frac{1}{2} \chi_p^M C_p^2)}{\partial t} 
+
\nabla \cdot \left( \mathbf{m}_{\chi_p^M} \frac{1}{2} C_p^2 \right).
\end{eqnarray}
Thanks to the consistency of volume fraction conservation, the volume fraction equation, i.e., the terms with under-brace in Eq.(\ref{Eq Component energy Left}), is recovered. Similarly, the derivation of the left-hand side (LHS) of Eq.(\ref{Eq Kinetic energy}) is 
\begin{eqnarray}\label{Eq Kinetic energy Left}
\mathbf{u} \cdot \left( \frac{\partial (\rho \mathbf{u})}{\partial t}
+
\nabla \cdot ( \mathbf{m} \otimes \mathbf{u}) \right)
=
\frac{\partial e_K }{\partial t}
+
\nabla \cdot \left( \mathbf{m}  \frac{1}{2} \mathbf{u} \cdot \mathbf{u} \right)
+
\frac{1}{2} \mathbf{u} \cdot \mathbf{u} 
\underbrace{ \left( \frac{\partial \rho }{\partial t}
+
\nabla \cdot \mathbf{m} \right) }_{0}
=
\frac{\partial e_K }{\partial t}
+
\nabla \cdot \left( \mathbf{m}  \frac{1}{2} \mathbf{u} \cdot \mathbf{u} \right).
\end{eqnarray}
The mass conservation equation, i.e., Eq.(\ref{Eq Mass}), is recovered in Eq.(\ref{Eq Kinetic energy Left}), thanks to the consistency of mass conservation and the consistency of mass and momentum transport. If the consistency conditions are violated, the terms with under-brace in Eq.(\ref{Eq Component energy Left}) and Eq.(\ref{Eq Kinetic energy Left}) are not zero any more. As a result, some additional terms will be added to the total energy of the multiphase and multicomponent system Eq.(\ref{Eq Total energy}). In other words, even though all the phases are in their thermodynamical equilibrium, all the components are homogeneous in their dissolvable region, and all the fluids are inviscid, the total energy of the multiphase and multicomponent system could still change due to the additional terms introduced by the inconsistency.  Such a behavior of the total energy is not physically plausible. Thus, the consistency conditions are playing an essential role in the physical behavior of the system. 

\subsection{Galilean invariance}\label{Sec Galilean invariance}
The proposed $N$-phase-$M$-component model satisfies the Galilean invariance, thanks to the consistency conditions. In this section, we call the reference frame Frame $(\mathbf{x},t)$, and another frame that is moving with respect to the reference frame with a constant velocity $\mathbf{u}_0$ Frame $(\mathbf{x}',t')$. All the variables with $'$ are measured in Frame $(\mathbf{x}',t')$. The Galilean transform is
\begin{equation}\label{Eq Galilean transform}
\mathbf{x}'=\mathbf{x}-\mathbf{u}_0 t,
\quad
t'=t,
\quad
\mathbf{u'}=\mathbf{u}-\mathbf{u}_0,
\quad
f'=f,
\quad
\nabla' f' = \nabla f,
\quad
\nabla'^2 f'=\nabla^2 f,
\quad
\frac{\partial f'}{\partial t'}=\frac{\partial f}{\partial t}+\mathbf{u}_0 \cdot \nabla f,
\end{equation}
where $f$ is a scalar function depending on time and space. With the help of the Galilean transform Eq.(\ref{Eq Galilean transform}), the governing equations defined in Frame $(\mathbf{x}',t')$ are
\begin{equation}\label{Eq Divergence-free GI}
\nabla' \cdot \mathbf{u}'=\nabla \cdot (\mathbf{u}-\mathbf{u}_0)=0,
\end{equation}
\begin{eqnarray}\label{Eq Phase-Field GI}
\frac{\partial \phi'_p }{\partial t'}
+
\nabla' \cdot (\mathbf{u}'\phi'_p)
-
\nabla' \cdot \left( \sum_{q=1}^N M'^\phi_{p,q} \nabla' \xi'_q \right)
=
\frac{\partial \phi_p }{\partial t}
+
\mathbf{u}_0 \cdot \nabla \phi_p
+
\nabla \cdot \left( (\mathbf{u}-\mathbf{u}_0)\phi_p \right)
-
\nabla \cdot \left( \sum_{q=1}^N M_{p,q}^\phi \nabla \xi_q \right)
=
0,
\end{eqnarray}
\begin{eqnarray}\label{Eq Phase-Field flux GI}
\mathbf{m}_{\phi'_p} 
=
\mathbf{u}' \phi'_p-\sum_{q=1}^N M'_{p,q} \nabla' \xi'_q
= 
(\mathbf{u}  -\mathbf{u}_0 ) \phi_p-\sum_{q=1}^N M_{p,q} \nabla \xi_q 
=
\mathbf{m}_{\phi_p}-\mathbf{u}_0 \phi_p,
\end{eqnarray}
\begin{eqnarray}\label{Eq Volume fraction flux GI}
\mathbf{m}_{\chi'_p}
=
\frac{1}{2} (\mathbf{u}'+\mathbf{m}_{\phi'_p})
=
\frac{1}{2} (\mathbf{u}-\mathbf{u}_0+\mathbf{m}_{\phi_p}-\mathbf{u}_0 \phi_p)
=
\mathbf{m}_{\chi_p}-\mathbf{u}_0 \chi_p,
\end{eqnarray}
\begin{eqnarray}\label{Eq chi_p^M flux GI}
\mathbf{m}_{\chi'^M_p}
=
\sum_{q=1}^N I_{p,q}^M \mathbf{m}_{\chi'_q}
=
\sum_{q=1}^M I_{p,q}^M \mathbf{m}_{\chi_q}-\mathbf{u}_0 \sum_{q=1}^M I_{p,q}^M \chi_q
=
\mathbf{m}_{\chi_p^M}-\mathbf{u}_0 \chi_p^M,
\end{eqnarray}
\begin{eqnarray} \label{Eq Component GI}
\frac{\partial (\chi'^M_p C'_p) }{\partial t'}
+
\nabla' \cdot (\mathbf{m}_{\chi'^M_p} C'_p) 
-
\nabla' \cdot (D'_p \nabla' C'_p)
=\\
\nonumber
\frac{\partial (\chi_p^M C_p) }{\partial t}
+
\mathbf{u}_0 \cdot \nabla (\chi_p^M C_p)
+
\nabla \cdot \left( (\mathbf{m}_{\chi_p^M}-\mathbf{u}_0 \chi_p^M) C_p \right) 
-
\nabla \cdot (D_p \nabla C_p)
=0,
\end{eqnarray}
\begin{eqnarray}\label{Eq Component flux GI}
\mathbf{m}_{C'_p}
=
\mathbf{m}_{\chi'^M_p} C'_p
-
D'_p \nabla' C'_p
=
(\mathbf{m}_{\chi_p^M}-\mathbf{u}_0 \chi_p^M) C_p
-
D_p \nabla C_p
=
\mathbf{m}_{C_p}
-
\mathbf{u}_0 \chi_p^M C_p ,
\end{eqnarray}
\begin{eqnarray}\label{Eq Mass flux GI}
\mathbf{m}'
=
\sum_{p=1}^N \rho_p^\phi \mathbf{m}_{\chi'_p}
+
\sum_{p=1}^M \rho_p^C \mathbf{m}_{C'_p}
=
\sum_{p=1}^N \rho_p^\phi (\mathbf{m}_{\chi_p}-\mathbf{u}_0 \chi_p)
+
\sum_{p=1}^M \rho_p^C (\mathbf{m}_{C_p}-\mathbf{u}_0 \chi_p^M C_p )
=
\mathbf{m}
- 
\mathbf{u}_0 \rho,
\end{eqnarray}
\begin{eqnarray}\label{Eq Mass GI}
\frac{\partial \rho'}{\partial t'}
+
\nabla' \cdot \mathbf{m}'
=
\frac{\partial \rho}{\partial t}
+
\mathbf{u}_0 \cdot \nabla \rho
+
\nabla \cdot (\mathbf{m}- \mathbf{u}_0 \rho)
=
0,
\end{eqnarray}
\begin{eqnarray}\label{Eq Momentum GI}
\frac{\partial (\rho' \mathbf{u}')}{\partial t'}
+
\nabla' \cdot (\mathbf{m}' \otimes \mathbf{u}')
+
\nabla' P'
-
\nabla' \cdot [\mu'(\nabla' \mathbf{u}'+\nabla' \mathbf{u}'^T)]
-
\rho' \mathbf{g}
-
\mathbf{f}'_s
=\\
\nonumber
\frac{\partial (\rho (\mathbf{u}-\mathbf{u}_0))}{\partial t}
+
\mathbf{u}_0 \cdot \nabla (\rho (\mathbf{u}-\mathbf{u}_0))
+
\nabla \cdot \left( (\mathbf{m}-\rho \mathbf{u}_0) \otimes (\mathbf{u}-\mathbf{u}_0) \right)\\
\nonumber
+
\nabla P
-
\nabla \cdot [\mu(\nabla (\mathbf{u}-\mathbf{u}_0)+\nabla (\mathbf{u}-\mathbf{u}_0)^T)]
-
\rho \mathbf{g}
-
\mathbf{f}_s
=\\
\nonumber
\frac{\partial (\rho \mathbf{u})}{\partial t}
+
\nabla \cdot (\mathbf{m} \otimes \mathbf{u})
-
\mathbf{u}_0 \underbrace{ \left( \frac{\partial \rho}{\partial t}
+
\nabla \cdot \mathbf{m} 
\right) }_{0}
+
\nabla P
-
\nabla \cdot [\mu(\nabla \mathbf{u}+\nabla \mathbf{u}^T)]
-
\rho \mathbf{g}
-
\mathbf{f}_s
=
0.
\end{eqnarray}
The Phase-Field equation Eq.(\ref{Eq Phase-Field GI}), the component equation Eq.(\ref{Eq Component GI}), the mass conservation equation Eq.(\ref{Eq Mass GI}), the momentum equation Eq.(\ref{Eq Momentum GI}), and the divergence-free condition Eq.(\ref{Eq Divergence-free GI}) in Frame $(\mathbf{x}',t')$ are identical to their correspondences, i.e., Eq.(\ref{Eq Phase-Field}), Eq.(\ref{Eq Component}), Eq.(\ref{Eq Mass}), Eq.(\ref{Eq Momentum}) and Eq.(\ref{Eq Divergence-free GI}), in Frame $(\mathbf{x},t)$. Thus the Galilean invariance is satisfied. It should be noted that the consistency of mass conservation and the consistency of mass and momentum transport play critical roles in the Galilean invariance of the momentum equation, see the under-braced term in Eq.(\ref{Eq Momentum GI}). If either of the consistency conditions is violated, the under-braced term in Eq.(\ref{Eq Momentum GI}) is non-zero, and consequently the Galilean invariance of the momentum equation is failed.

The Galilean invariance implies that a homogeneous velocity field is an admissible solution for the momentum equation. Consider the case where $\mathbf{u}'=\mathbf{0}$, the divergence-free condition is satisfied and the momentum equation in Frame $(\mathbf{x}',t')$ requires that
\begin{equation}\label{Eq Mechanical equilibrium GI}
-\nabla' P'+\rho' \mathbf{g} + \mathbf{f}'_s=\mathbf{0},
\end{equation}
which corresponds to the mechanical equilibrium. From the Galilean transform, we have the same mechanical equilibrium, i.e.,
\begin{equation}\label{Eq Mechanical equilibrium}
-\nabla P + \rho \mathbf{g} + \mathbf{f}_s=\mathbf{0},
\end{equation}
and $\mathbf{u}=\mathbf{u}_0$ in Frame $(\mathbf{x},t)$. So the divergence-free condition is true in Frame $(\mathbf{x},t)$. The momentum equation in Frame $(\mathbf{x},t)$ becomes
\begin{equation}
\frac{\partial (\rho \mathbf{u}_0) }{\partial t}
+
\nabla \cdot (\mathbf{m} \otimes \mathbf{u}_0)
=
\nabla \cdot [\mu(\nabla \mathbf{u}_0+\nabla \mathbf{u}_0^T)]
\end{equation}
The right-hand side (RHS) is zero due to $\nabla \mathbf{u}_0 =\mathbf{0}$. The left-hand side(LHS) is again zero since the consistency of mass conservation and the consistency of mass and momentum transport recover the mass conservation equation Eq.(\ref{Eq Mass}). Consequently, the homogeneous velocity $\mathbf{u}_0$ is an admissible solution.

\subsection{Summary of the N-Phase-M-Component model}\label{Sec Summary model}
A consistent and conservative $N$-phase-$M$-component flow model is proposed, which allows different densities and viscosities of individual pure phases and components, different surface tensions between every two phases, and different diffusivities between phases and components. The proposed model satisfies all the consistency conditions, conserves the mass of individual pure phases, conserves the amount of individual components inside their dissolvable regions, and conserves the momentum of the multiphase and multicomponent system. It also satisfies the energy law and the Galilean invariance. These properties of the model are not dependent on the material properties of the phases and components considered. 

The proposed model can be applied to study some multiphase flows where the miscibilities of each pair of phases are different. For example, to model a three-phase system where Phases 01 and 02 are miscible while Phases 01 and 03 are immiscible, as well as Phases 02 and 03, we can define a 2-phase-1-component system, where Phase 1 without Component 1 represents Phase 01, Phase 1 with a specific amount, e.g., $1\mathrm{mol/m^3}$, of Component 1 represents Phase 02, and Phase 2 represents Phase 03. We can set $I_{1,1}^M=1$ and $I_{1,2}^M=0$. As a result, Phase 1, with/without Component 1, is immiscible with Phase 2, and the immisciblities between Phases 01 and 03 and between Phases 02 and 03 are properly modeled. Component 1 is only dissolvable in Phase 1. When Phase 1 with Component 1 moves to locations where Phase 1 has no Component 1, Component 1 starts to be transported by both convection and diffusion to those locations, which models the miscible behavior of Phases 01 and 02.
However, the proposed model is not ready for the problems where the miscible phases have significantly different surface tensions with respect to a phase that they are immiscible with or where a phase is miscible with several other phases that are immiscible with each other, since neither the Marangoni effect due to the presence of the components nor the change of phase volumes due to the transport of components has been considered.

The proposed model is flexible to model the cross-interface transport of a component which is dissolvable in both sides of the interface. Although we consider that the diffusive flux is continuous across the phase interfaces if the component is dissolvable in both of the phases, the proposed model works in the scenario where the component is not allowed to cross the phase interfaces even though it is dissolvable in both sides of the interface. For example, Component 01 is dissolvable in both Phases $p$ and $q$ but is unable to cross the phase interfaces between Phases $p$ and $q$. We can set Component 1 with $I_{1,p}^M=1$ and $I_{1,q}^M=0$, and Component 2 with $I_{2,p}^M=0$ and $I_{2,q}^M=1$. Consequently, Component 1 is only allowed to be dissolved in Phase $p$ but not in Phase $q$, and Component 2 is the opposite. Neither Component 1 nor 2 can cross the phase interface between Phases $p$ and $q$. Consequently, the concentration of Component 01 in its dissolvable region $\chi_{01}^M C_{01}$ is represented by $\chi_1^M C_1+\chi_2^M C_2$.

In addition to the $N$-phase-$M$-component model, more importantly, the present study proposes a general framework that physically connects the phases, components, and fluid flows, with the help of the proposed consistency conditions. Thus, it can be implemented to incorporate many other models for locating the phases or transporting the components. One can freely choose another physical model, instead of Eq.(\ref{Eq Phase-Field}), to govern the order parameters. Also, one can replace the convection-diffusion equation Eq.(\ref{Eq Component p, Omega_q}) with other transport equation that more accurately produces the dynamics of the components, and obtain the component equation following the diffuse domain approach and the consistency of volume fraction conservation, as described in Section \ref{Sec Component}. These result in changing the definitions of the Phase-Field flux and component flux in Eq.(\ref{Eq Phase-Field Conservation}) and Eq.(\ref{Eq Component Conservation}), while the rest of the equations, e.g., the density of the fluid mixture Eq.(\ref{Eq Density}), the consistent mass flux Eq.(\ref{Eq Mass flux}), and the momentum equation Eq.(\ref{Eq Momentum}) remain unchanged. The Galilean invariance of the momentum equation and the kinetic energy equation Eq.(\ref{Eq Kinetic energy}) are always valid, regardless of the concrete definitions of the Phase-Field flux and component flux which depend on the models for the order parameters and components. However, the free energy, the component energy, and the Galilean invariance of the order parameters and components will be affected.

\section{Numerical method}\label{Sec Numerical method}
The numerical scheme for the proposed $N$-phase-$M$-component model is introduced in this section. We use the numerical operators defined in our previous work \citep{Huangetal2020} for spatial discretizations. Specifically, the collocated grid arrangement is used, where all the dependent variables are stored at the cell centers, and the fluxes and gradients are stored at the cell faces. In addition, there are normal components of the velocity being stored at the cell faces, and the cell-face velocity is divergence-free in the discrete sense using the central difference at the cell centers. A variable $f$ at $(x_i,y_j)$ is represented by $f_{i,j}$ or $[f]_{i,j}$, and therefore the volume of cell $(x_i,y_j)$ is denoted by $[\Delta \Omega]_{i,j}$. We may skip the subscript if the equations work for every cell.
The convective operators are approximated by the 5th-order WENO scheme \citep{JiangShu1996}, and the Laplace operator and the gradient operator are approximated by the 2nd-order central difference. We denote, for example, the discrete gradient operator by $\tilde{\nabla}$, to distinguish it from its corresponding continuous one. The linear interpolation from the nearest nodal values is represented by $\bar{f}$, and any other interpolations or approximations are denoted by $\tilde{f}$. The time derivative is approximated by $\frac{\gamma_t f^{n+1} -\hat{f}}{\Delta t}$, where $f^{n+1}$ denotes the value of $f$ at time level $n+1$, $\Delta t$ is the time step size, and $\gamma_t$ and $\hat{f}$ are scheme-dependent. Unless otherwise specified, the 2nd-order backward difference scheme is used. Therefore, $\gamma_t=1.5$ and $\hat{f}=2f^{n}-0.5f^{n-1}$. $f^{*,n+1}$ is the extrapolation along the time direction and the 2nd-order one is $f^{*,n+1}=2f^{n}-f^{n-1}$.
Implementations of different kinds of boundary conditions are also described in detail and we refer interested readers to \citep{Huangetal2020}. 
For a clear presentation, we consider only two dimensions. The extension to higher dimensions is straightforward, and all the analyses and discussions in this section are valid for higher-dimensional cases. 

\subsection{Scheme for the Phase-Field equation}\label{Sec Scheme for the Phase-Field equation}
The scheme for solving the Phase-Field equation Eq.(\ref{Eq Phase-Field}) is elaborated in our previous work \citep{Huangetal2020N}, which is modified from the one in \citep{Dong2018}. 
The scheme is formally 2nd-order accurate in both time and space, and satisfies the consistency of reduction in the discrete level, 
the mass conservation of individual pure phases, i.e., $\sum_{i,j} [\phi_p^n \Delta \Omega]_{i,j}=\sum_{i,j} [\phi_p^0 \Delta \Omega]_{i,j}$ for all $p$, 
and the summation constraint, i.e., $\sum_{p=1}^N \phi_p=2-N$, at all the cells and time levels. Special cares have been paid to the convection term in Eq.(\ref{Eq Phase-Field}) to avoid generating fictitious phases, local voids or overfilling. After solving the equation, the consistent and conservative boundedness mapping algorithm \citep{Huangetal2020B} is applied to guarantee that $\{\phi_p\}_{p=1}^N$ is in $[-1,1]$ while preserving all the aforementioned properties of the solution. The resultant fully-discrete Phase-Field equation is
\begin{equation}\label{Eq Phase-Field fully-discrete}
\frac{\gamma_t \phi_p^{n+1}-\hat{\phi}_p}{\Delta t}
+
\tilde{\nabla} \cdot \tilde{\mathbf{m}}_{\phi_p}
=
0,
\quad
1\leqslant p \leqslant N,
\end{equation} 
where $\tilde{\mathbf{m}}_{\phi_p}$ is the discrete Phase-Field flux of Phase $p$ and its definition is available in \citep{Huangetal2020B}.

Once $\{\phi_p^{n+1}\}_{p=1}^N$ and $\{\tilde{\mathbf{m}}_{\phi_p}\}_{p=1}^N$ are obtained from Eq.(\ref{Eq Phase-Field fully-discrete}), the volume fractions $\{\chi_p^{n+1}\}_{p=1}^N$ and volumetric fluxes $\{\tilde{\mathbf{m}}_{\chi_p}\}_{p=1}^N$ are computed from Eq.(\ref{Eq Volume fraction}) and Eq.(\ref{Eq Volume fraction flux}), respectively, and we can proceed to solve the component equation Eq.(\ref{Eq Component}).

\subsection{Scheme for the component equation}\label{Sec Scheme for the component equation} 
The scheme for solving the component equation Eq.(\ref{Eq Component}) is
\begin{equation}\label{Eq Component fully-discrete}
\frac{\gamma_t \chi_p^{M,n+1} C_p^{n+1} -\widehat{\chi_p^M C_p}}{\Delta t}
+
\tilde{\nabla} \cdot (\tilde{\mathbf{m}}_{\chi_p^M} \tilde{C}_p^{*,n+1})
=
\tilde{\nabla} \cdot ( D_p \tilde{\nabla} C_p^{n+1} ),
\quad 1 \leqslant p \leqslant M,
\end{equation} 
where $\{\chi_p^M\}_{p=1}^M$, $\{\tilde{\mathbf{m}}_{\chi_p^M}\}_{p=1}^M$, and $\{D_p\}_{p=1}^M$ are computed from Eq.(\ref{Eq chi_p^M}), Eq.(\ref{Eq chi_p^M flux}), and Eq.(\ref{Eq D_p}), respectively, using $\{\chi_p^{n+1}\}_{p=1}^N$ and $\{\tilde{\mathbf{m}}_{\chi_p}\}_{p=1}^N$ from Section \ref{Sec Scheme for the Phase-Field equation}. 

The 2nd-order backward difference scheme is applied in Eq.(\ref{Eq Component fully-discrete}) for the temporal discretization but $C^{n+1}$ is replaced by $C^{*,n+1}$ in the convection term. Note that the difference between $C^{n+1}$ and $C^{*,n+1}$ is $O(\Delta t^2)$, so the scheme is still formally 2nd-order accurate in time. The application of the 2nd-order spatial discretization results in formally 2nd-order accuracy of the scheme in space.
The scheme Eq.(\ref{Eq Component fully-discrete}) conserves the total amount of Component $p$ in its dissolvable region, i.e., $\sum_{i,j} [(\chi_p^M C_p)^{n+1} \Delta \Omega]_{i,j}=\sum_{i,j} [(\chi_p^M C_p)^{0} \Delta \Omega]_{i,j}$ for all $p$, if there is no source of the components at the boundary of the domain. After multiplying $\Delta \Omega$ to Eq.(\ref{Eq Component fully-discrete}) and then summing it over all the cells, both the convection and diffusion terms vanish due to the property of the discrete divergence operator, see the proof in \citep{Huangetal2020N,Huangetal2020}. As a result, $\sum_{i,j} [(\chi_p^M C_p) \Delta \Omega]_{i,j}$ doesn't change as time advances. 
The consistency of reduction in the discrete level is also satisfied by the scheme Eq.(\ref{Eq Component fully-discrete}). Suppose $C_p$ is zero up to time level $n$ and there is no source of the component at the boundary of the domain, then the convection term and $\widehat{\chi_p^M C_p}$ vanish. The resulting equation is a discretized homogeneous Helmholtz equation. Thus, $C_p$ remains to be zero at time level $n+1$. In other words, if Component $p$ is absent at $t=0$ and it doesn't have any sources at the boundary of the domain, it won't appear at $\forall t>0$. Therefore the consistency of reduction is satisfied.

The discrete component fluxes are
\begin{equation}\label{Eq Discrete component flux}
\tilde{\mathbf{m}}_{C_p}
=
\tilde{\mathbf{m}}_{\chi_p^M} \tilde{C}_p^{*,n+1}
-
D_p \tilde{\nabla} C_p^{n+1},
\quad
1 \leqslant p \leqslant M,
\end{equation}
and it also satisfies the consistency of reduction since it becomes zero if $C_p$ is zero up to time level $n+1$.

Once $\{\chi_p^{M,n+1} C_p^{n+1}\}_{p=1}^M$ and $\{\tilde{\mathbf{m}}_{C_p}\}_{p=1}^M$ are available, the density, viscosity and mass flux are computed from Eq.(\ref{Eq Density}), Eq.(\ref{Eq Viscosity}) and Eq.(\ref{Eq Mass flux}), respectively, and we can proceed to solve the momentum equation.

\subsection{Scheme for the momentum equation with the divergence-free constraint}\label{Sec Scheme for the momentum equation}
The scheme for the momentum equation with the divergence-free constraint is the same as those in \citep{Huangetal2020} and it has been successfully applied to two- and multi-phase incompressible immiscible flows \citep{Huangetal2020, Huangetal2020N, Huangetal2020CAC, Huangetal2020B}. It is a projection scheme with formally 2nd-order accuracy in both time and space, where the divergence-free constraint is enforced exactly by the cell-face velocity, i.e., $\tilde{\nabla} \cdot \mathbf{u}=0$. The momentum is conserved, i.e., $\sum_{i,j} [(\rho \mathbf{u})^n \Delta \Omega]_{i,j}=\sum_{i,j} [(\rho \mathbf{u})^0\Delta \Omega]_{i,j}$, if the interfacial tension is not counted, and the momentum conservation does not depend on the explicit form of the mass flux in the inertial term. The interfacial force, i.e., $\mathbf{f}_s$ in Eq.(\ref{Eq Surface force}), can be computed by either the balanced-force method, which achieves a better mechanical equilibrium in the discrete level, or the conservative method, which fully conserves the momentum equation \citep{Huangetal2020N}. The proof of the momentum conservation of the scheme is available in \citep{Huangetal2020N}. The scheme satisfies the consistency of reduction and recovers exactly the fully-discretized single-phase Navier-Stokes equation inside each pure phase if the discrete mass flux $\tilde{\mathbf{m}}$ is reduction consistent. This is true since both $\{\tilde{\mathbf{m}}_{\phi_p}\}_{p=1}^N$ and $\{\tilde{\mathbf{m}}_{C_p}\}_{p=1}^M$ are reduction consistent and the discrete mass flux is computed from them with Eq.(\ref{Eq Mass flux}). The consistency of mass and momentum transport requires that the mass flux in the inertial term of the momentum equation should be the one satisfying the consistency of mass conservation. This statement should be held at the discrete level. In the following, we show that the discrete mass flux $\tilde{\mathbf{m}}$ from Eq.(\ref{Eq Mass flux}) using $\{\tilde{\mathbf{m}}_{\phi_p}\}_{p=1}^N$ and $\{\tilde{\mathbf{m}}_{C_p}\}_{p=1}^M$ satisfies the consistency of mass conservation. 

Based on the definitions of volume fractions Eq.(\ref{Eq Volume fraction}) and volumetric fluxes Eq.(\ref{Eq Volume fraction flux}), the fully-discrete Phase-Field equation Eq.(\ref{Eq Phase-Field fully-discrete}), and the divergence-free constraint, we obtain
\begin{equation}\label{Eq chi_p fully-discrete}
\frac{\gamma_t \chi_p^{n+1}-\hat{\chi_p} }{\Delta t}
+
\tilde{\nabla} \cdot \tilde{\mathbf{m}}_{\chi_p}
=
\frac{1}{2} \left(
\underbrace{\frac{\gamma_t-\overbrace{\hat{1}}^{\gamma_t} }{\Delta t}}_{0}
+
\underbrace{\tilde{\nabla} \cdot \mathbf{u}}_{0}
+
\underbrace{ 
\frac{\gamma_t \phi_p^{n+1}-\hat{\phi_p} }{\Delta t}
+
\tilde{\nabla} \cdot \tilde{\mathbf{m}}_{\phi_p}
}_{0}
\right)
=
0,
\end{equation}
which is the discrete counterpart of Eq.(\ref{Eq chi_p}). As a result, the consistency of volume fraction conservation is satisfied in the discrete level by noticing that Eq.(\ref{Eq chi_p fully-discrete}) can be derived from Eq.(\ref{Eq Component fully-discrete}) with homogeneous $C_p$.
Based on the definition of the discrete component fluxes Eq.(\ref{Eq Discrete component flux}) and the fully-discrete component equation Eq.(\ref{Eq Component fully-discrete}), we obtain
\begin{equation}\label{Eq Component fully-discrete Conservation}
\frac{\chi_p^{M,n+1} C_p^{n+1}-\widehat{\chi_p^M C_p}}{\Delta t}
+
\tilde{\nabla} \cdot \tilde{\mathbf{m}}_{C_p}
=
\frac{\chi_p^{M,n+1} C_p^{n+1}-\widehat{\chi_p^M C_p}}{\Delta t}
+
\tilde{\nabla} \cdot (\tilde{\mathbf{m}}_{\chi_p^M} \tilde{C}_p^{*,n+1})
-
\tilde{\nabla} \cdot (D_p \tilde{\nabla} C_p^{n+1})
=
0,
\end{equation}
which is the discrete counterpart of Eq.(\ref{Eq Component Conservation}). Thanks to Eq.(\ref{Eq chi_p fully-discrete}) and Eq.(\ref{Eq Component fully-discrete Conservation}), we obtain
\begin{equation}\label{Eq Mass fully-discrete}
\frac{\gamma_t \rho^{n+1}-\hat{\rho}}{\Delta t}
+
\tilde{\nabla} \cdot \tilde{\mathbf{m}}
=
\sum_{p=1}^N 
\rho_p^\phi 
\underbrace{\left(  
\frac{\gamma_t \chi_p^{n+1}-\hat{\chi_p}}{\Delta t}
+
\tilde{\nabla} \cdot \tilde{\mathbf{m}}_{\chi_p} 
\right)}_{0}
+
\sum_{p=1}^M 
\rho_p^C  
\underbrace{\left(  
\frac{\gamma_t \chi_p^{M,n+1} C_p^{n+1}-\widehat{\chi_p^M C_p}}{\Delta t}
+
\tilde{\nabla} \cdot \tilde{\mathbf{m}}_{C_p}
\right)}_{0}
=0,
\end{equation}
which is the discrete counterpart of the mass conservation equation Eq.(\ref{Eq Mass}).
Therefore, the consistency of mass conservation is satisfied by the discrete mass flux $\tilde{\mathbf{m}}$, and thus, the consistency of mass and momentum transport is satisfied in the momentum equation as well. As discussed in Section \ref{Sec Galilean invariance}, a constant vector field should be an admissible solution of the momentum equation if the mechanical equilibrium Eq.(\ref{Eq Mechanical equilibrium}) is satisfied. This is held, independent of the cell size, the time step size, the numbers of phases and components, and the material properties of them, by the present scheme for the momentum equation, as long as the consistency of mass conservation and the consistency of mass and momentum transport are satisfied in the discrete level. The proof is available in \citep{Huangetal2020N}. 

\subsection{Summary of the scheme}\label{Sec Summary scheme}
The proposed scheme is decoupled, semi-implicit, and formally 2nd-order accurate in both time and space. It conserves the mass of individual pure phases, the total amount of the components inside their corresponding dissolvable regions, and, as a result, the total mass of the fluid mixture. The momentum is conserved if the conservative method for the interfacial forces is applied. All the consistency conditions, i.e., the consistency of reduction, the consistency of volume fraction conservation, the consistency of mass conservation, and the consistency of mass and momentum transport, are satisfied in the discrete level. The scheme also eliminates any generation of local void or overfilling. Therefore, the scheme is consistent and conservative. The Galilean invariance and the energy law of the discrete system are to be explored numerically in the next section. 

One of the advantages of the decoupled semi-implicit scheme is that one solves the complicated system part by part and only needs to solve linear systems in every part. This avoids solving a huge non-linear coupled system from a fully implicit scheme and is probably more efficient in a single time step. However, the scheme is only conditionally stable and the time step size $\Delta t$ is restricted. Since all the convection terms are treated explicitly, the CFL condition needs to be considered, i.e., $\Delta t \leqslant \Delta t_{\mathrm{CFL}} \sim \frac{h}{U_m}$ where $h$ is the grid size and $U_m$ is the scale of the maximum velocity. Another restriction comes from the surface tension \citep{Brackbilletal1992,Francoisetal2006}, i.e., $\Delta t \leqslant \Delta t_{\sigma} \sim \sqrt{\frac{h^3}{4\pi} \min (\frac{\rho_p+\rho_q}{\sigma_{p,q}})_{p \neq q}}$. The dominant part of the viscous term, i.e., $\nabla \cdot (\mu \nabla \mathbf{u})$, has been treated implicitly, which greatly removes the dependency of stability on the Reynolds number. As stated in \citep{FerzigerPeric2002}, explicitly treating the remaining part of the viscous term is reasonable because it contributes in only a minor way to the viscous force, and a more careful analysis is available in \citep{Badalassietal2003}. The aforementioned criteria have been commonly used in computational fluid dynamics, and interested readers can refer to \citep{FerzigerPeric2002,Brackbilletal1992,Francoisetal2006,Sussmanetal1994,Sussmanetal2007,Badalassietal2003,Dong2012,Dong2018}. The effect of $M_0^\phi$ in Eq.(\ref{Eq Mobility}) on the stability of the scheme has been discussed in \citep{Dong2018}, and it is observed that a large $M_0^\phi$ could lead to instability. The choice of $M_0^\phi$ in the present study is based on the practices in \citep{Dong2018,Huangetal2020N}, so that it won't be the dominant factor of numerical instability but provides reasonably good results. 
In addition to stability, accuracy is another important aspect when determining the time step size, since most interests are in the dynamical behavior of the multiphase and multicomponent problems. Therefore, a smaller time step size is necessary to produce accurate dynamical data.
The efficiency of the scheme is not a major concern at the moment. Most attention is paid on developing a scheme that honors the physical properties of the proposed model. 

\section{Results}\label{Sec Results}
In this section, we first investigate the convergence behaviors of the component equation Eq.(\ref{Eq Component}), derived from the diffuse domain approach \citep{Lietal2009}, and its numerical scheme Eq.(\ref{Eq Component fully-discrete}). The other part of the proposed scheme, i.e., the one excluding the component equation, has been analyzed and numerically validated in our previous works \citep{Huangetal2020,Huangetal2020CAC,Huangetal2020N,Huangetal2020B} and we don't repeat them here.
Then, we demonstrate that the model is capable of accurately capturing different circumstances of cross-interface transports of a component that is dissolvable in both sides of the interface.
With the help of the rising bubble problem in \citep{Hysingetal2009}, the ability of the proposed model and scheme to produce sharp-interface solution is demonstrated, the effect of the consistency of volume fraction conservation is illustrated, and the formal order of accuracy of the entire scheme and its convergence to sharp-interface solution is quantified.
The Galilean invariance in the discrete level is validated by a 4-phase-2-component setup including significant differences of material properties. 
The mass conservation, momentum conservation, and energy law at the discrete level are discussed with the horizontal shear layer. 
Finally, two problems are performed to demonstrate the capability of the proposed model and scheme in complicated and challenging problems including various critical factors of multiphase and multicomponent flows.

Unless otherwise specified, we chose $\eta=0.01$ and $M_0^\phi=10^{-7}$, use $h$ to denote the cell/grid size, apply the balanced-force method to discretize the interfacial force in the momentum equation, 
set the initial velocity to be zero,
and call $\chi_p^M C_p$, which represents the concentration of Component $p$ in its dissolvable region, the concentration of Component $p$ for convenience.

\subsection{Convergence tests for the diffuse domain approach}\label{Sec Convergence tests}
The formal order of accuracy of the scheme for the multiphase incompressible flows, including the Phase-Field equation Eq.(\ref{Eq Phase-Field}) and the momentum equation Eq.(\ref{Eq Momentum}), and the convergence behavior of its numerical solution to the sharp-interface one have been analyzed and numerically investigated in our previous works \citep{Huangetal2020,Huangetal2020CAC,Huangetal2020N,Huangetal2020B}. The scheme is formally 2nd-order accurate in both time and space, and the numerical solution of the model converges to the sharp-interface solution with an accuracy between 1.5th- to 2nd-order. The present case focuses on the newly introduced component equation Eq.(\ref{Eq Component}) derived from the diffuse domain approach \citep{Lietal2009}. Li et al. \citep{Lietal2009} used the matched asymptotic analysis based on the interfacial thickness $\eta$ and showed that the diffuse domain approach recovers, at the leading order, the original equation defined in a time-dependent domain and the boundary condition at that domain boundary. Their numerical results showed the convergence of the approach to the solution of the original problem. In this section, we demonstrate the convergence of the diffuse domain approach using a diffusion problem with an analytical solution.

The domain considered is $[1\times1]$ with periodic boundaries along the $x$ axis and with free-slip boundaries along the $y$ axis. The number of cells along each axis is ranging from $16$ to $256$. The number of time steps performed is $40$ for cell size $1/16$ and is doubled when the cell size is halved. As a result, the time step size is proportional to cell size. Two inviscid phases with matched density are considered. Phase 1 is at the bottom below $y_0=0.7$ and Phase 2 is at the top. Since our major focus is on the convergence behavior of the diffuse domain approach, the densities of both pure phases are $1$ and neither surface tension nor gravity is taken into account. Component 1, whose density and viscosity are both $1$, is dissolvable in Phase 1 only, with diffusivity $0.1$. Its concentration is zero initially, and at the bottom wall, its concentration is constantly equal to $1$. 

The concentration of Component 1 follows the diffusion equation with a Dirichlet boundary at $y=0$ and a homogeneous Neumann boundary at $y=y_0$. Such a diffusion problem can be solved by separation of variables and the exact solution is
\begin{equation}\label{Eq Convergence Exact}
C_1^S(y,t)=C_0 \left(1+\sum_{r=0}^\infty \frac{-2}{\lambda_r L} \exp \left( -D \lambda_r^2 t \right) \sin \left( \lambda_r (y-y_1) \right) \right),
\quad
\lambda_r=\frac{1+2r}{2L} \pi,
\end{equation}
where $y_1$ and $y_2$ are the locations of the Dirichlet and homogeneous Neumann boundaries, $L$ is the length of the domain, i.e., $L=y_2-y_1$, and $C_0$ is the concentration at the Dirichlet boundary. Specifically in the present case, $y_1=0$, $y_2=y_0=0.7$ and $C_0=1$. It should be noted that Eq.(\ref{Eq Convergence Exact}) works in $[y_1,y_0]$, which is $[0,0.7]$. The sharp interface solution of the concentration of Component 1 in $[0,1]$ is $H_1 C_1^S$, where $H_1$ is the exact indicator function of Phase 1 and it is $H(y_0-y)$ with $H(y)$ the Heaviside function. However, in the diffuse domain approach, the exact indicator function $H_1$ is replaced by the smoothed indicator function $\chi_1^M$, which is $\frac{1}{2} \left( 1+\tanh\left( \frac{y_0-y}{\sqrt{2}\eta} \right) \right)$ in our case. As a result, the best solution to be expected from the diffuse domain approach is $\chi_1^M C_1^S$, and we call it the semi-sharp-interface solution. In the following, we are going to compare the numerical results to both the sharp-interface solution (S) and the semi-sharp-interface solution (SS) under different circumstances. 

\begin{figure}[!t]
	\centering
	\includegraphics[scale=.5]{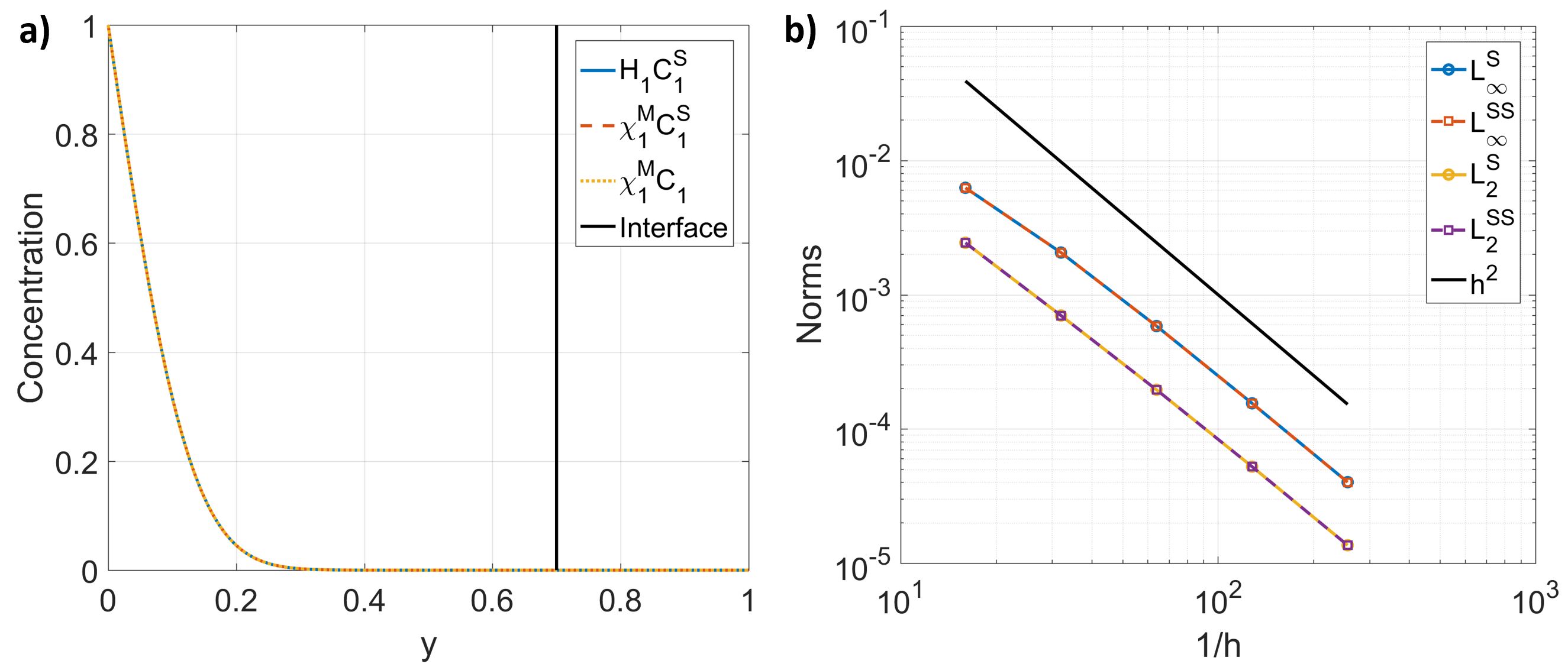}
	\caption{Results of the convergence tests at $t=0.05$. a) Concentration profiles. b) Errors for different grid sizes. \label{Fig Convergence-005} }
\end{figure}

In the first case, we consider the moment at $t=0.05$, at which the edge of the boundary layer of the concentration is far away from the phase interface. Thus, the phase interface should have no effect on the solution, and the sharp-interface and semi-sharp-interface solutions are identical, i.e., $H_1 C_1^S=\chi_1^M C_1^S$. Under this circumstance, the accuracy of the numerical solution should be the same as its formal order of accuracy, which is 2nd-order accurate in both time and space. Our results, shown in Fig.\ref{Fig Convergence-005} match the analysis very well. 
Fig.\ref{Fig Convergence-005} \textbf{a)} shows the concentration profiles of the sharp-interface solution $H_1 C_1^S$, the semi-sharp-interface solution $\chi_1^M C_1^S$ and the numerical solution $\chi_1^M C_1$, and they overlap with each other. The edge of the boundary layer of the concentration is at about $y=0.3$, which is far enough away from the phase interface at $y=0.7$ to avoid its effect on the solution. For a clear presentation, we only include the result from $256$ cells in each direction. Fig.\ref{Fig Convergence-005} \textbf{b)} shows the convergence behavior, quantified by the $L_\infty$ and $L_2$ errors. The point-wise error is defined as the numerical solution minus the sharp-interface or the semi-sharp-interface solution. The $L_\infty$ error is the maximum of the absolute point-wise error and the $L_2$ error is the root-mean-square of the point-wise error. The superscripts ``S'' and ``SS'' refer to the sharp-interface and semi-sharp-interface solutions, respectively. We can observe that both $L_\infty$ and $L_2$ errors are converging to zero with 2nd-order accuracy, as expected. The differences between $L_\infty^S$ and $L_\infty^{SS}$ and between $L_2^S$ and $L_2^{SS}$ are indistinguishable, which implies that the sharp-interface solution is the same as the semi-sharp-interface solution. Since the time step size is propositional to the cell size, we can conclude that the scheme is formally 2nd-order accurate in both time and space.

\begin{figure}[!t]
	\centering
	\includegraphics[scale=.5]{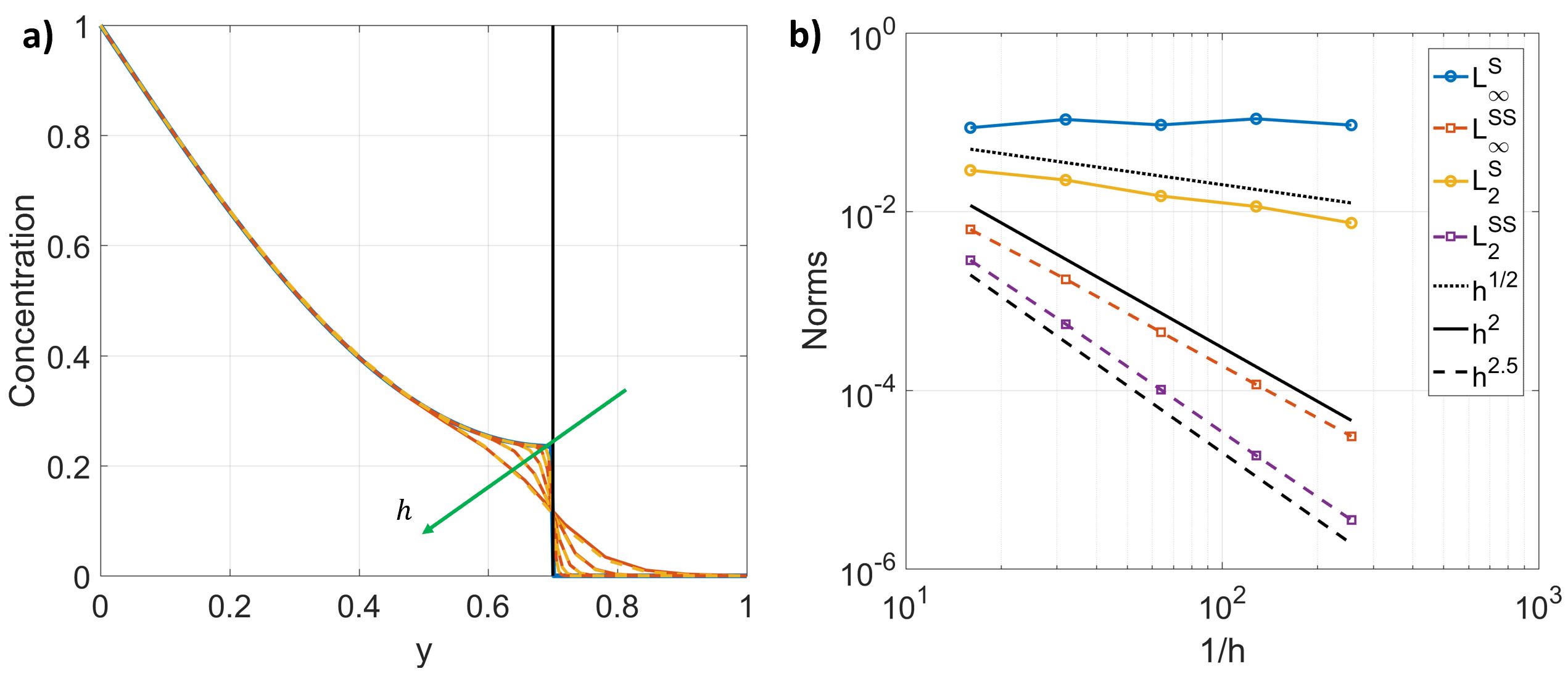}
	\caption{Results in the convergence tests at $t=1$. a) Profiles of the concentrations, blue solid line: sharp-interface solution, red solid lines: semi-sharp-interface solutions with different cell sizes, yellow dashed lines: numerical solutions with different cell sizes, black solid line: interface. b) Errors for different grid sizes. \label{Fig Convergence-100} }
\end{figure}

\begin{figure}[!t]
	\centering
	\includegraphics[scale=.5]{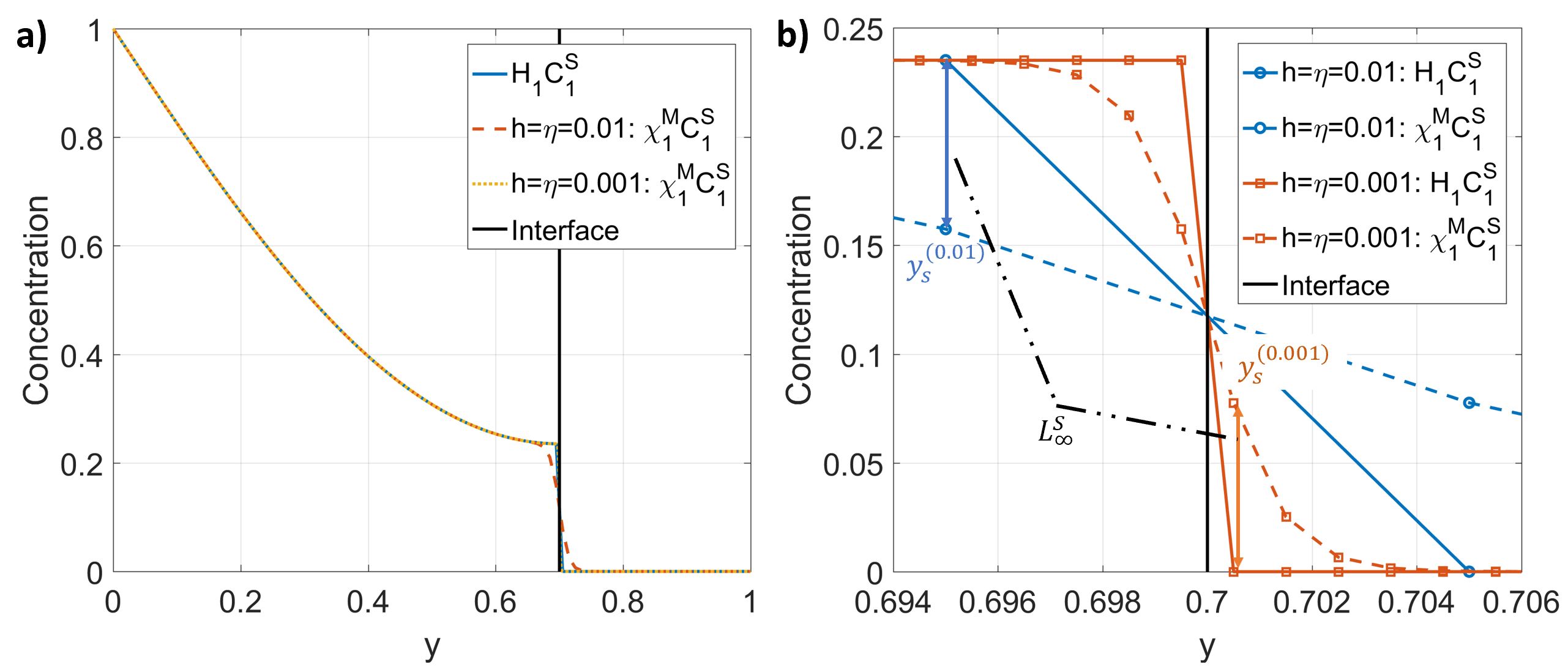}
	\caption{The sharp-interface and semi-sharp-interface solutions with $h=\eta=0.01$ and $h=\eta=0.001$ in the convergence tests at $t=1$. a) Concentration profiles. b) Concentration profiles close to the interface. \label{Fig Convergence-100-SS} }
\end{figure}

\begin{figure}[!t]
	\centering
	\includegraphics[scale=.5]{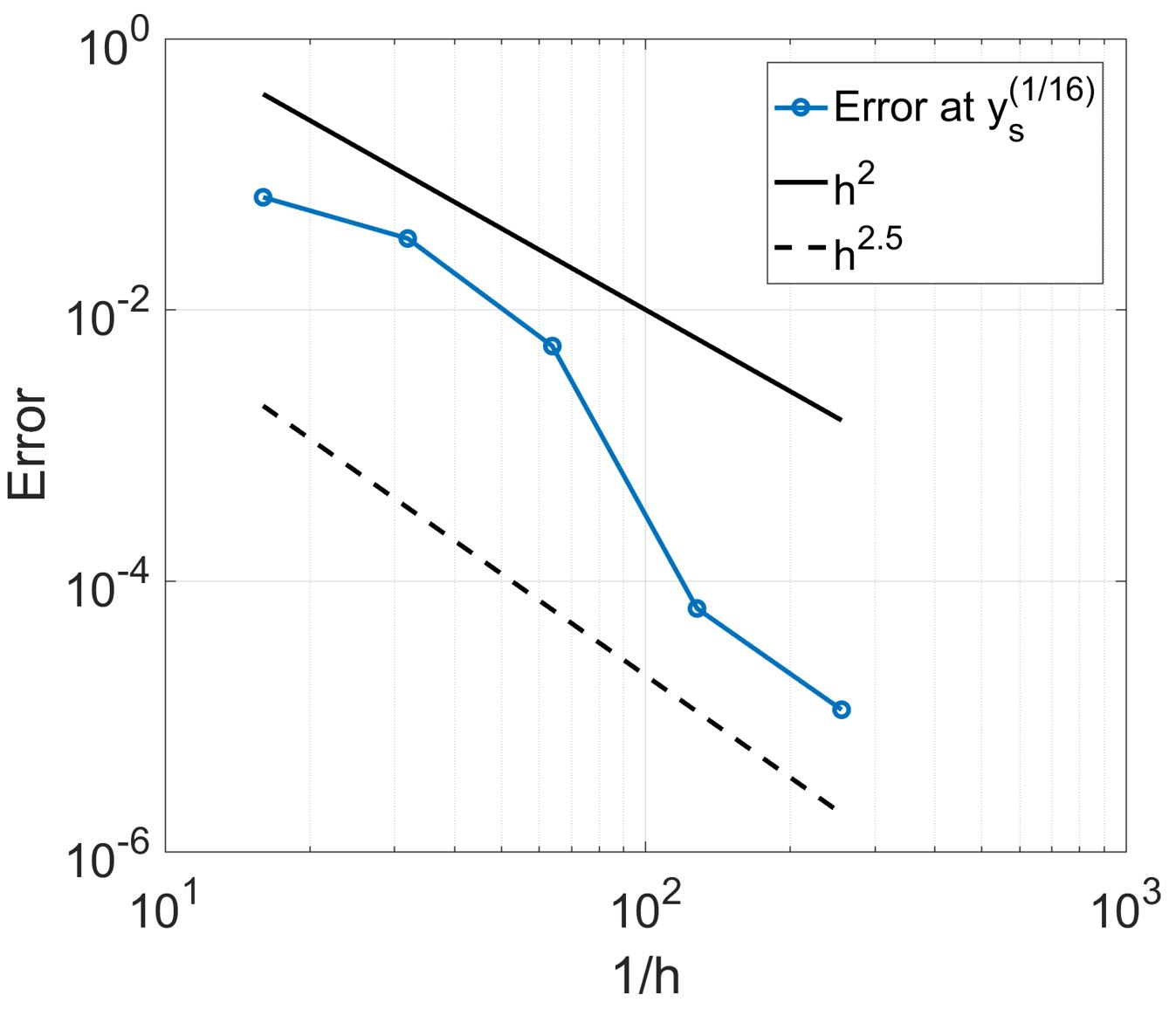}
	\caption{The point-wise errors between the numerical and sharp-interface solutions at $y_s^{(1/16)}$ in the convergence tests at $t=1$. \label{Fig Convergence-100-Point} }
\end{figure}

In the second case, we consider the moment at $t=1$, at which the edge of the boundary layer of the concentration has reached the phase interface. As a result, the sharp-interface solution and the semi-sharp-interface solution are different. We chose $\eta$ equal to $h$, so that the numerical solution converges to the semi-sharp-interface and to the sharp-interface solutions simultaneously as $h$ tends to zero. Fig.\ref{Fig Convergence-100} \textbf{a)} shows the profiles of the sharp-interface solution, the semi-sharp-interface solution, and the numerical solutions with different cell sizes. 
It is clear that the numerical solution, as well as the semi-sharp-interface solution, is approaching the sharp-interface solution, as $h$ and $\eta$ are getting smaller. Fig.\ref{Fig Convergence-100} \textbf{b)} shows the norms of the errors and we observe that $L_\infty^{SS}$ is converging to zero with 2nd-order accuracy and $L_2^{SS}$ converges even faster with 2.5th-order accuracy. However, $L_\infty^{S}$ doesn't converge to zero while the convergence rate of $L_2^{S}$ is very slow at about 0.5th-order accuracy. Since the numerical solution converges to the semi-sharp-interface solution, the difference between the semi-sharp-interface solution and the sharp-interface solution dominantly contributes to the convergence behaviors of $L_\infty^S$ and $L_2^S$. However, those behaviors of $L_\infty^S$ and $L_2^S$ are not consistent with what is observed in Fig.\ref{Fig Convergence-100} \textbf{a)}. To explain this, we use Fig.\ref{Fig Convergence-100-SS}, where the sharp-interface solution and the semi-sharp-interface solutions with $h=\eta=0.01$ and with $h=\eta=0.001$ are shown. 
It is obvious that the difference between the sharp-interface solution and the semi-sharp-interface solution majorly appears at the interfacial region, and the difference is much smaller after $h$ and $\eta$ are $10$ times smaller. However, when we compute the $L_\infty$ errors between the two solutions under the corresponding cell sizes, they are $0.07764$ for $h=\eta=0.01$ and $0.07762$ for $h=\eta=0.001$, which doesn't change much. After zooming into the interfacial region shown in Fig.\ref{Fig Convergence-100-SS} \textbf{b)}, the maximum difference between the sharp-interface and semi-sharp-interface solutions is at the grid point closest to the interface and its location is denoted by $y_s^{(h)}$. Although the point-wise error at $y_s^{(0.01)}$ with $h=\eta=0.001$ is smaller than that with $h=\eta=0.01$, $y_s^{(0.001)}$ is closer to the interface than $y_s^{(0.01)}$. Consequently, the point-wise error at $y_s^{(0.001)}$ is larger than that at $y_s^{(0.01)}$ with $h=\eta=0.001$, and is similar to that at $y_s^{(0.01)}$ with $h=\eta=0.01$. As a result, the $L_\infty$ error between the sharp-interface and semi-sharp-interface solutions doesn't change much as $h$ and $\eta$ reduce. Due to the behavior observed in Fig.\ref{Fig Convergence-100-SS}, we consider the convergence of the point-wise error between the numerical and sharp-interface solutions at a specific location. Fig.\ref{Fig Convergence-100-Point} shows the point-wise error at $y_s^{(1/16)}$, and by using this measure, the error converges to zero with at least 2nd-order accuracy.

\subsection{Two diffusion problems}\label{Sec Two diffusion problems}
Two diffusion problems are performed to demonstrate that the proposed model is capable of modeling different circumstances of cross-interface transports of a component that is dissolvable in both sides of the interface.
The domain considered is $[1\times1]$ with periodic boundaries along the $x$ axis and free-slip boundaries along the $y$ axis. The domain is discretized by $128\times128$ cells and the time step size is $\Delta t=10^{-3}$. Phase 1, whose pure phase density is $1000$ and viscosity is $10^{-3}$, is at the bottom below $y=0.3$, while the rest of the domain is filled by Phase 2, whose pure phase density and viscosity are $1$ and $2\times10^{-5}$, respectively. The surface tension between the phases is $0.0728$ and no gravity is considered. Component 01 with density $50$ and viscosity $10^{-4}$ is dissolvable in both of the phases with diffusivity $0.02$ inside Phase 1 and $0.08$ inside Phase 2. Initially, there is no Component 01 inside the domain, while its concentrations are $1$ at the bottom wall and $0.1$ at the top wall. 

In the first case, Component 01 is unable to cross the phase interface between Phases 1 and 2. As discussed in Section \ref{Sec Summary model}, we need to define Component 1 and Component 2 such that Component 1 is dissolvable only in Phase 1 and Component 2 is dissolvable only in Phase 2. Consequently, the concentration of Component 01 is represented by $\chi_{01}^M C_{01}=\chi_1^M C_1 + \chi_2^M C_2$. The sharp-interface steady solution is a step function jumping from $1$ to $0.1$ at $y=0.3$.
In the second case, Component 01 is able to cross the phase interface between Phases 1 and 2. Then we only need to define a single component, i.e., Component 1, sharing the same material properties with Component 01 and dissolving in both Phases 1 and 2. As a result, the concentration of Component 01 is $\chi_{01}^M C_{01}=\chi_1^M C_1$. The sharp-interface steady solution is two straight segments with different slopes connecting at $y=0.3$, i.e.,
\begin{equation}
C^S(y)=\left\{
\begin{array}{ll}
-\frac{36}{19}y+1, \quad y \leqslant 0.3\\
-\frac{9}{19}(y-1)+0.1, \quad y \geq 0.3
\end{array}
\right.
\end{equation}

\begin{figure}[!t]
	\centering
	\includegraphics[scale=.5]{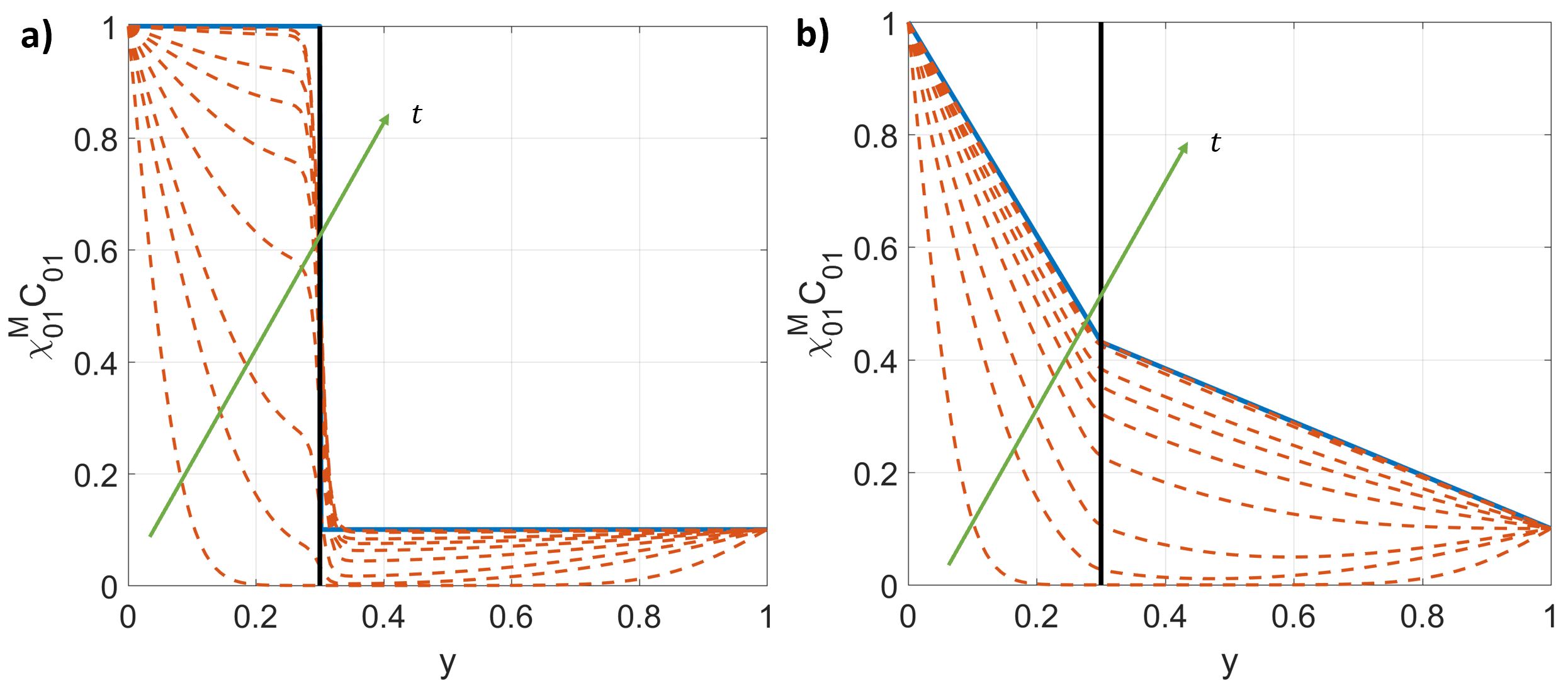}
	\caption{Profiles of the concentrations, blue solid line: the sharp-interface steady solution, red dashed lines: the numerical solution at $t=0.1,0.5,1,2,3,4,5,8,10$, black solid line: interface. a) results of the first case of the two diffusion problems. b) results of the second case of the two diffusion problems.  \label{Fig Two-Diffusion} }
\end{figure}

The profiles of $\chi_{01}^M C_{01}$ at selected moments in the two cases are shown in Fig.\ref{Fig Two-Diffusion}, along with the corresponding sharp-interface steady solutions. 
As time goes on, the numerical solutions successfully approach the corresponding sharp-interface steady solutions, even though the background phases have a density ratio of $1000$ and a viscosity ratio of $50$. Due to the diffuse domain approach, we can observe that the profiles change smoothly across the interface instead of a sharp transition. The results in this section demonstrate that the proposed model is capable of modeling whether a component is allowed to cross a phase interface when it is dissolvable in both sides of the interface.

\subsection{Rising bubble}\label{Sec Rising bubble}
In this section, the rising bubble problem in \citep{Hysingetal2009} is considered to demonstrate the capability of the proposed model and scheme of producing sharp-interface solutions, to illustrate the effect of the proposed consistency of volume fraction conservation, and to quantify the convergence of the entire scheme.
The domain considered is $[1\times2]$, and the top and bottom boundaries are no-slip while the right and left ones are free-slip. A circular bubble is initially at $(0.5,0.5)$ with a radius $0.25$, while the other part of the domain is filled with a liquid. The bubble has a density $1$ and a viscosity $0.1$, while the density and viscosity of the liquid are $1000$ and $10$, respectively. Their surface tension is $1.96$, and the gravity is $0.98$, pointing downward. The motion of the bubble is quantified by the trajectory of its center, i.e., $y_c=\int_{\Omega} y \chi_b d\Omega/\int_{\Omega} \chi_b d\Omega$, and the rising speed, i.e., $v_c=\int_{\Omega} v \chi_b d\Omega/\int_{\Omega} \chi_b d\Omega$, where $\chi_b$ is the volume fraction of the bubble and $v$ is the velocity along the $y$ direction. Here, the mid-point rule is used to compute the integrals. Additional two components are added. Component 1, having a density $0.5$ and viscosity $0.05$, is only dissolvable in Phase 1 with a diffusivity $0.01$. Component 2 is only dissolvable in Phase 2 with a diffusivity $0.05$ and has a density $1600$ and viscosity $14$. 

\begin{figure}[!t]
	\centering
	\includegraphics[scale=.45]{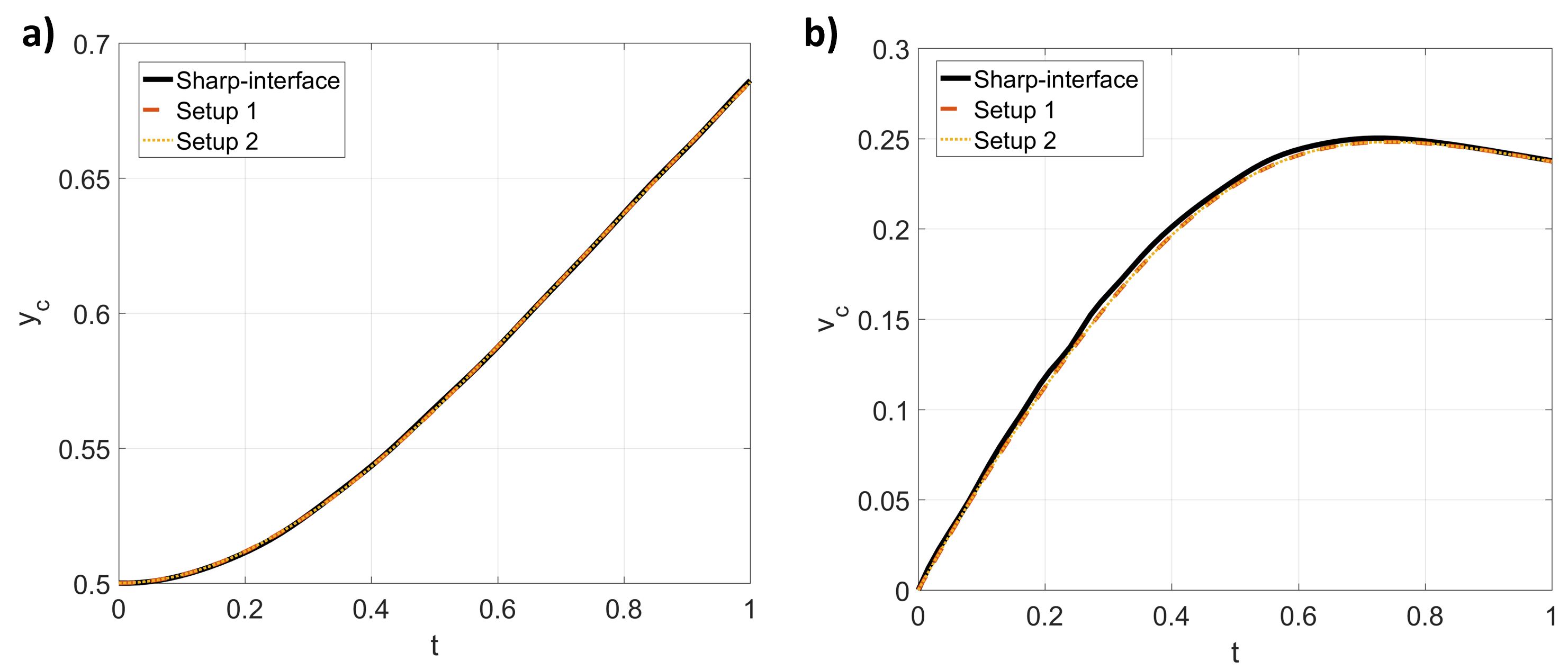}
	\caption{Results of the rising bubble problem under Setups 1 and 2. a) The trajectory of the bubble center versus time. b) The rising speed of the bubble versus time. The sharp-interface results are from \citep{Hysingetal2009}. \label{Fig RB} }
\end{figure}

We first compare the results from the proposed model and scheme with the sharp-interface one in \citep{Hysingetal2009} using the following two setups.
\begin{itemize}
\item \textit{Setup 1:}
The bubble is represented by Phase 1, while the liquid is Phase 2. The components are absent, by giving their concentrations to be zeros at $t=0$. Therefore, the densities, viscosities, and surface tension of the phases are the same as the given values for the bubble and liquid, and $\chi_b$ equals $\chi_1$. The domain is discretized by a grid size of $h=\frac{1}{128}$. The time step size is $\Delta t=0.128h$ and $\eta$ is the same as $h$.

\item \textit{Setup 2:} 
The bubble is represented by Phase 1 with Component 1 having a homogeneous concentration 1 dissolved in it. On the other hand, the liquid is composed of Phase 2 and Component 2 with a homogeneous concentration 0.5. Thus, the concentrations are initially $1$ and $0.5$ for Components 1 and 2, respectively. The material properties of Phase 1 become $0.5$ for the density and $0.05$ for the viscosity, while they are $200$ for the density and $3$ for the viscosity of Phase 2. The rest is the same as Setup 1.
\end{itemize}
As pointed out in Section \ref{Sec Summary model}, any pure phase can be modeled as another pure phase dissolving components with homogeneous concentrations. For example, one can model a salt water as a pure phase or as a solution of water, which is the pure phase (or the background fluid), and salt, which is the component, with a homogeneous concentration. Therefore, Setups 1 and 2 should produce the same results. Fig.\ref{Fig RB} shows the results from Setups 1 and 2, along with the sharp-interface results from \citep{Hysingetal2009}. The results from both setups are indistinguishable from each other and agree well with the sharp-interface results. In addition, under Setup 1, the concentrations of the components, which are zero initially, are always zero during the computation. This confirms the consistency of reduction of the model and scheme, as discussed in Sections \ref{Sec Consistency analysis} and \ref{Sec Scheme for the component equation}, respectively.

\begin{figure}[!t]
	\centering
	\includegraphics[scale=.6]{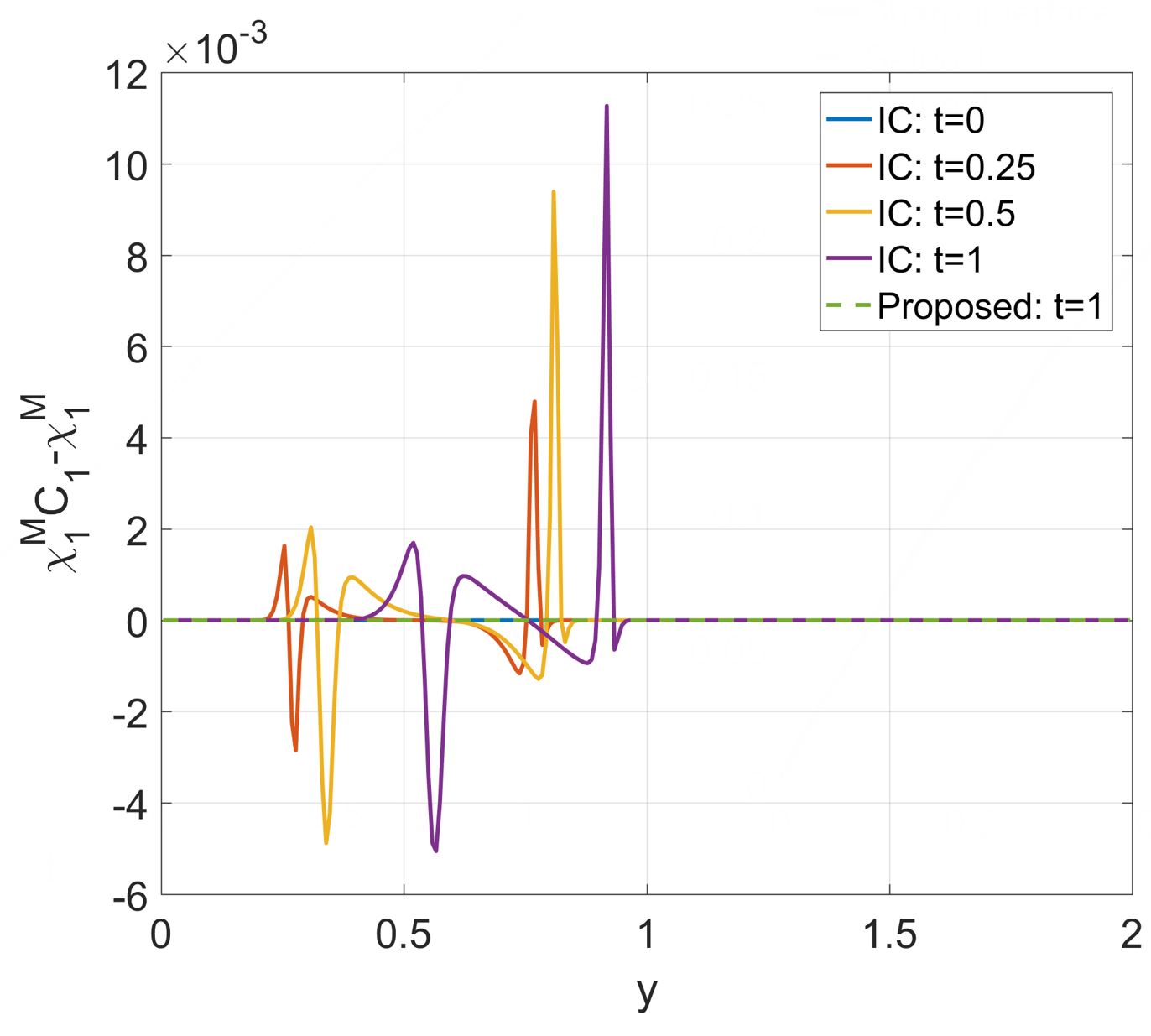}
	\caption{Differences between $\chi_1^M C_1$ and $\chi_1^M$ under Setup 2 at $x=0.5$. IC: the concentrations are solved from Eq.(\ref{Eq Component H}) with $\{H_p^M=\chi_p^M\}_{p=1}^M$. Proposed: the concentrations are solved from the proposed Eq.(\ref{Eq Component}) that satisfies the consistency of volume fraction conservation. \label{Fig RB-Profile} }
\end{figure}

Next, the effect of the newly proposed consistency condition, i.e., the consistency of volume fraction conservation is considered. The effects of the other consistency conditions have been discussed in, e.g.,  \citep{Huangetal2020,Huangetal2020CAC,Huangetal2020N,Huangetal2020B} for the consistencies of mass conservation and mass and momentum transport, and \citep{BoyerMinjeaud2014,LeeKim2015,Dong2017,Dong2018,Abadi2018,Huangetal2020N,Huangetal2020B} for the consistency of reduction. Setup 2 is repeated but the concentrations of the components are solved from Eq.(\ref{Eq Component H}), instead of the proposed Eq.(\ref{Eq Component}), and $\{H_p^M\}_{p=1}^M$ are directly replaced by $\{\chi_p^M\}_{p=1}^M$, like the original diffuse domain approach \citep{Lietal2009}, without considering the consistency of volume fraction conservation. Results from solving Eq.(\ref{Eq Component H}) are labeled as ``IC'' which means inconsistent. Since Setups 1 and 2 are equivalent, the profile of $\chi_p^M C_p$ should be the same as $\chi_p^M (C_p|_{t=0})$. Fig.\ref{Fig RB-Profile} shows the difference between $\chi_1^M C_1$ and $\chi_1^M$ at $x=0.5$ at different moments. Due to violating the consistency of volume fraction conservation, it is observed from the ``IC'' results that the concentration of Component 1 is suffering from unphysical oscillations around the boundary of the bubble. Even worse, the oscillations are growing with time, which probably becomes an origin of numerical instability in highly dynamical problems. On the other hand, the difference is less than $10^{-11}$ at $t=1$ from the proposed consistent model and scheme. The concentration of Component 2 behaviors in a similar manner, and therefore they are not shown here.

\begin{figure}[!t]
	\centering
	\includegraphics[scale=.45]{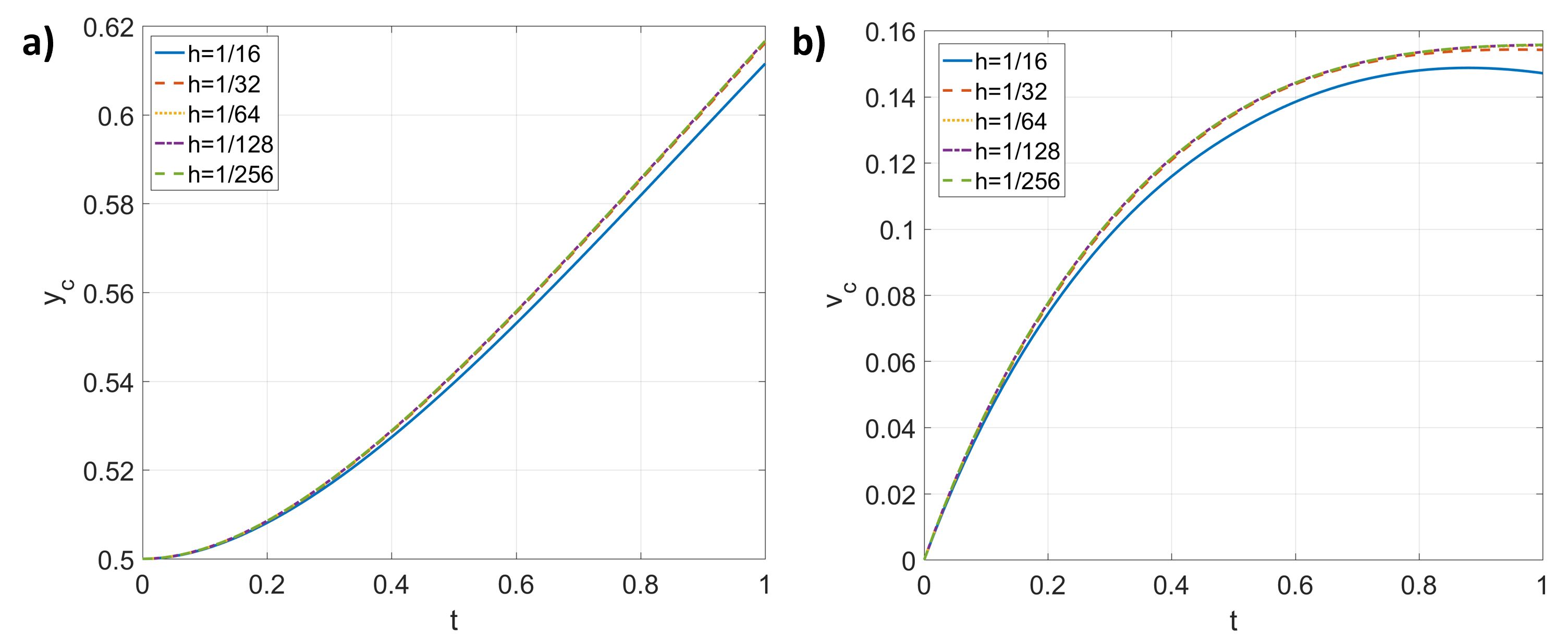}
	\includegraphics[scale=.45]{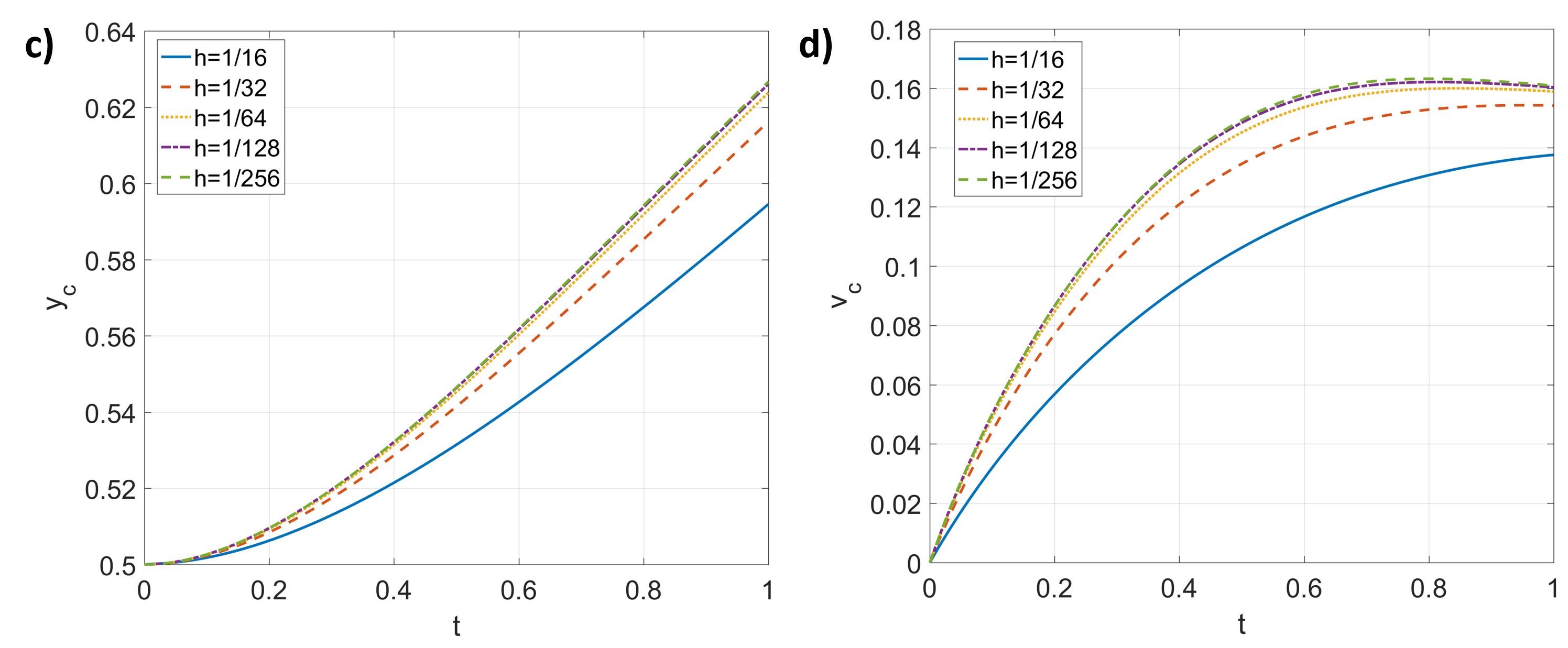}
	\caption{Results of the rising bubble problem under Setups 3 and 4 using different grid sizes. a) The trajectory of the bubble center versus time under Setup 3. b) The rising speed of the bubble versus time under Setup 3. c) The trajectory of the bubble center versus time under Setup 4. d) The rising speed of the bubble versus time under Setup 4.\label{Fig RB-Convergence} }
\end{figure}

\begin{figure}[!t]
	\centering
	\includegraphics[scale=.6]{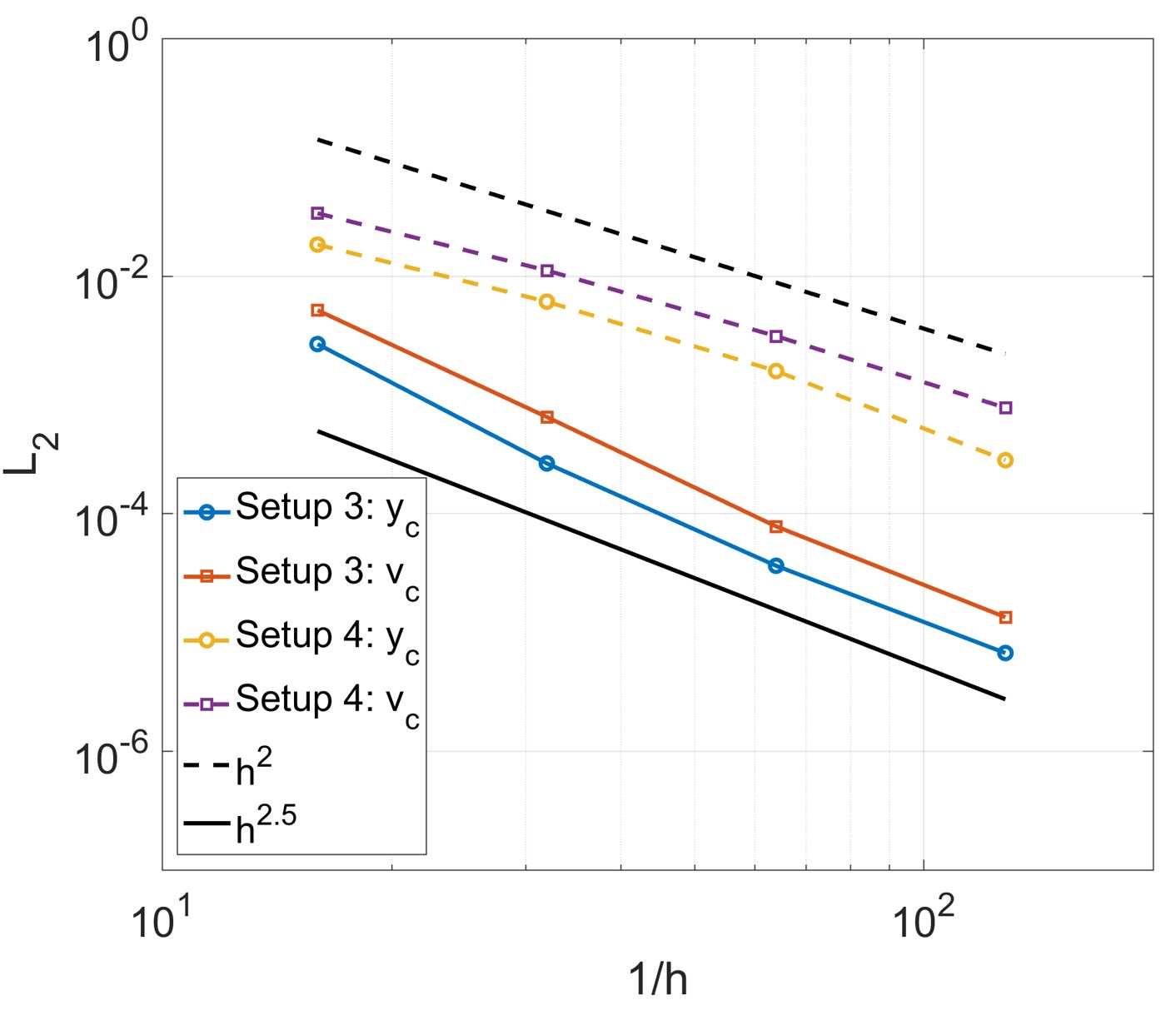}
	\caption{The $L_2$ norms of $y_c$ and $v_c$ minus their corresponding reference solutions form the finest grid versus the grid sizes under Setups 3 and 4. \label{Fig RB-L2} }
\end{figure}

Finally, the convergence of the entire scheme is considered using Setups 3 and 4, modified from Setup 2, such that the concentrations of the components are not homogeneous initially. 
\begin{itemize}
\item \textit{Setup 3}: 
The concentration of Component 1 is unity inside a circle at $(0.5,0.5)$ with a radius $0.1$ but zero elsewhere initially. The concentration of Component 2 is $0.5$ above $y=0.85$ but zero elsewhere initially. The grid size is ranging from $\frac{1}{16}$ to $\frac{1}{256}$, and $\eta$ is fixed to be $\frac{1}{32}$. The solution from the finest grid is considered as the reference solution.
\item \textit{Setup 4}: 
The same as Setup 3 but $\eta$ is changing with the grid size, i.e., $\eta=h$.
\end{itemize}
The solutions from different grid sizes in Setup 3 are approximating the same exact solution of the model with a fixed $\eta$. Therefore, its convergence behavior represents the formal order of accuracy of the proposed scheme. On the other hand, the solutions in Setup 4 converge to a series of solutions that approach the one with zero $\eta$, i.e., the sharp-interface solution. It usually converges slower than the formal order of accuracy of the scheme but is more related to practical applications. Fig.\ref{Fig RB-Convergence} shows the results of Setups 3 and 4, and convergences of the results are evident after successively refining the grid size under both setups. Fig.\ref{Fig RB-L2} shows the $L_2$ norms, i.e., the root mean square, of $y_c$ and $v_c$ minus their corresponding reference solutions from the finest grid. The convergence under Setup 3 is close to but better than 2nd order, which validates the formally 2nd-order accuracy of the proposed scheme. The convergence in Setup 4 is a little slower than that in Setup 3 but still close to 2nd order. Similar studies have been performed using the multiphase model, i.e., the one proposed in the present work without the component part, in \citep{Huangetal2020N,Huangetal2020B}, and the present results are consistent with those studies, as well as those in Section \ref{Sec Convergence tests} of the present work.

The present case demonstrates that the proposed model and scheme are able to produce sharp-interface solutions. The proposed consistency of volume fraction conservation is essential to eliminate unphysical fluctuations of components around phase interfaces. The formal order of accuracy of the entire scheme is 2nd-order, and the convergence behavior towards the sharp interface solution is reasonably satisfactory with a rate close to 2nd order.

\subsection{Validation of the Galilean invariance}\label{Sec Validation Galilean invariance}
In this section, we perform a numerical experiment to validate that the numerical results from the proposed scheme is Galilean invariant.
The domain considered is $[1\times1]$ with double-periodic boundaries. The domain is discretized by $[100\times100]$ cells and the time step size is $\Delta t=10^{-4}$. We consider four phases and two components. 
Phase 1, whose pure phase density is $10000$ and viscosity is $10^{-3}$, is inside a circle of radius $0.15$ at $(0.3,0.7)$. 
Phase 2, whose pure phase density is $1000$ and viscosity is $10^{-1}$, is inside a circle of radius $0.1$ at $(0.75,0.6)$. 
Phase 3, whose pure phase density is $100$ and viscosity is $10^{-2}$, is inside a horizontal channel between $0.2 \leqslant y \leqslant 0.4$. 
Phase 4, whose pure phase density is $1$ and viscosity is $10^{-4}$, fills the rest of the domain. 
The interfacial tensions are $\sigma_{1,2}=0.04$, $\sigma_{1,3}=0.0728$, $\sigma_{2,3}=0.055$, $\sigma_{1,4}=0.04$, $\sigma_{2,4}=0.055$, and $\sigma_{3,4}=0.055$, and no gravity is considered.
Component 1, whose density is $5$ and viscosity is $10^{-3}$, is dissolvable in Phases 2 and 4, with diffusivity $10^{-3}$ and $10^{-2}$, respectively. It is initially inside a circle of radius $0.05$ at $(0.5,0.55)$ with a homogeneous concentration 1.
Component 2, whose density is $1$ and viscosity is $10^{-2}$, is dissolvable in Phases 3 and 4, with diffusivity $10^{-1}$ and $10^{-2}$, respectively. It is initially inside a circle of radius $0.05$ at $(0.4,0.3)$ with a homogeneous concentration 1.
It should be noted that the density ratio considered in the present case is $10,000$ and the viscosity ratio is about $1,000$.

The first observer is fixed with the domain of interest and the velocities are zero initially. Since the interfaces are either circular or plenary, they should neither move nor deform. As a result, the first observer observes diffusions of Components 1 and 2. The configurations observed by the first observer is labeled as ``S''.
The second observer moves with respect to the domain of interest at a constant velocity $-\mathbf{u}_0=-\{1,0\}$, which is equivalent to give $\mathbf{u}_0$ as the initial velocity in this case. The second observer observes both diffusions of Components 1 and 2, and translations of all the interfaces without deformations. The configurations observed by the second observer is labeled as ``V''. With given $\mathbf{u}_0$, the two configurations ``S'' and ``V'' should be identical at $t=1$ due to the Galilean invariance.

\begin{figure}[!t]
	\centering
	\includegraphics[scale=.45]{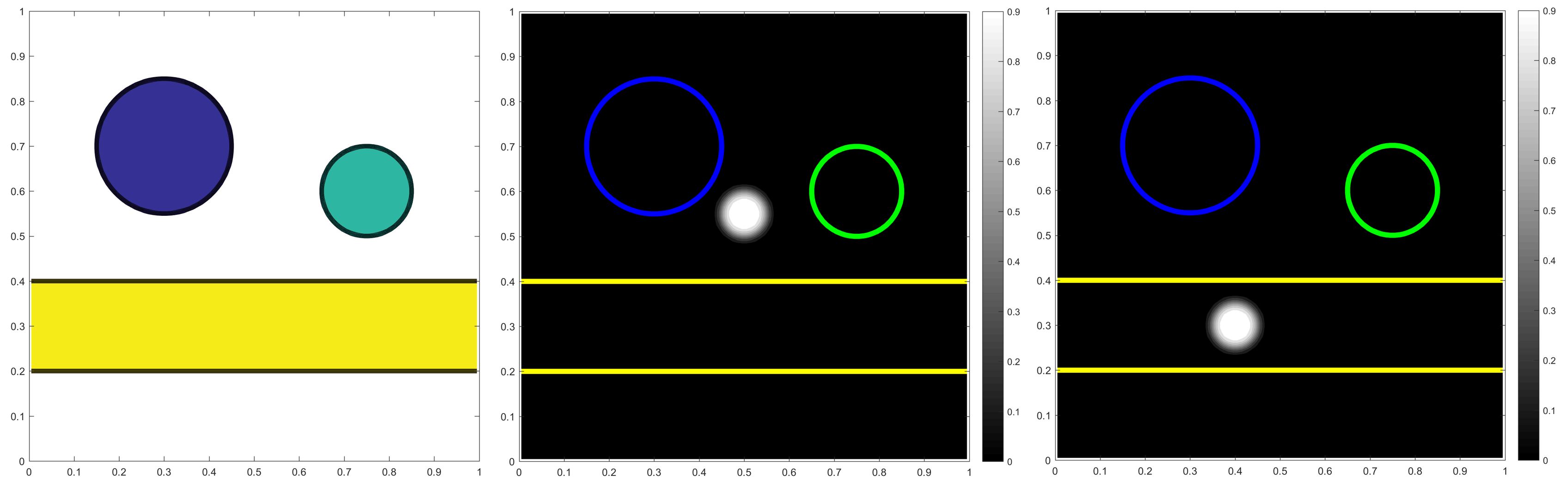}
	\includegraphics[scale=.45]{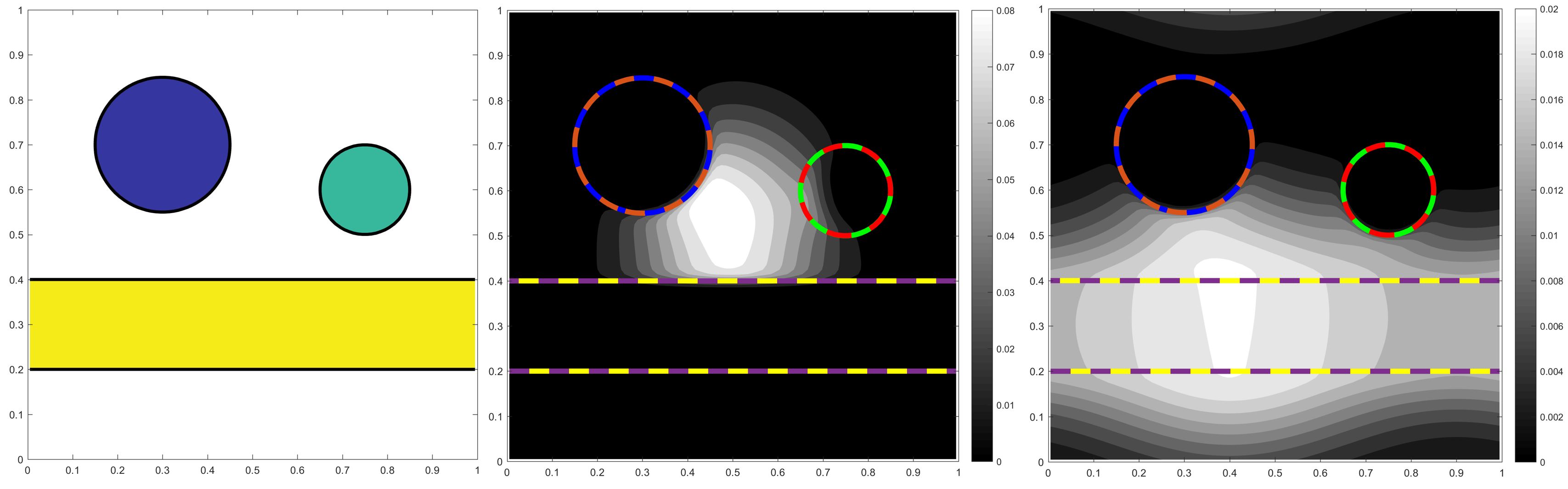}
	\includegraphics[scale=.45]{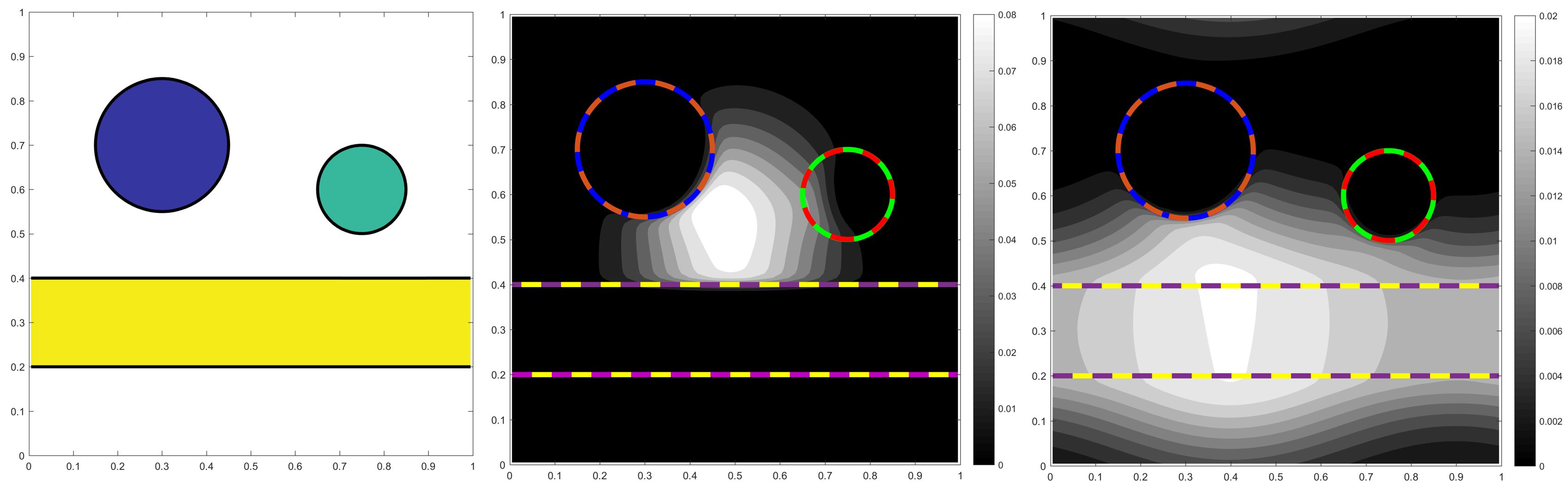}
	\caption{Results in the validation of Galilean invariance. Top row: configurations at $t=0$. Middle row: configuration S at $t=1$. Bottom row: configuration V at $t=1$. Left column: conflagrations of the 4 phases, blue: Phase 1, green: Phase 2, yellow: Phase 3, and White: Phase 4. Middle column: configuration of Component 1. Right column: configuration of Component 2. In the middle and right columns, blue solid line: interface of Phase 1 at $t=0$, orange dashed line: interface of Phase 1 at $t=1$, green solid line: interface of Phase 2 at $t=0$, red dashed line: interface of Phase 2 at $t=1$, yellow solid line: interface of Phase 3 at $t=0$, purple dashed line: interface of Phase 3 at $t=1$.   \label{Fig Galilean-Configurations} }
\end{figure}

Fig.\ref{Fig Galilean-Configurations} shows the configurations of different phases and components at $t=0$ and at $t=1$ observed by the two observers. 
In configuration ``S'', the interfaces neither move nor deform, and the components are diffusing only in their corresponding dissolvable regions. The locations of the interfaces at $t=1$ is on top of those at $t=0$. Component 1, which is initially in Phase 4 only, has diffused into Phase 2 but not to Phases 1 and 3, since Component 1 is not dissolvable in Phases 1 and 3. Component 2, which is dissolvable in Phases 3 and 4, has diffused across the interface between Phases 3 and 4 but not to Phases 1 and 2.
In configuration ``V'', the interfaces at $t=1$ return to their original locations without any deformation, and the behaviors of the components are the same as those in configuration ``S''. The results in configuration ``V'' has little difference from those in configuration ``S''.

\begin{figure}[!t]
	\centering
	\includegraphics[scale=.45]{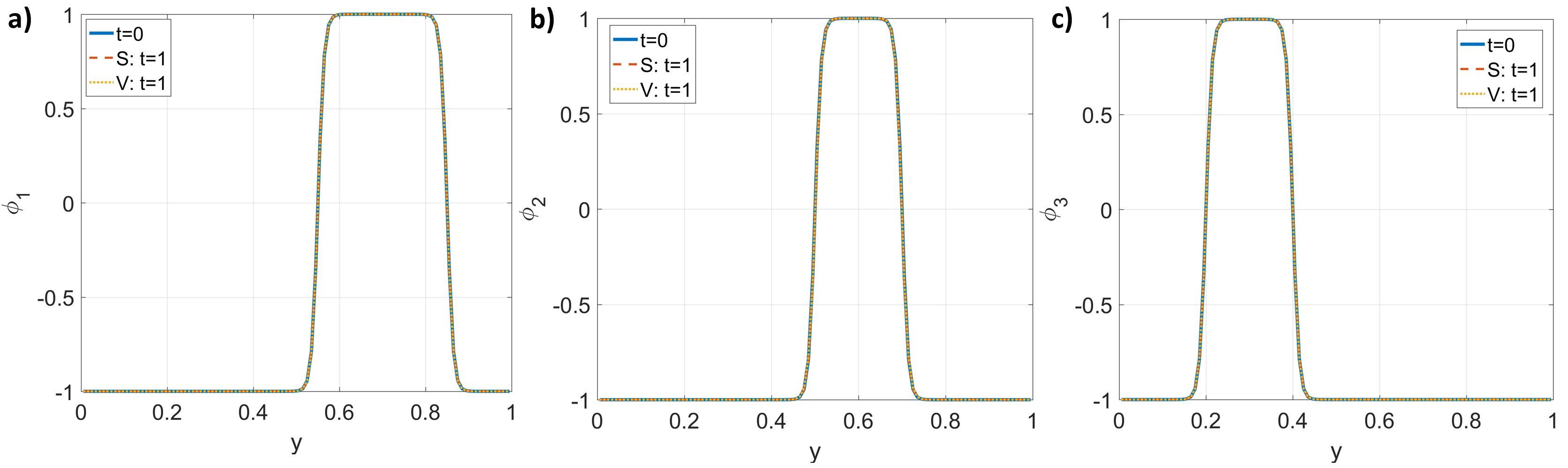}
	\caption{Profiles of the 3 phases in the validation of Galilean invariance. a) Profiles of Phase 1 at $x=0.3$. b) Profiles of Phase 2 at $x=0.75$. c) Profiles of Phase 3 at $x=0.9$.\label{Fig Galilean-Profile-Phase} }
\end{figure}

\begin{figure}[!t]
	\centering
	\includegraphics[scale=.5]{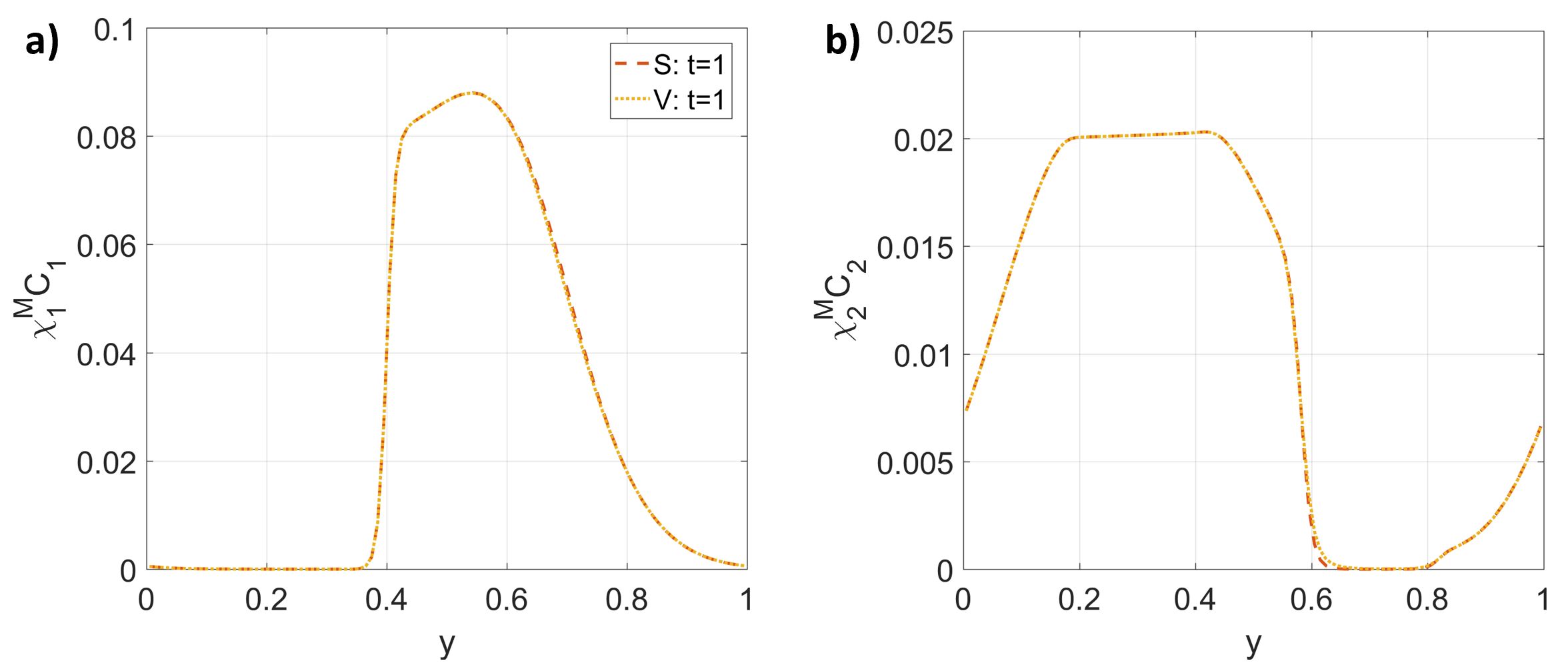}
	\caption{Profiles of the 2 components in the validation of Galilean invariance. a) Profiles of Component 1 at $x=0.5$. b) Profiles of Component 2 at $x=0.7$. \label{Fig Galilean-Profile-Component} }
\end{figure}

Fig.\ref{Fig Galilean-Profile-Phase} and Fig.\ref{Fig Galilean-Profile-Component} further demonstrate the Galilean invariance of the results. Fig.\ref{Fig Galilean-Profile-Phase} shows the profiles of Phase 1 at $x=0.3$, Phase 2 at $x=0.75$, and Phase 3 at $x=0.9$, at $t=0$ and $t=1$ from both configurations ``S'' and ``V''. The corresponding profiles are overlapping with each other. Fig.\ref{Fig Galilean-Profile-Component} shows the profiles of Component 1 at $x=0.5$ and Component 2 at $x=0.4$ at $t=1$ from both configurations ``S'' and ``V''. Again, the corresponding profiles are indistinguishable from each other. Consequently, the Galilean invariance is preserved by the scheme.

\subsection{Horizontal shear layer}\label{Sec Horizontal shear layer}
We discuss the mass conservation, momentum conservation, and energy law using the result from the horizontal shear layer.
The domain considered is $[1\times1]$ with double-periodic boundaries. The domain is discretized by $[128\times128]$ cells, and the time step size is $\Delta t=0.1h$. 
Phase 1, whose pure phase density is $10$ and viscosity is $10^{-2}$, is initially in $0.25 \leqslant y \leqslant 0.75$. 
Phase 2, whose pure phase density is $1$ and viscosity is $10^{-3}$, fills the rest of the domain. 
The surface tension between them is $0.1$ and no gravity is considered. $\eta$ is $\frac{1}{30\sqrt{2}}$ in this case.
Component 1, whose density is $20$ and viscosity is $10^{-3}$, is dissolvable in Phase 2 only, with diffusivity $10^{-3}$. It initially occupies $0 \leqslant y \leqslant 0.125$ and $0.875 \leqslant y \leqslant 1$, with a homogeneous concentration 1.
The horizontal velocity is initially $1$ inside Phase 1 while it is $-1$ inside Phase 2. A sinusoidal vertical velocity perturbation is added, whose amplitude is 0.05, and wavelength is $1$. In this case, we use both the balanced-force method (B) and the conservative method (C) to discretize the interfacial force. 

\begin{figure}[!t]
	\centering
	\includegraphics[scale=.5]{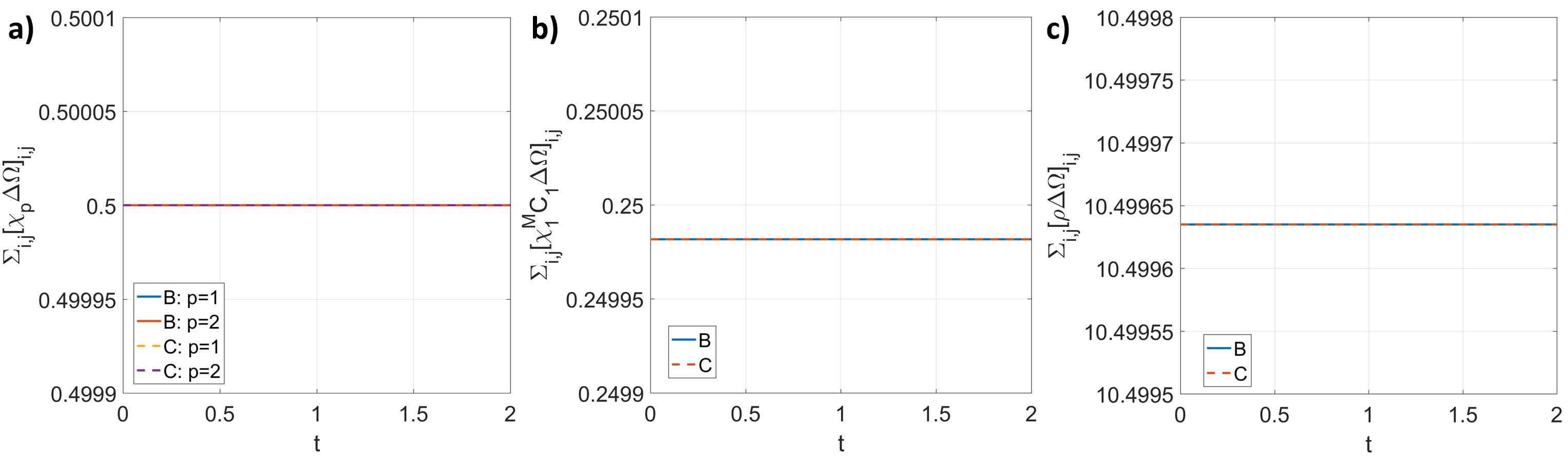}
	\caption{Time histories of the quantities related to mass conservation in the horizontal shear layer. a) Time histories of total volumes of individual phases. b) Time histories of the total amount of Component 1 in its dissolvable region. c) Time histories of the total mass of the fluid mixture. \label{Fig HSL-Mass} }
\end{figure}

Fig.\ref{Fig HSL-Mass} shows the time histories of the total volumes of individual phases, the total amount of Component 1 inside its dissolvable region, and the total mass of the fluid mixture. It is clear that all the quantities shown in Fig.\ref{Fig HSL-Mass} have little change as time goes on, and we've checked that the changes of them with respect to their initial values are below $10^{-12}$, which matches our analysis in Section \ref{Sec Scheme for the component equation} and in \citep{Huangetal2020N,Huangetal2020B}. The results demonstrate that the proposed scheme conserves the mass of individual pure phases, i.e., $\left\{\rho_p^\phi\sum_{i,j}[\chi_p \Delta \Omega]_{i,j}\right\}_{p=1}^N$, conserves the amount of each component in its dissolvable region, i.e., $\left\{\sum_{i,j}[ \chi_p^M C_p \Delta \Omega]_{i,j}\right\}_{p=1}^M$ , and, thus, conserves the mass of the fluid mixture, i.e., $\sum_{i,j}[\rho \Delta \Omega]_{i,j}=\sum_{p=1}^N \rho_p^\phi \sum_{i,j}[\chi_p \Delta \Omega]_{i,j}+ \sum_{p=1}^M \rho_p^C \sum_{i,j}[\chi_p^M C_p \Delta \Omega]_{i,j}$ from Eq.(\ref{Eq Density}), in the discrete level. 

\begin{figure}[!t]
	\centering
	\includegraphics[scale=.5]{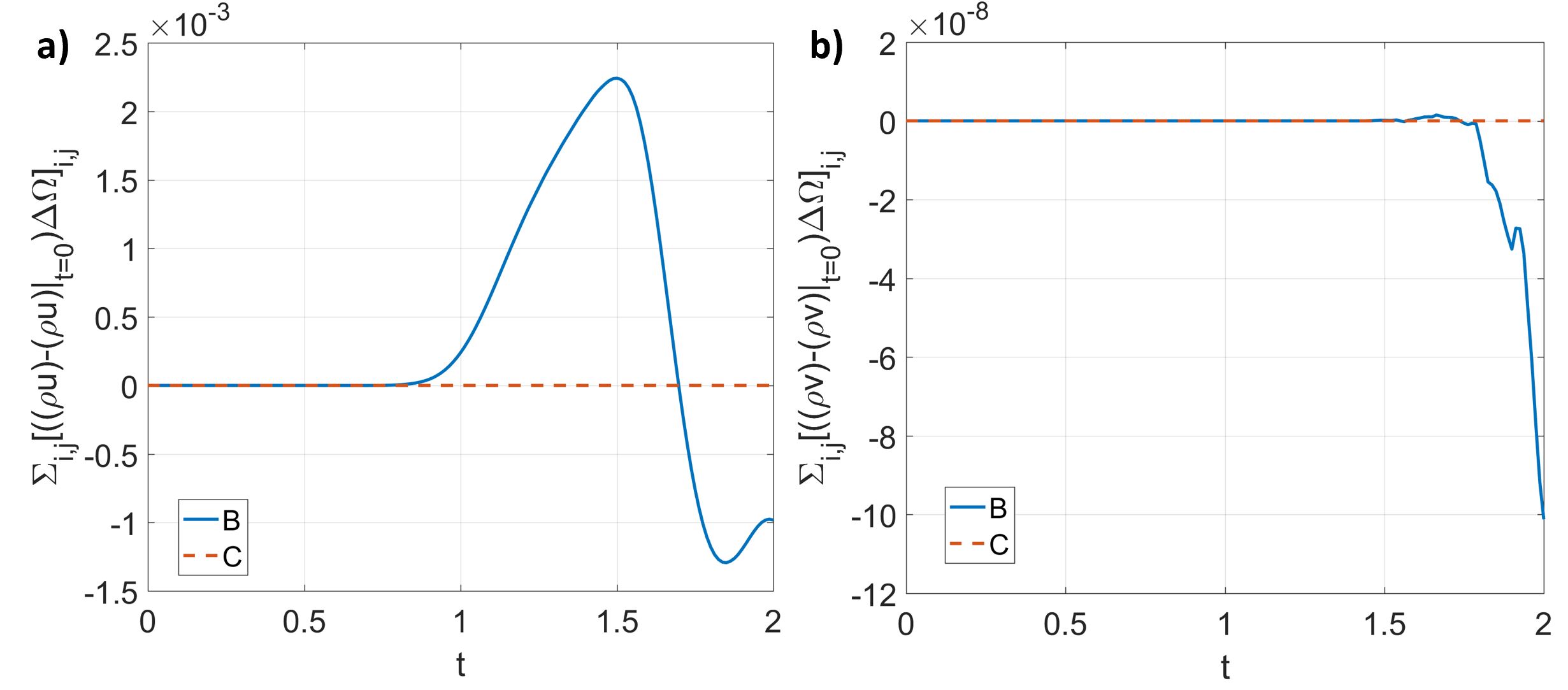}
	\caption{Time histories of the momentum in the horizontal shear layer. a) Time histories of the $x$ component of the momentum. b) Time histories of the $y$ component of the momentum. \label{Fig HSL-Momentum} }
\end{figure}

Fig.\ref{Fig HSL-Momentum} shows the time histories of the changes of the momentum. Using the conservative method, the momentum is conserved at the discrete level. On the other hand, using the balanced-force method doesn't conserve the momentum exactly. However, the change of the momentum is small, which is less than $0.25\%$ of its initial value in this case. Therefore, the momentum is essentially conserved by the balanced-force method. The obtained results are consistent with our previous work in \citep{Huangetal2020,Huangetal2020CAC,Huangetal2020N,Huangetal2020B} for multiphase flows without components, and they demonstrate that including components doesn't change the property of momentum conservation of the scheme.

\begin{figure}[!t]
	\centering
	\includegraphics[scale=.5]{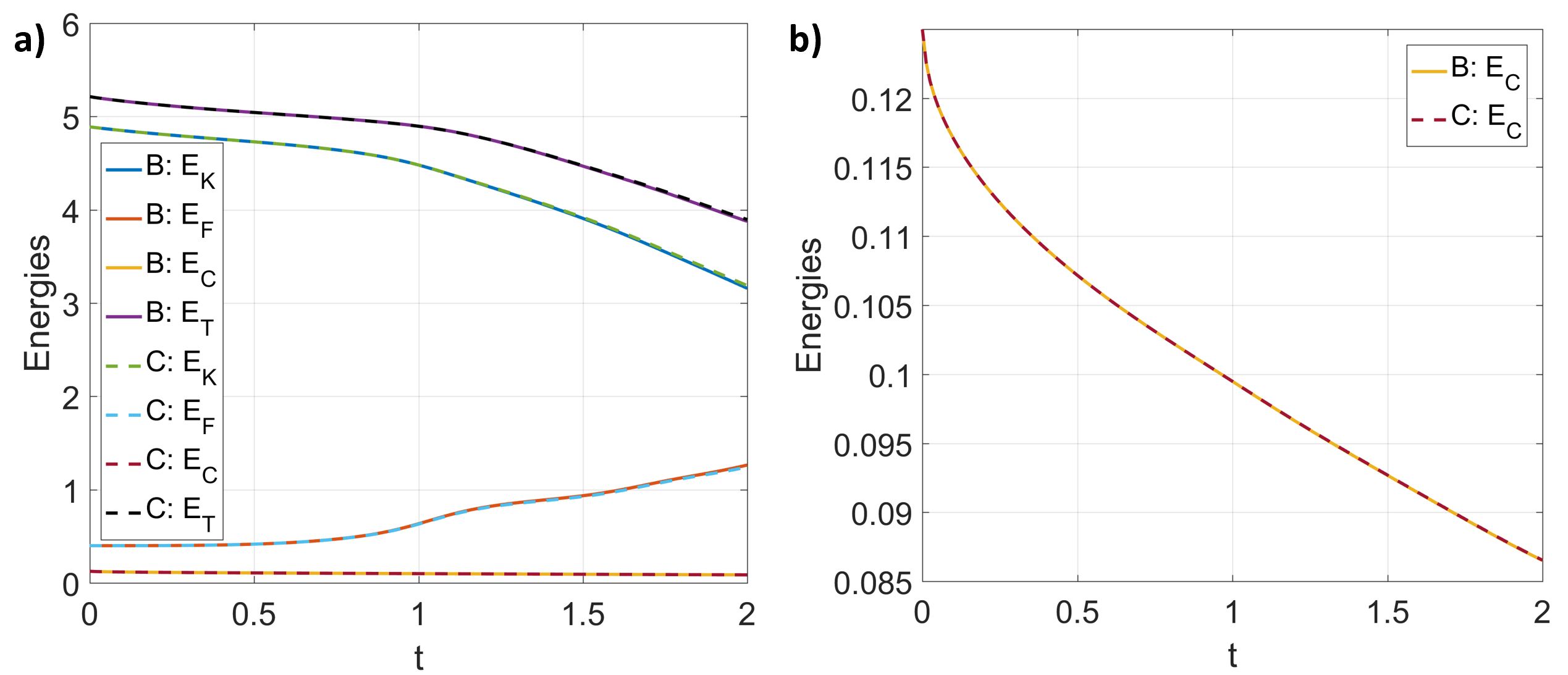}
	\caption{Time histories of the energies in the horizontal shear layer. a) Time histories of the kinetic energy, free energy, component energy, and total energy. b) Time histories of the component energy. \label{Fig HSL-Energy} }
\end{figure}

Fig.\ref{Fig HSL-Energy} shows the time histories of the kinetic energy $E_K$, the free energy $E_F$, the component energy $E_C$ and the total energy $E_T=E_K+\frac{1}{2}E_F+E_C$. Both the balanced-force method and the conservative method give similar results. In this case, the contribution from the component energy is negligibly small. We can observe from Fig.\ref{Fig HSL-Energy} \textbf{a)} that the kinetic energy is decaying due to both the viscous effect and the energy transfer to the free energy. Therefore, the free energy is increased. The total energy is always decaying, demonstrating that the energy law Eq.(\ref{Eq Total energy}) is honored at the discrete level by the scheme. Fig.\ref{Fig HSL-Energy} \textbf{b)} shows that the component energy is also decaying all the time.

\subsection{Miscible falling drop}\label{Sec Miscible falling drop}
We consider a liquid drop, initially in the air, falling down into another liquid at the bottom, and gradually mixing with the bottom liquid.
This is a three-phase case where the liquid drop (Phase 01) is miscible with the other bottom liquid (Phase 02) while the air (Phase 03) is not miscible with either of the liquids (Phases 01 or 02). 
The densities and viscosities of Phases 01, 02, and 03 are $\rho_{01}=3000\mathrm{kg/m^3}$, $\mu_{01}=1.5\times10^{-3}\mathrm{Pa \cdot s}$, $\rho_{02}=1000\mathrm{kg/m^3}$, $\mu_{02}=1\times10^{-3}\mathrm{Pa \cdot s}$, $\rho_{03}=1\mathrm{kg/m^3}$, and $\mu_{03}=2\times10^{-5}\mathrm{Pa \cdot s}$. 
The surface tension between the immiscible pairs is $0.0728\mathrm{N/s}$ and the diffusivity between the miscible pair is $1\times10^{-5}\mathrm{m^2/s}$. The gravity is pointing downward with a magnitude $9.8\mathrm{m/s^2}$. 

As mentioned in Section \ref{Sec Summary model}, this three-phase case can be turned into a 2-phase-1-component setup, where Phase 01 is represented by Phase 1 with $1\mathrm{mol/m^3}$ Component 1, Phase 02 is represented by pure Phase 1, Phase 03 is represented by pure Phase 2, and Component 1 is only dissolvable in Phase 1, i.e., $\{I_{p,q}^M\}=\{1,0\}$. As a result, we have $\rho_1^\phi=1000\mathrm{kg/m^3}$, $\mu_1^\phi=1\times10^{-3}\mathrm{Pa \cdot s}$, $\rho_2^\phi=1\mathrm{kg/m^3}$,  $\mu_2^\phi=2\times10^{-5}\mathrm{Pa \cdot s}$, $\sigma_{1,2}=0.0728\mathrm{N/s}$, $\rho_1^C=2000\mathrm{kg/mol}$, $\mu_1^C=5\times10^{-4}\mathrm{Pa \cdot s \cdot m^3/mol}$, $D_{1,1}=1\times10^{-5}\mathrm{m^2/s}$. The governing equations are non-dimensionalized by a length scale $0.01\mathrm{m}$, a density scale $1\mathrm{kg/m^3}$, an acceleration scale $1\mathrm{m/s^2}$, and a concentration scale $1\mathrm{mol/m^3}$. Consequently, Phase 01 is represented by Phase 1 with unity concentration of Component 1 in the results reported.

The domain considered is $[1 \times 1]$ with periodic boundaries along the $x$ axis and with free-slip boundaries along the $y$ axis. The domain is discretized by $[128\times128]$ cells and the time step size is $\Delta t=10^{-4}$. Initially, the circular drop of Phase 1 is at $(0.5, 0.75)$ with a radius $0.15$, and there is Component 1 with a homogeneous unity concentration inside it. The bottom of the domain below $y=0.3$ is filled with pure Phase 1. The rest of the domain is occupied by Phase 2. 

\begin{figure}[!t]
	\centering
	\includegraphics[scale=.2]{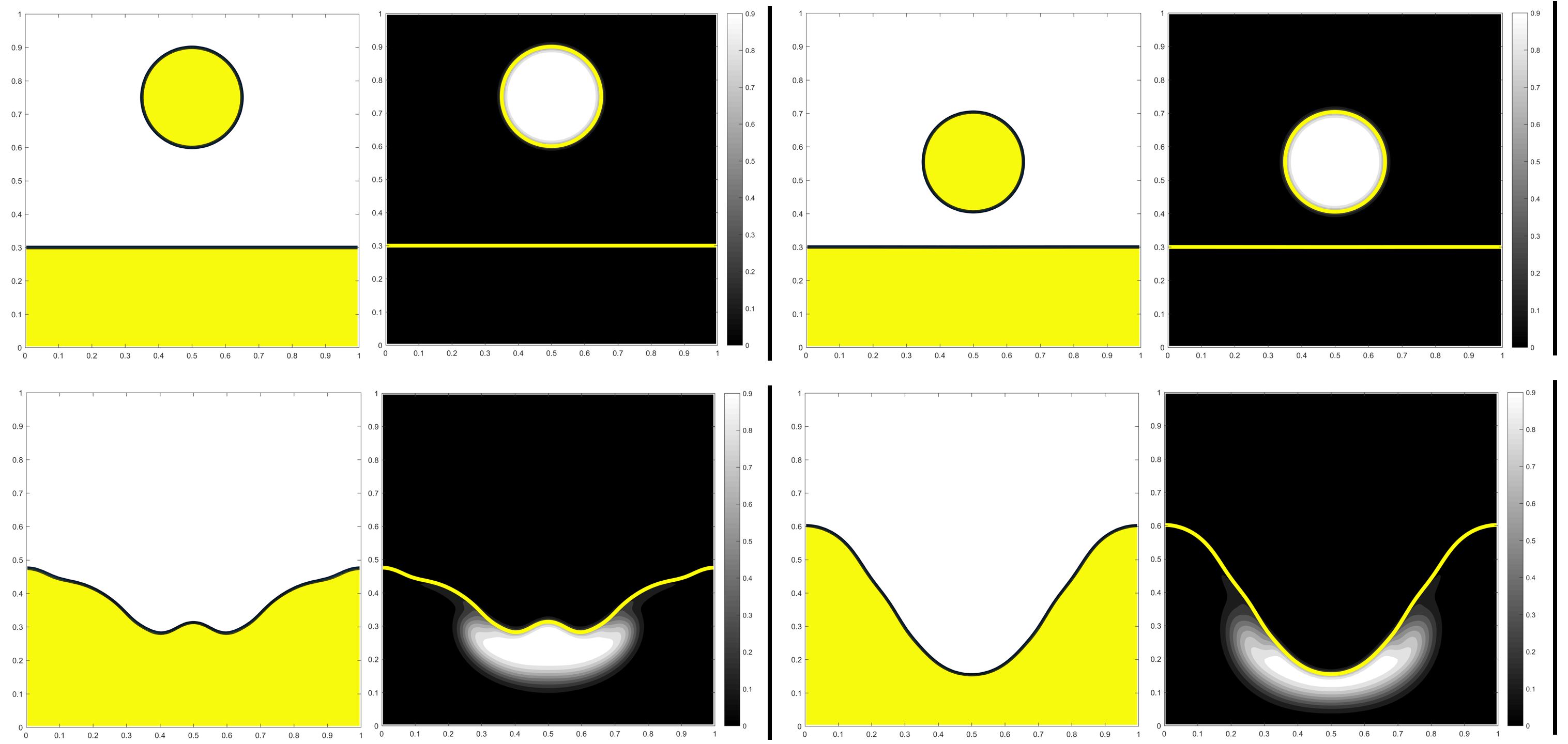}
	\includegraphics[scale=.2]{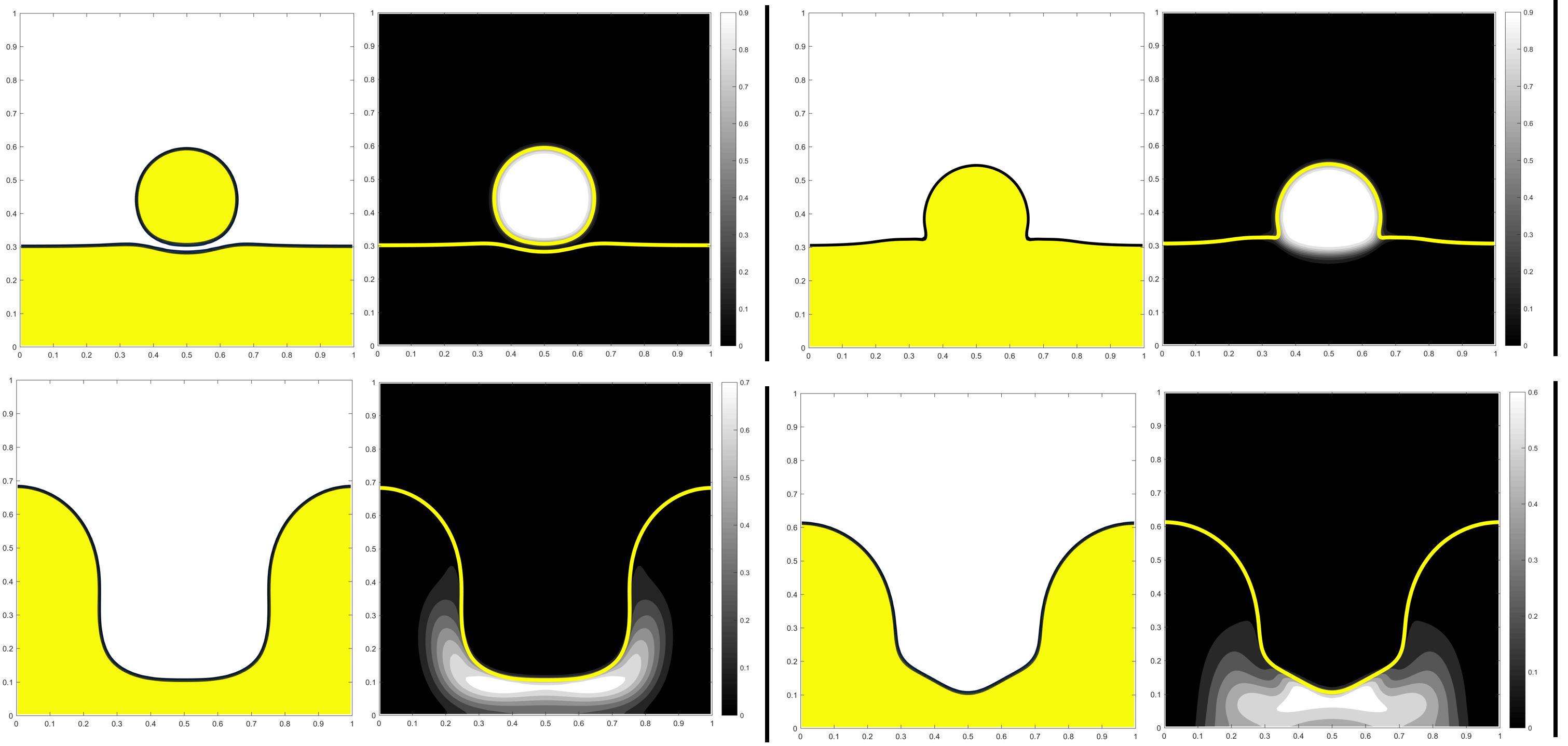}
	\includegraphics[scale=.2]{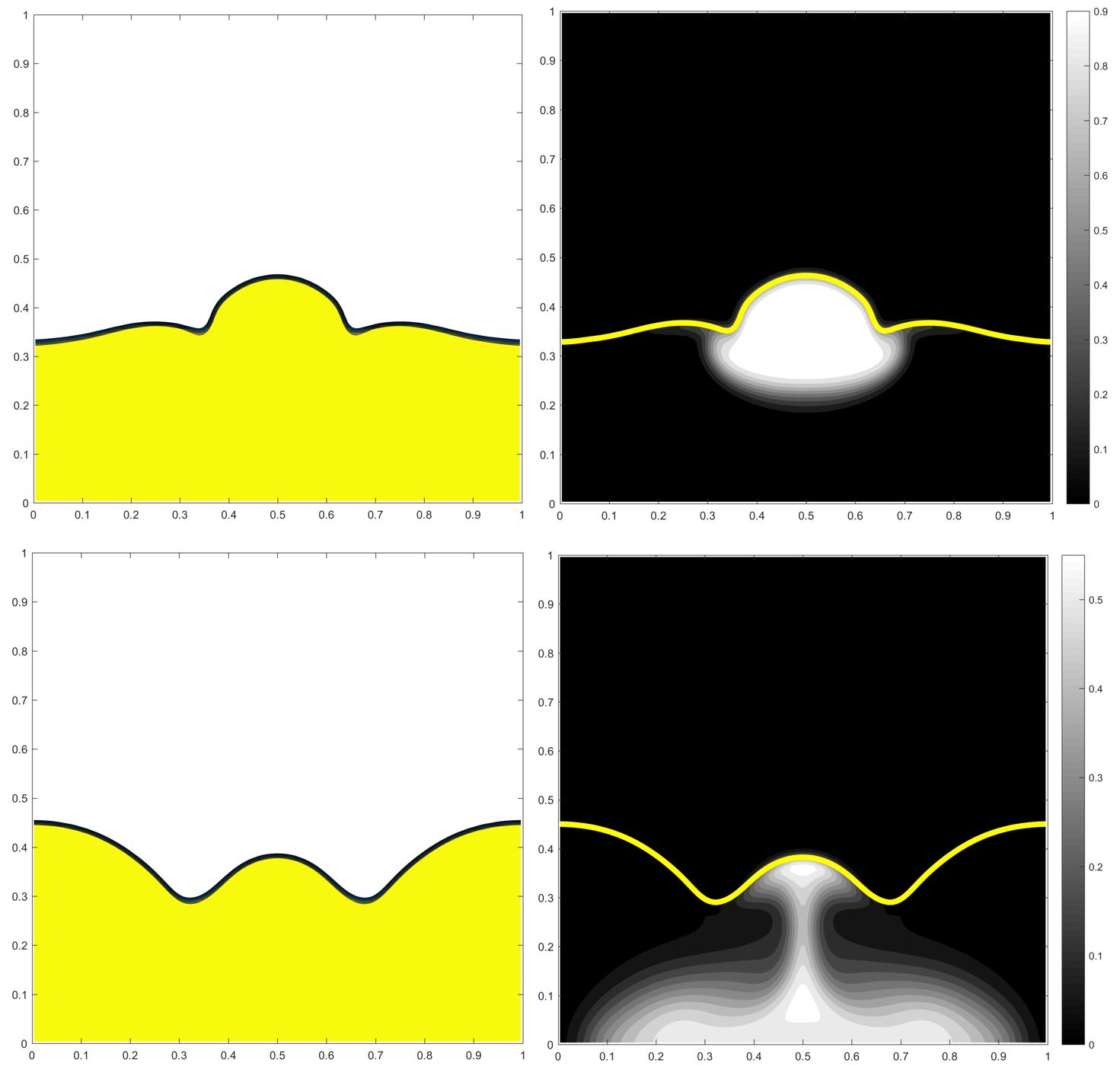}\\
	\includegraphics[scale=.2]{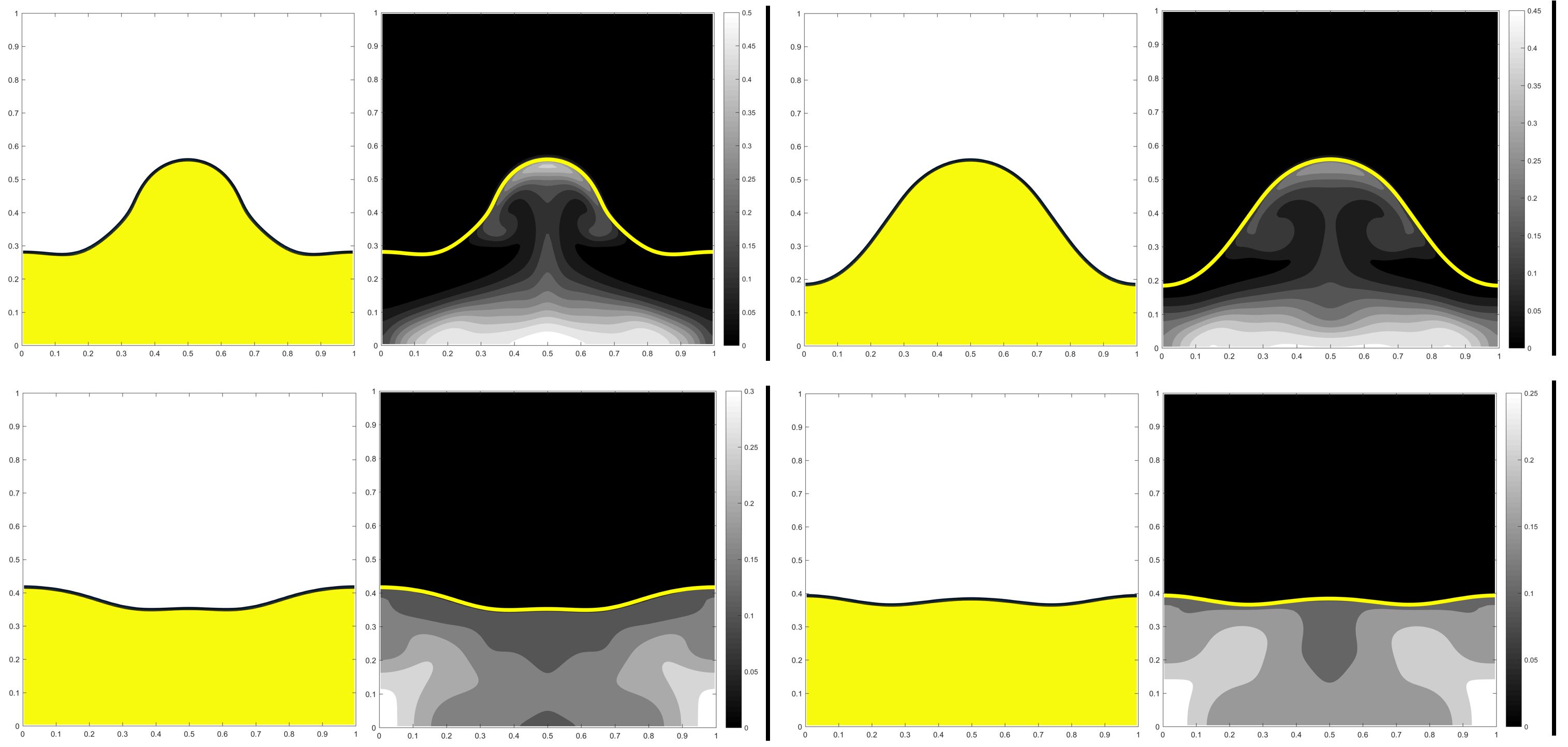}
	\includegraphics[scale=.2]{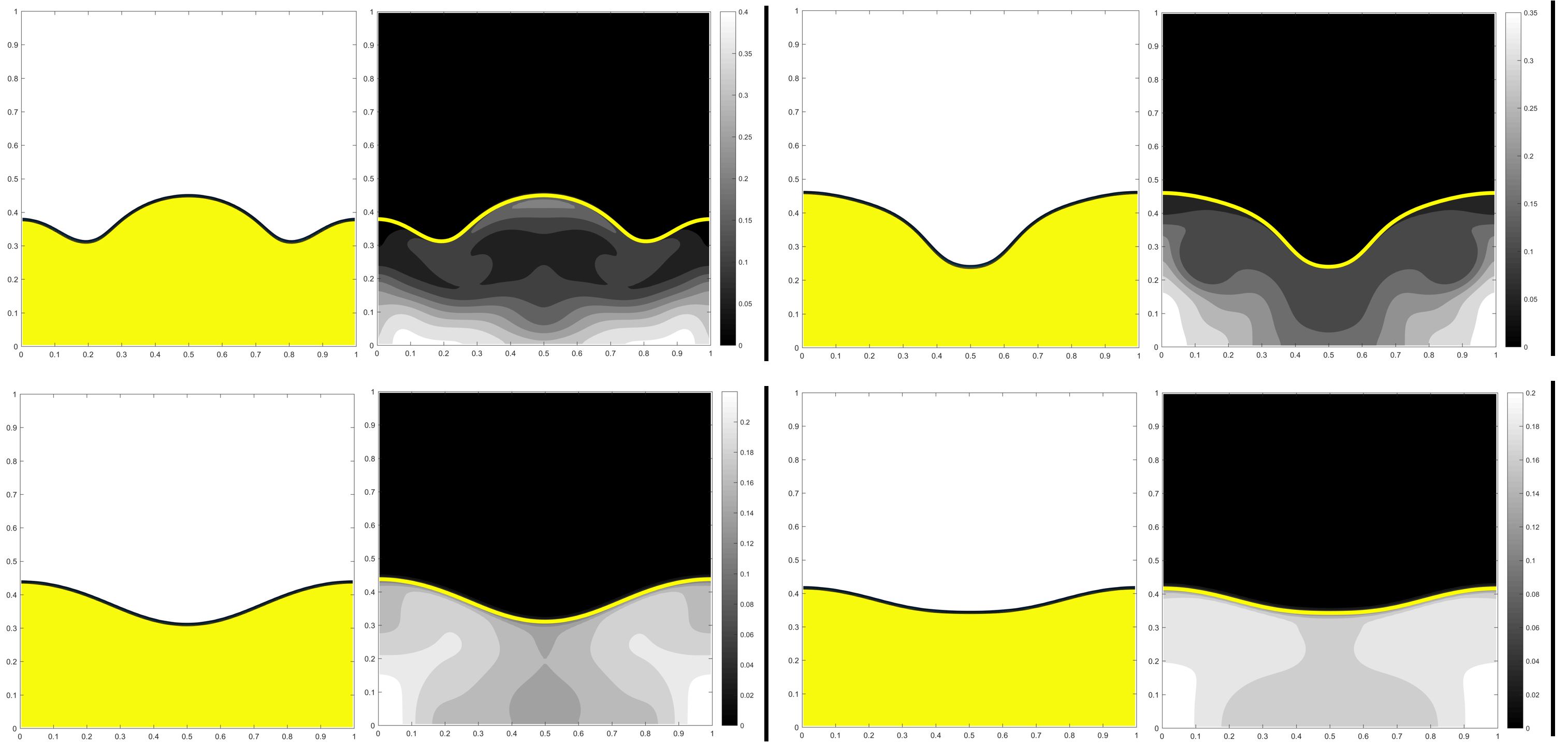}
	\includegraphics[scale=.2]{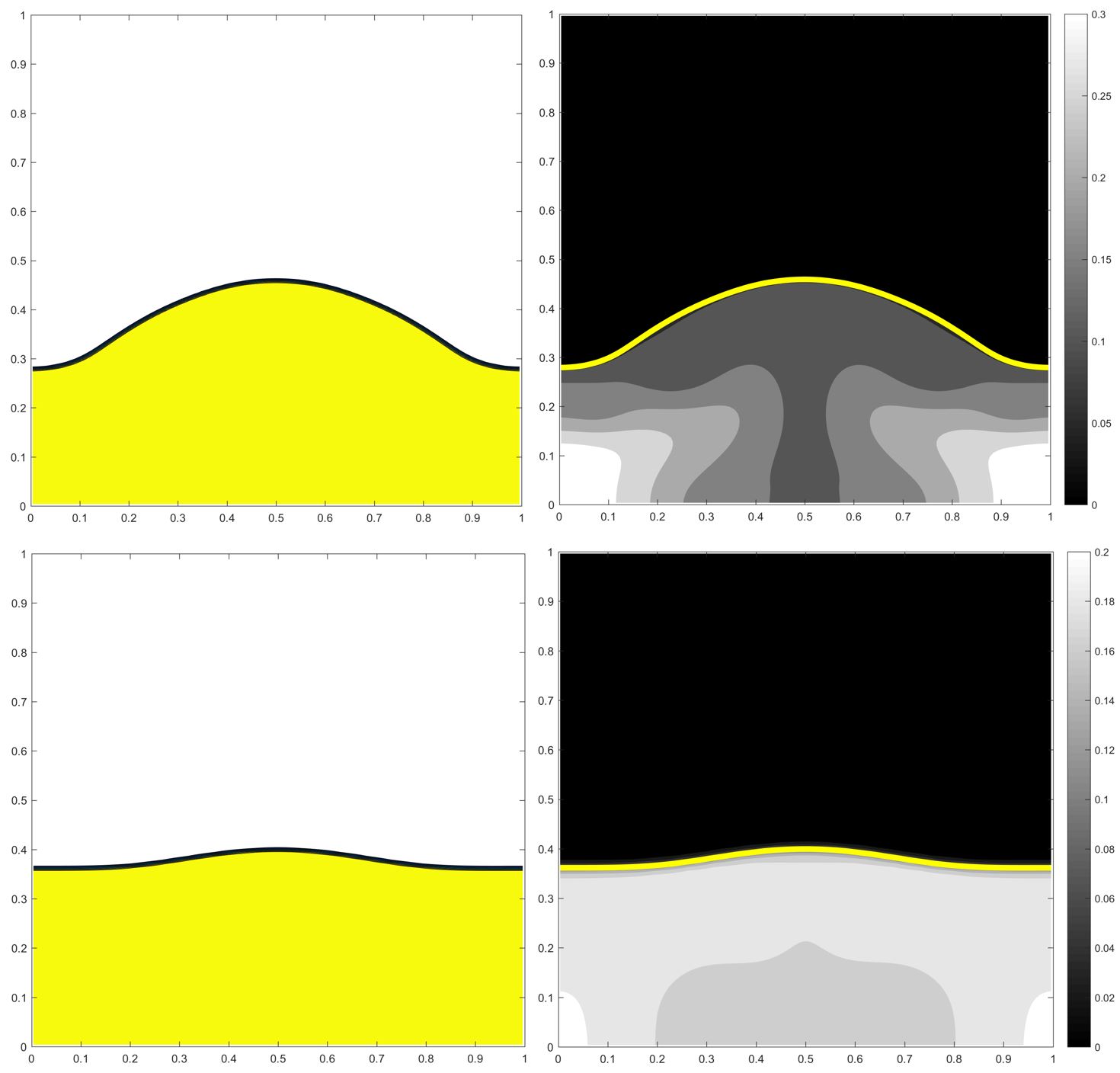}\\
	\includegraphics[scale=.2]{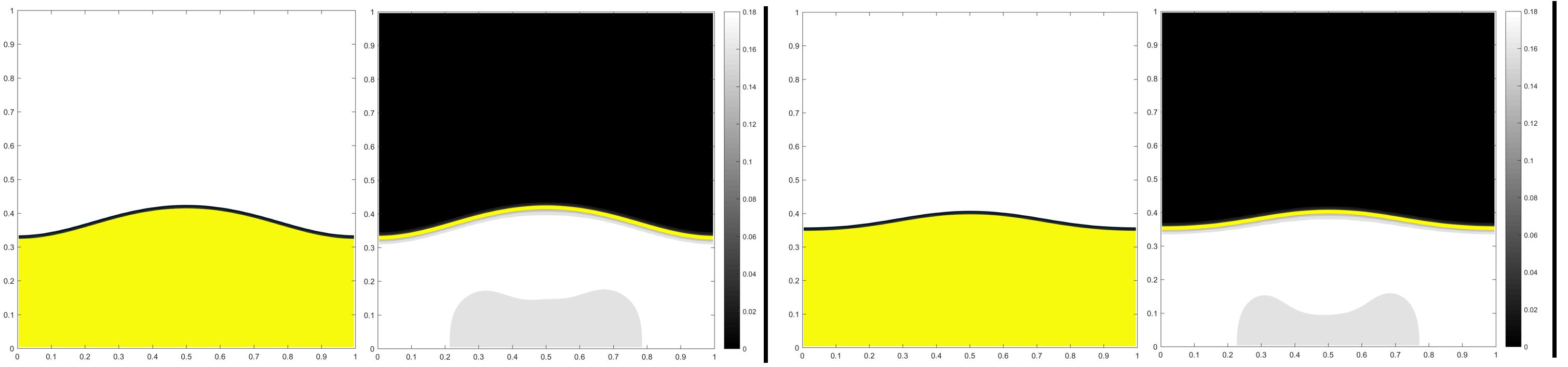}
	\includegraphics[scale=.2]{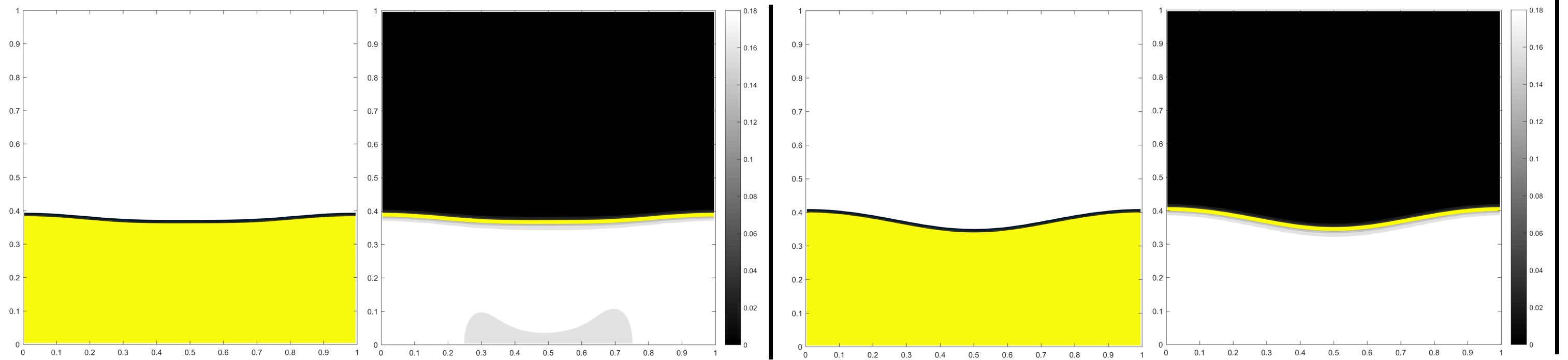}
	\includegraphics[scale=.2]{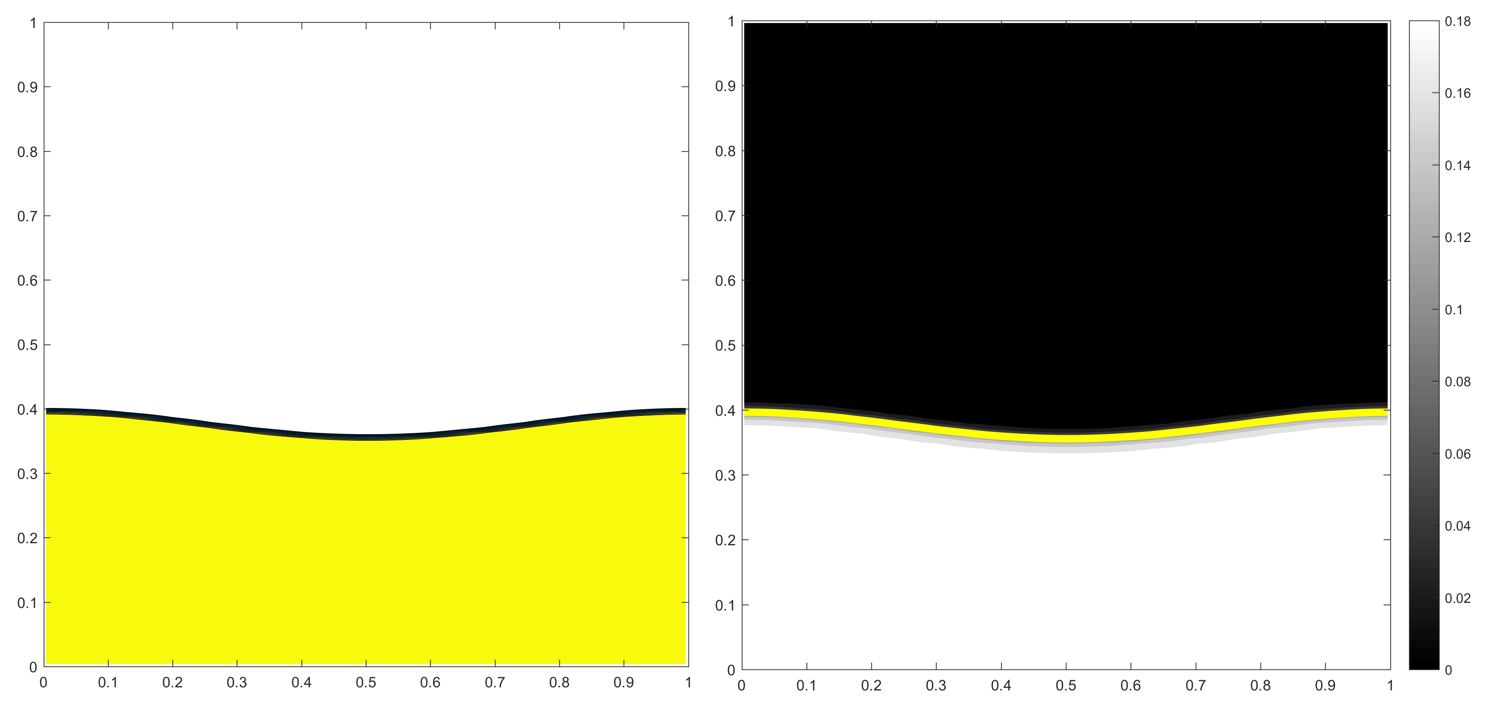}
	\caption{Results of the miscible falling drop, from left to right and top to bottom, $t=0.00$, $0.20$, $0.25$, $0.27$, $0.30$, $0.35$, $0.40$, $0.50$, $0.60$, $0.70$, $0.80$, $0.90$, $1.00$, $1.20$, $1.40$, $1.60$, $1.80$, $2.20$, $2.60$, $3.00$, $3.40$, $3.80$, $4.20$, $4.60$, $5.00$. Left of each panel: configuration of the phases, yellow: Phase 1, white: Phase 2. Right of each panel: configuration of Component 1, yellow solid lines: interfaces of Phase 1. \label{Fig MiscibleDrop} }
\end{figure}

\begin{figure}[!t]
	\centering
	\includegraphics[scale=.5]{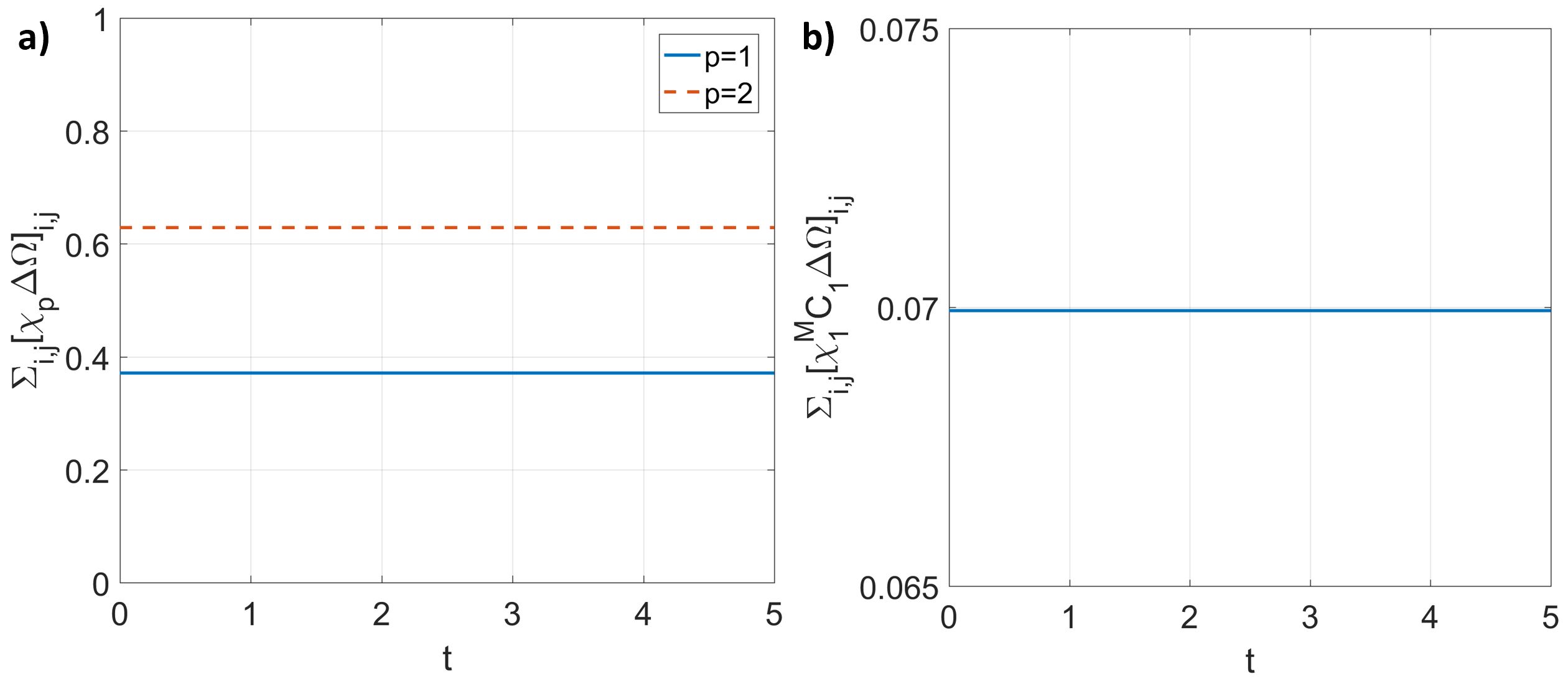}
	\caption{Time histories of the total volumes of individual phases and the total amount of Component 1 in its dissolvable region in the miscible falling drop. a) Time histories of the total volumes of individual phases. b) Time histories of the total amount of Component 1 in its dissolvable region. \label{Fig MiscibleDrop-Mass} }
\end{figure}

The results are shown in Fig.\ref{Fig MiscibleDrop} by the configurations of the phases and components at selected moments. Phase 1 is filled by the yellow color and Phase 2 is represented by the white color. Component 1 is shown by its concentration along with the yellow lines representing the interfaces between Phases 1 and 2. Due to both the heaviness of the drop, which is 3000 times heavier than its surrounded air, and the strong surface tension, the drop is falling without obvious deformation until it is close to the bottom liquid tank. At this moment, the bottom of the drop is flattened slightly. During the falling of the drop, Component 1 inside the drop preserves to be homogeneous. As a result, Phase 01 is well represented by the combination of Phase 1 and Component 1 with unity concentration. After the drop merges to the bottom tank, Component 1 starts to be transported, by both convection and diffusion, inside the liquid tank, and this process is modeling the mixing between Phases 01 and 02. We observe that Component 1 is first transported along the phase interface. Due to the inertia of the drop, the interface reaches a large ``U'' shape before it bounces back. Most of Component 1 is transported to the bottom of the domain. Since the vertical velocity close to the bottom free-slip boundary is small, and Component 1 is heaviest among all the phases and components, the major part of Component 1 stays at the bottom, and only small amount of it moves upward following the movement of the interface. As the interface moves upward, the Rayleigh-Taylor instability occurs. It should be noted that the appearance of Component 1 makes the fluid denser. Consequently, the fluid with a larger amount of Component 1, is penetrating to the fluid with less amount of Component 1, and the Rayleigh-Taylor instability is triggered. The interface keeps moving upward then downward while the amplitude is attenuated by the viscous effect. 
In the meanwhile, Component 1 becomes more homogeneous inside the bottom tank as time goes on. At the end of the simulation, we observe that the movement of the interface is not significant and Component 1 is distributed homogeneously inside Phase 1 (the bottom liquid tank).
Fig.\ref{Fig MiscibleDrop-Mass} shows the time histories of the total volumes of individual phases and the total amount of Component 1 in its dissolvable region. All the quantities in Fig.\ref{Fig MiscibleDrop-Mass} are conserved exactly even though the problem is highly complicated and dynamical.

In summary, the problem considered is very challenging. It includes many critical factors in multiphase and multicomponent flows, i.e., gravitation, surface tension, topological change, large deformation of the interface. In addition, the maximum density ratio is 3000, and the miscibilities are different between every two phases. Nevertheless, the proposed model and scheme are effective and robust for solving this problem.

\subsection{Falling drops with moving contact lines}\label{Sec Falling drops with moving contact lines}
To demonstrate the capability of the proposed model and scheme, we present the final example that includes three phases and three components, large density and viscosity ratios, multiphase interfacial tensions, and the effect of contact angles and moving contact lines. 

The densities and viscosities of the 3 pure phases are $\rho_1^\phi=1000\mathrm{kg/m^3}$, $\mu_1^\phi=10^{-3}\mathrm{Pa \cdot s}$, $\rho_2^\phi=500\mathrm{kg/m^3}$, $\mu_2^\phi=10^{-1}\mathrm{Pa \cdot s}$, $\rho_3^\phi=1\mathrm{kg/m^3}$, and $\mu_3^\phi=2\times10^{-5}\mathrm{Pa \cdot s}$. The interfacial tensions are $\sigma_{1,2}=0.04\mathrm{N/m}$, $\sigma_{1,3}=0.0728\mathrm{N/m}$, and $\sigma_{2,3}=0.055\mathrm{N/m}$, and the gravity is $\mathbf{g}=\{0,-9.8\}\mathrm{m/s^2}$. The contact angle between Phases 1 and 2 at the right boundary is $135^0$ and the other is $90^0$. 
The densities and viscosities of the 3 components are $\rho_1^C=500\mathrm{kg/mol}$, $\mu_1^C=5\times10^{-4}\mathrm{Pa \cdot s \cdot m^3/mol}$, $\rho_2^C=100\mathrm{kg/mol}$, $\mu_2^C=1\times10^{-3}\mathrm{Pa \cdot s \cdot m^3/mol}$, $\rho_3^C=1\mathrm{kg/mol}$, $\mu_3^C=1\times10^{-4}\mathrm{Pa \cdot s \cdot m^3/mol}$. Component 1 is only dissolvable in Phase 1 with diffusivity $1\times10^{-5}\mathrm{m^2/s}$, Component 2 is dissolvable in both Phases 1 and 2 with diffusivity $5\times10^{-4}\mathrm{m^2/s}$ and $5\times10^{-5}\mathrm{m^2/s}$, respectively, and Component 3 is dissolvable in both Phases 1 and 3 with diffusivity $5\times10^{-6}\mathrm{m^2/s}$ and $2\times10^{-5}\mathrm{m^2/s}$, respectively. The governing equations are non-dimensionalized with a length scale $0.01\mathrm{m}$, a density scale $1\mathrm{kg/m^3}$, an acceleration scale $1\mathrm{m/s^2}$, and a concentration scale $1\mathrm{mol/m^3}$.

The domain considered is $[1\times1]$ with no-slip boundaries. The domain is discretized by $[128\times128]$ cells and the time step size is $\Delta t=10^{-4}$. Initially, a drop of Phase 1 at $(0.3,0.75)$ with a radius $0.15$ is above a tank of Phase 1 filling $0 \leqslant y \leqslant 0.3$. A drop of Phase 2 is at $(0.75,0.6)$ with a radius $0.1$. The rest of the domain is filled by Phase 3. Components 1 and 2 are homogeneously distributed with unity concentration inside the drops of Phases 1 and 2, respectively. Component 3 is distributed between $x=0.5$ and $x=0.6$ with unity concentration.

\begin{figure}[!t]
	\centering
	\includegraphics[scale=.18]{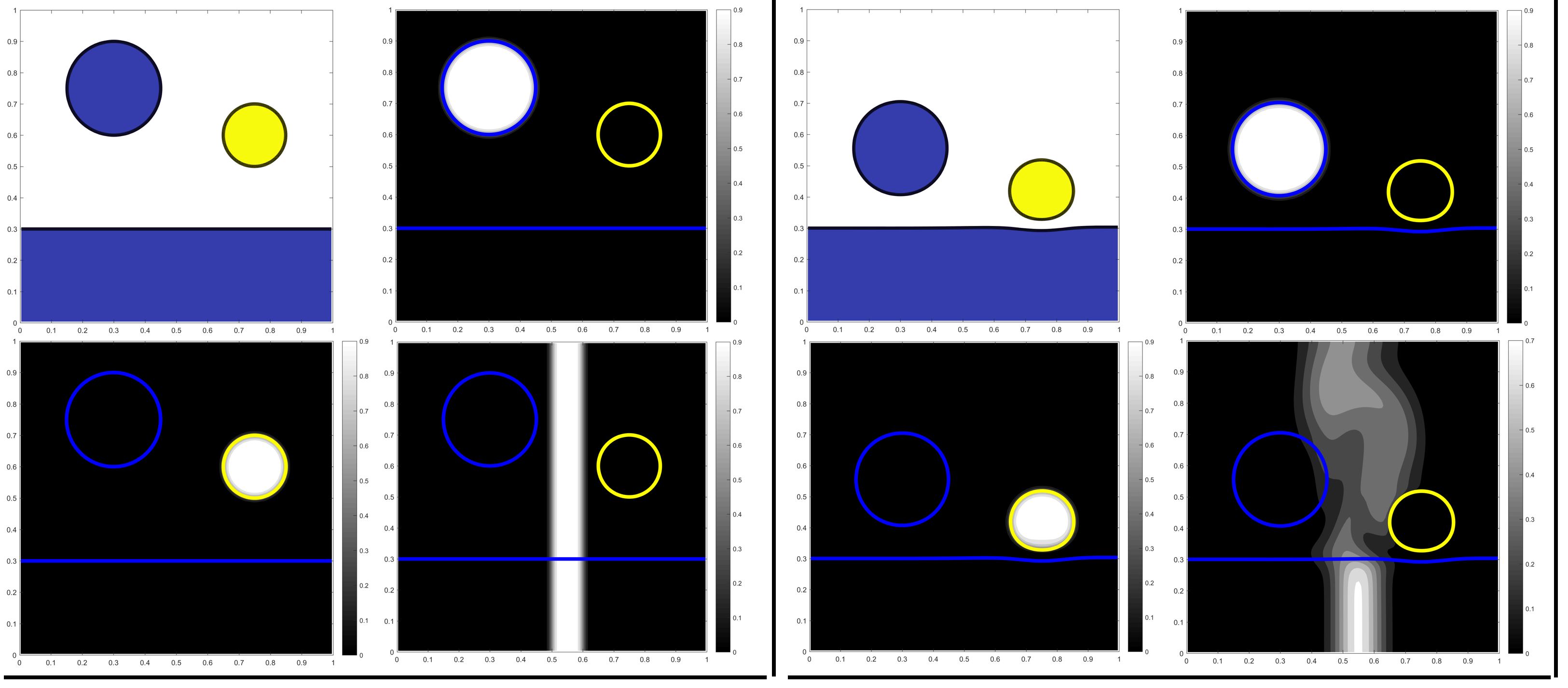}
	\includegraphics[scale=.18]{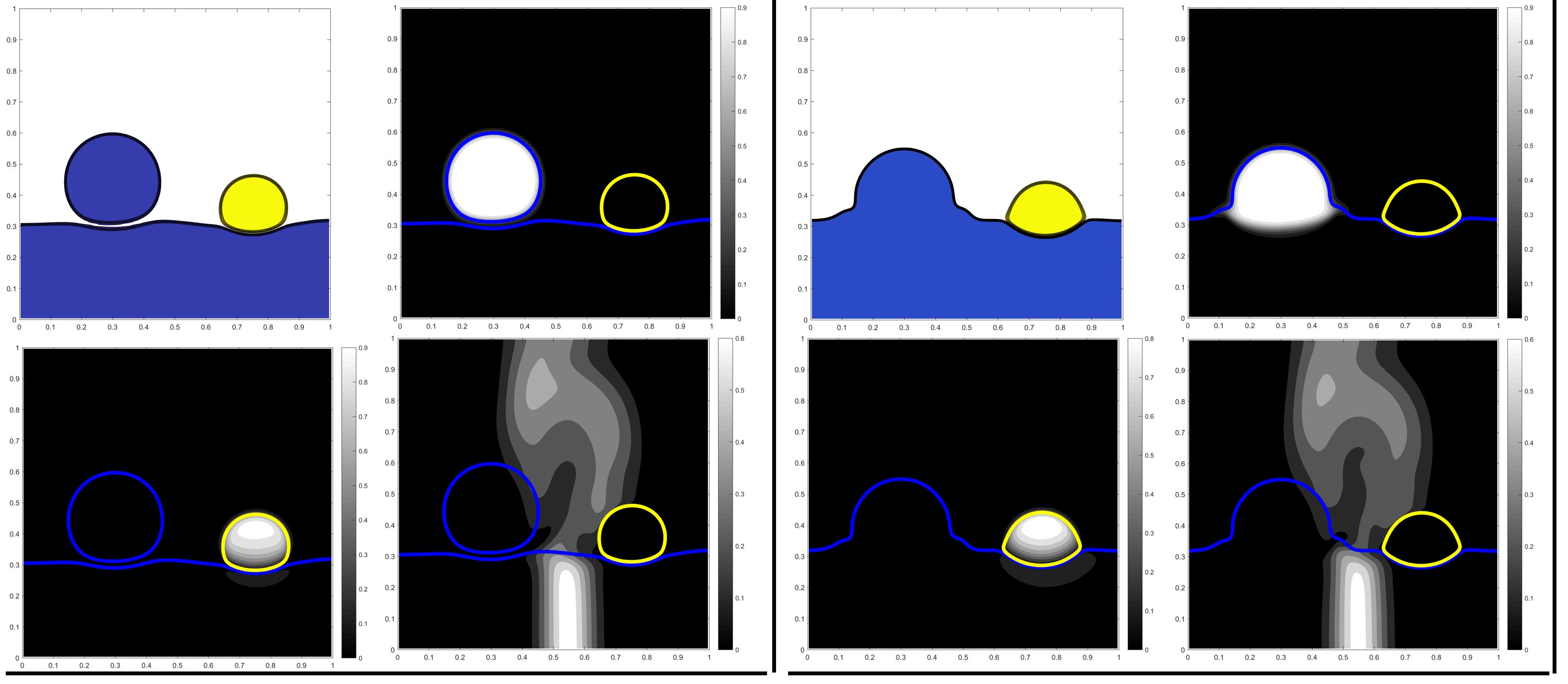}
	\includegraphics[scale=.18]{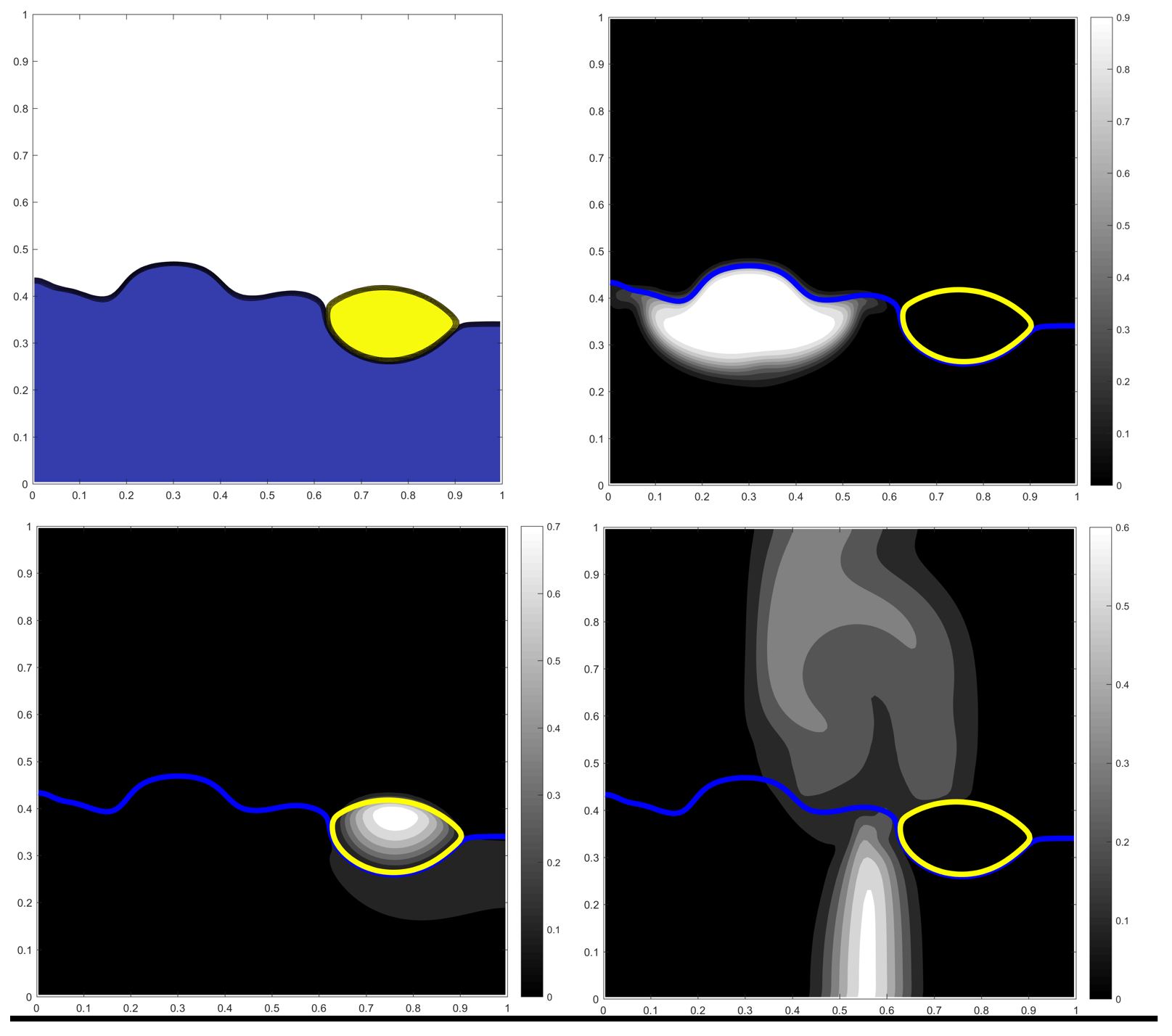}\\
	\includegraphics[scale=.18]{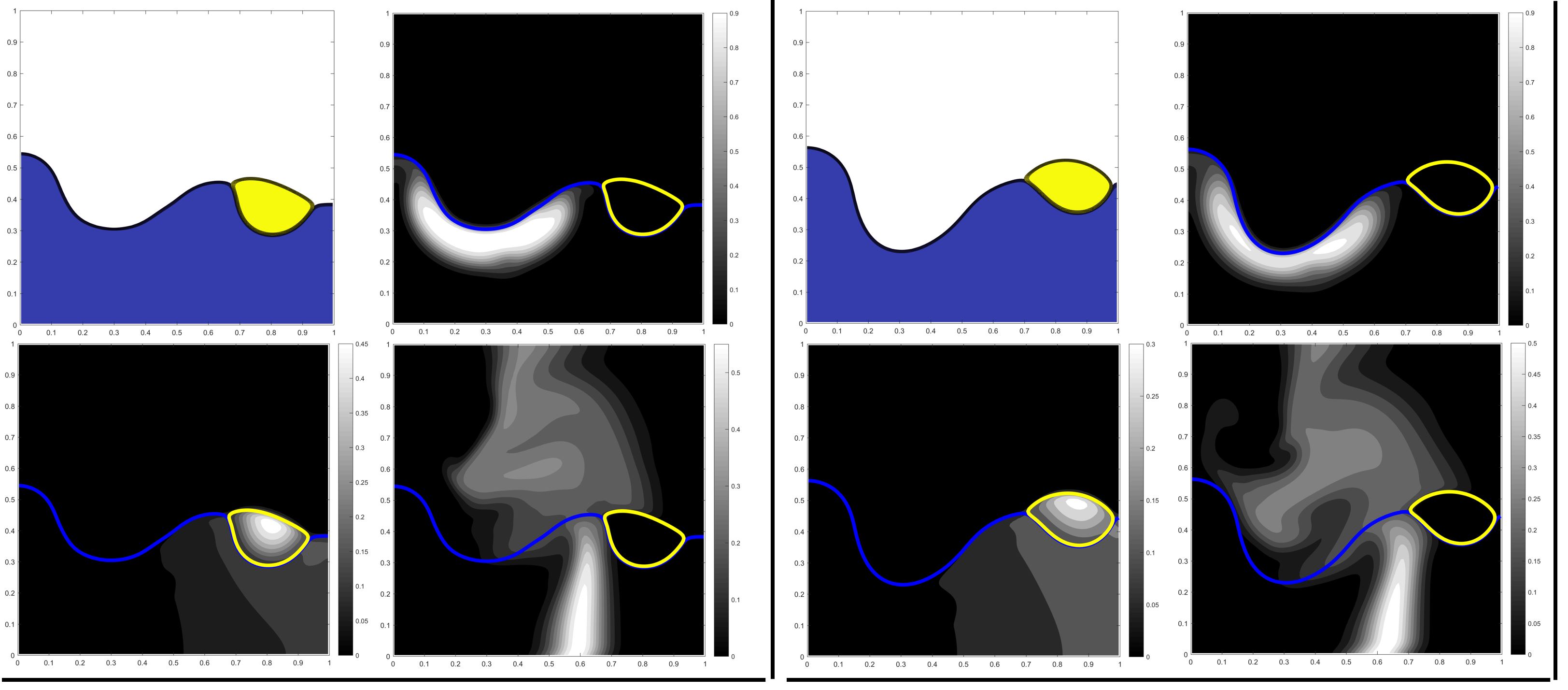}
	\includegraphics[scale=.18]{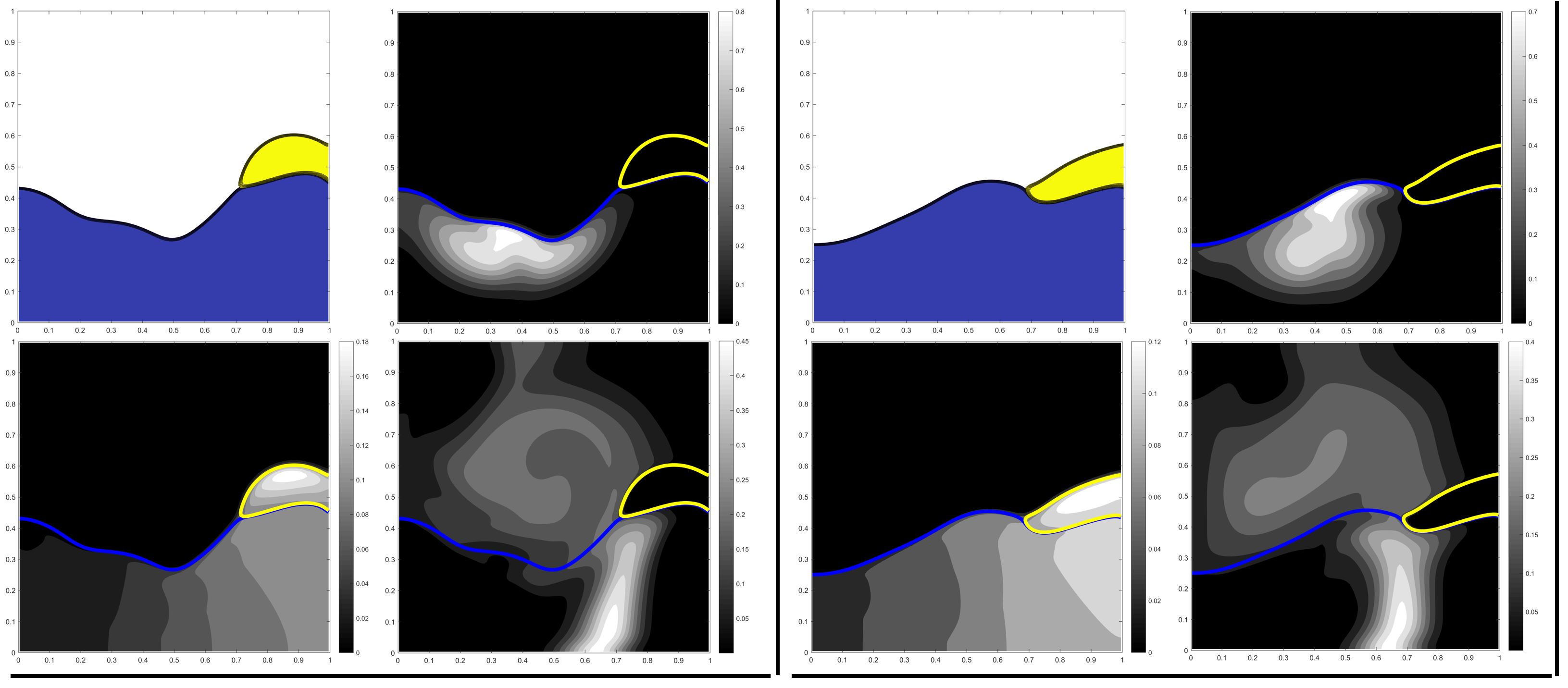}
	\includegraphics[scale=.18]{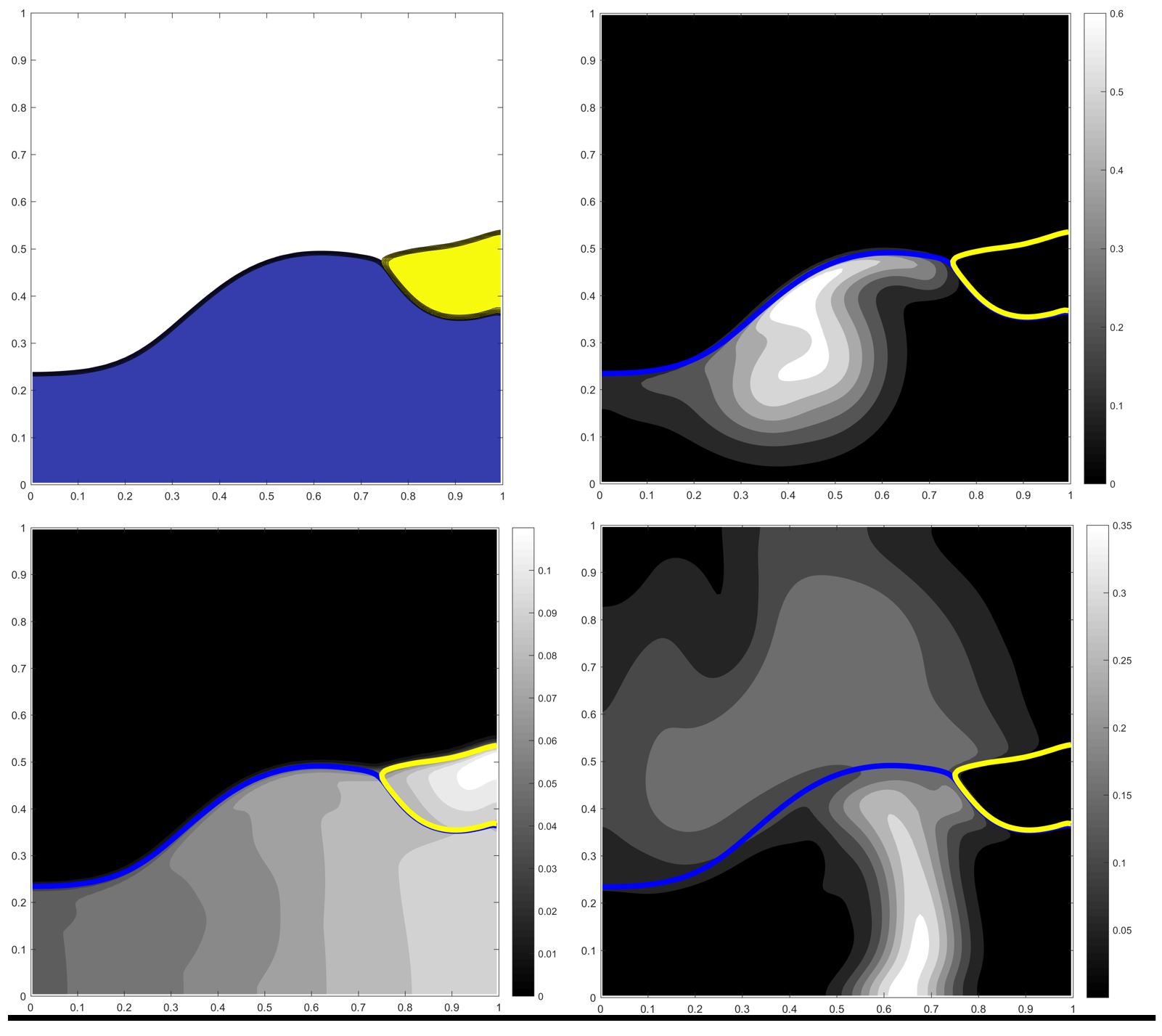}\\
	\includegraphics[scale=.18]{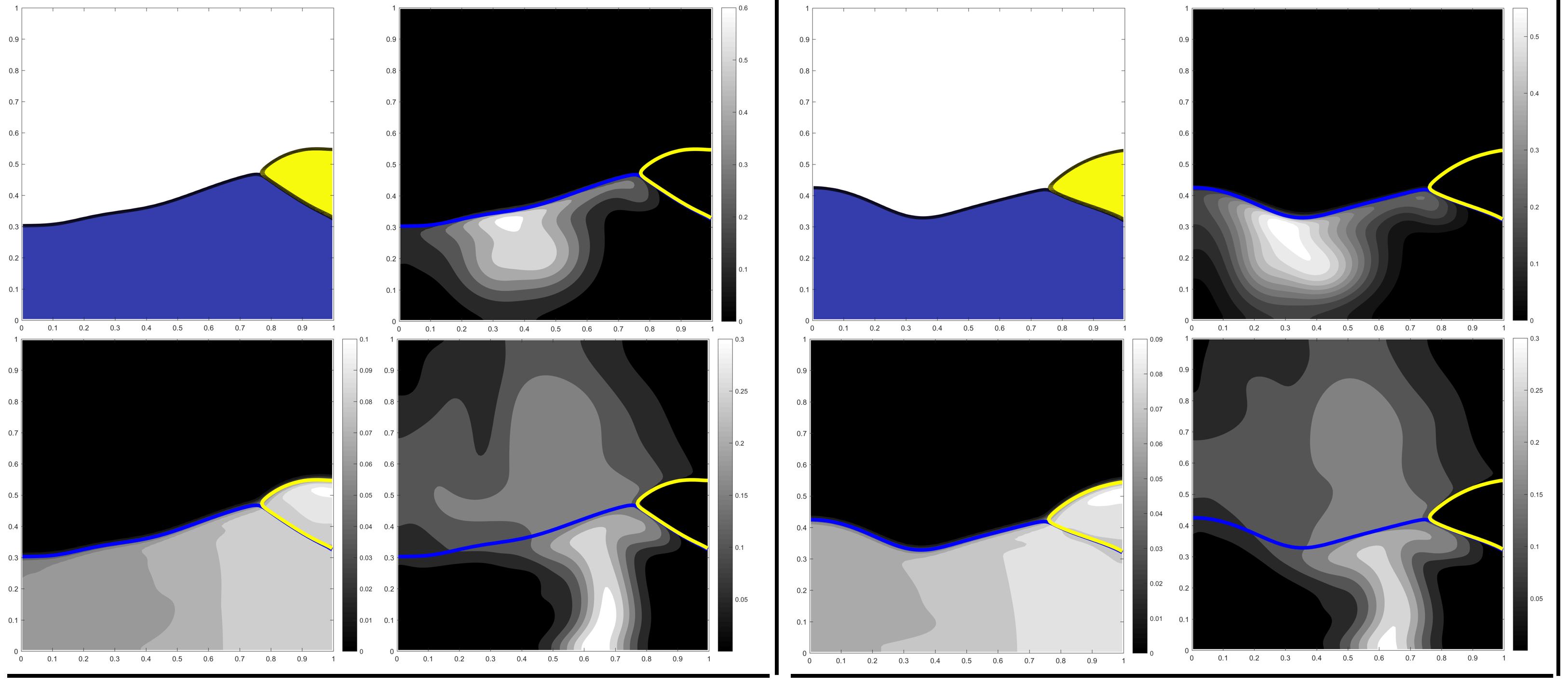}
	\includegraphics[scale=.18]{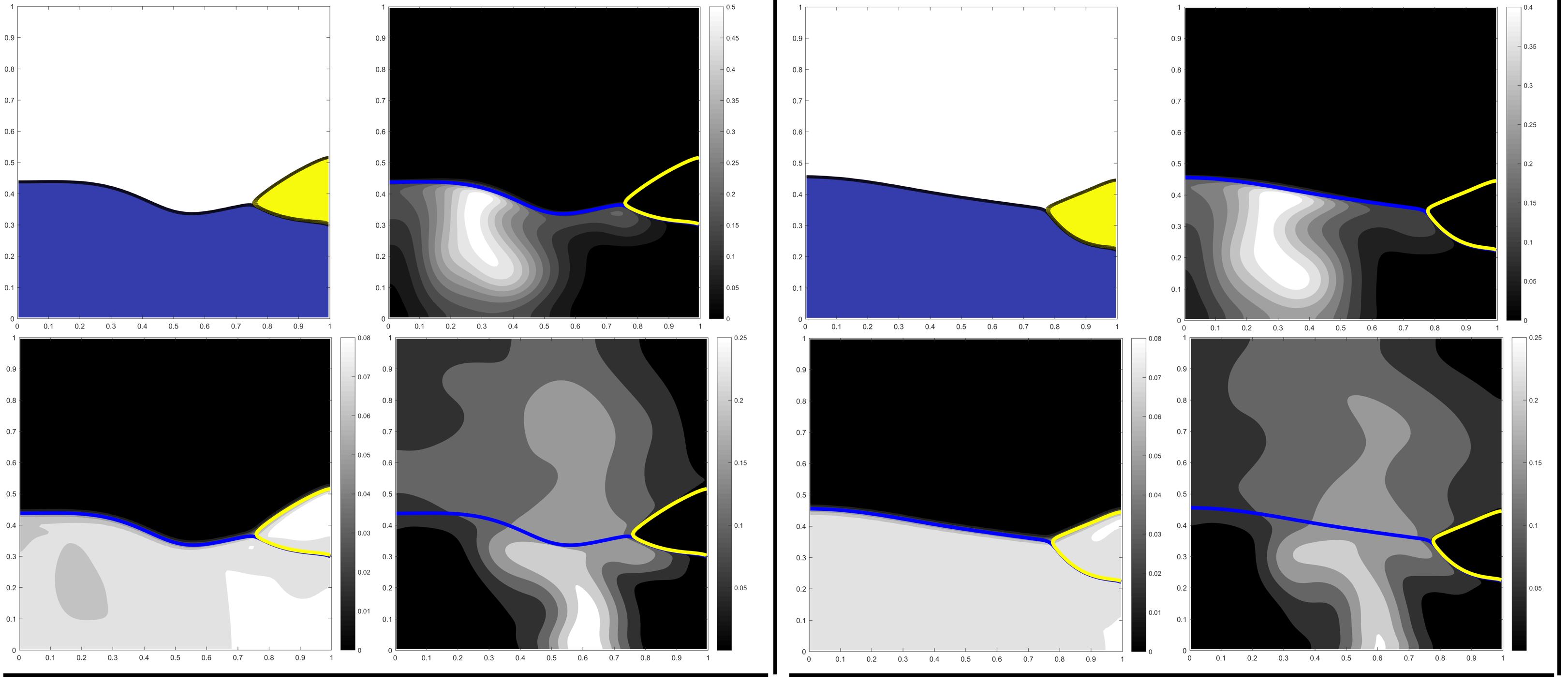}
	\includegraphics[scale=.18]{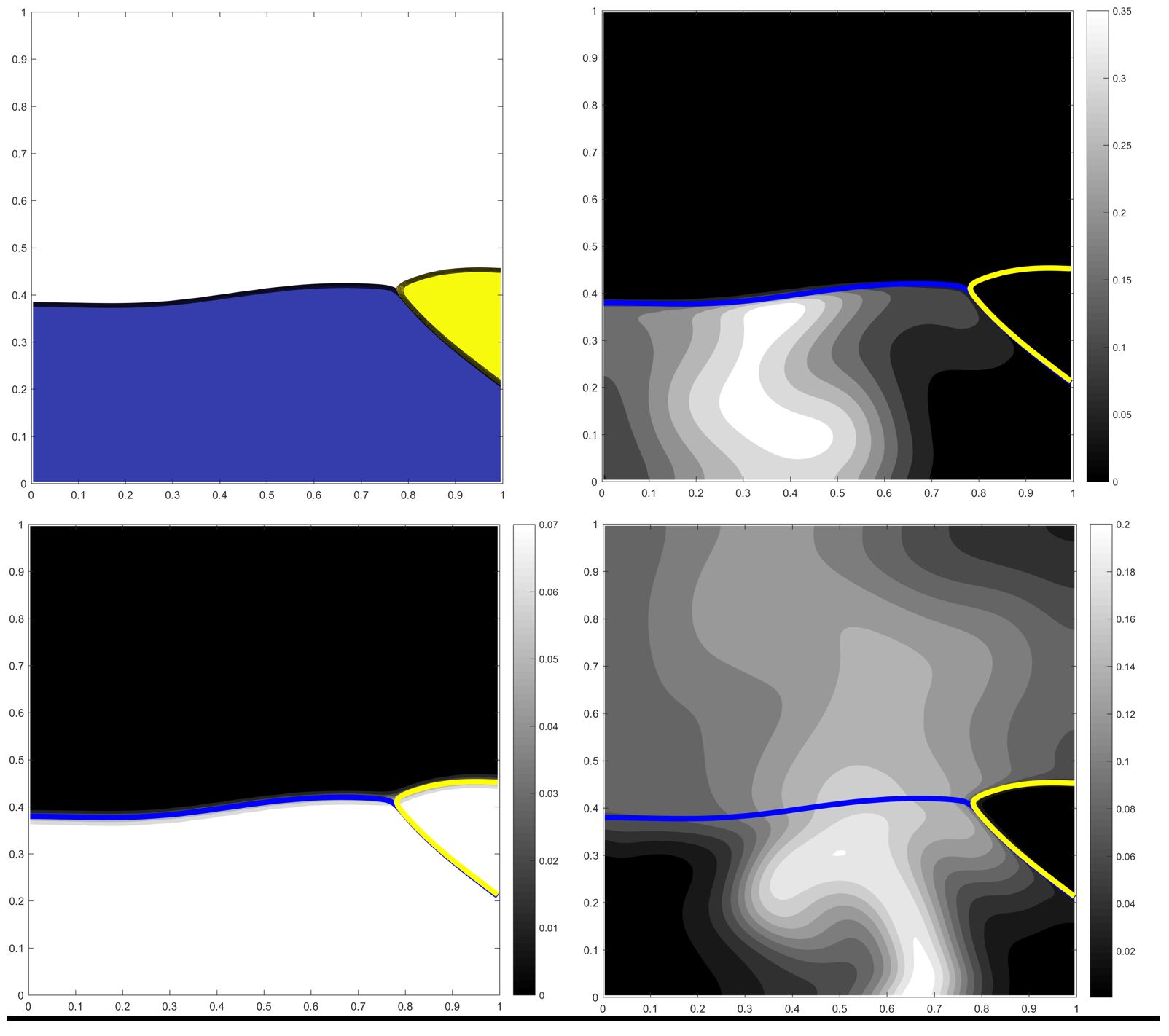}\\
	\includegraphics[scale=.18]{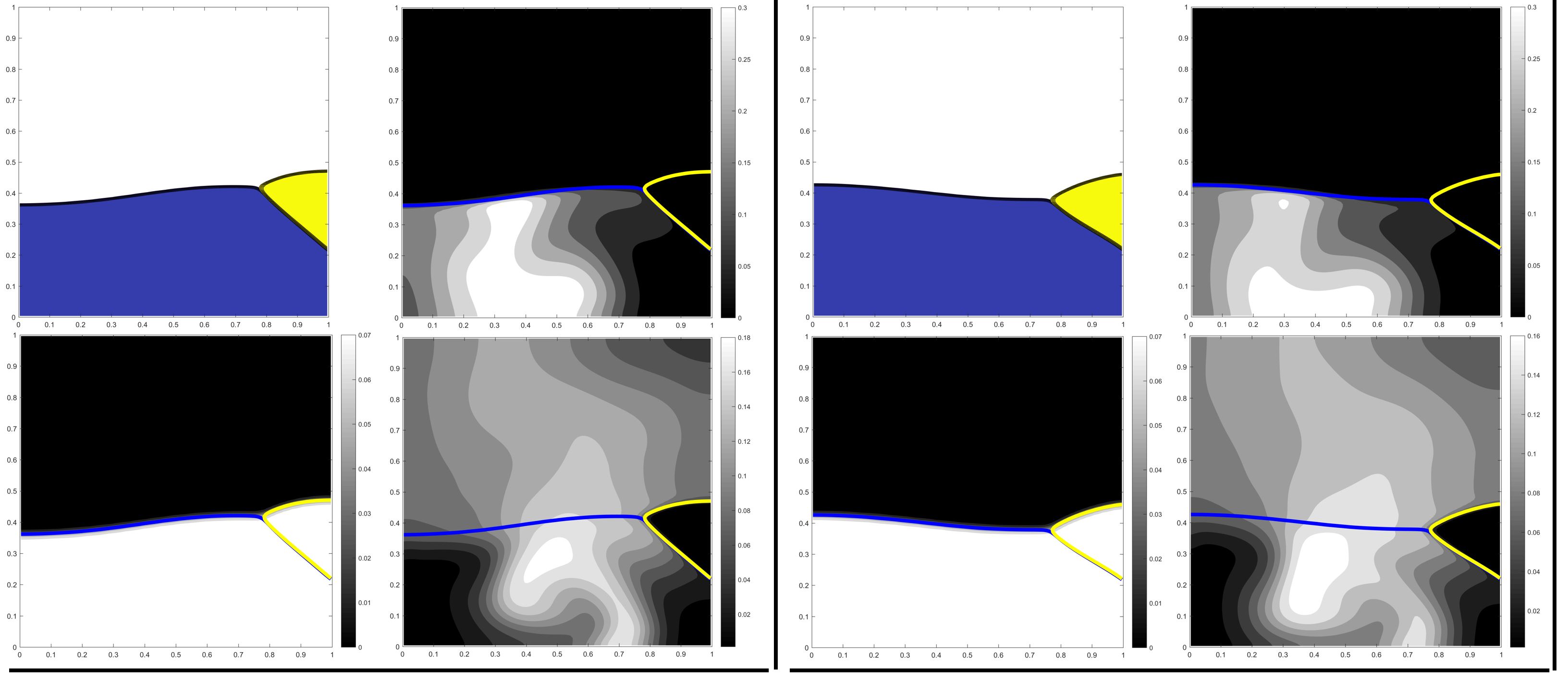}
	\includegraphics[scale=.18]{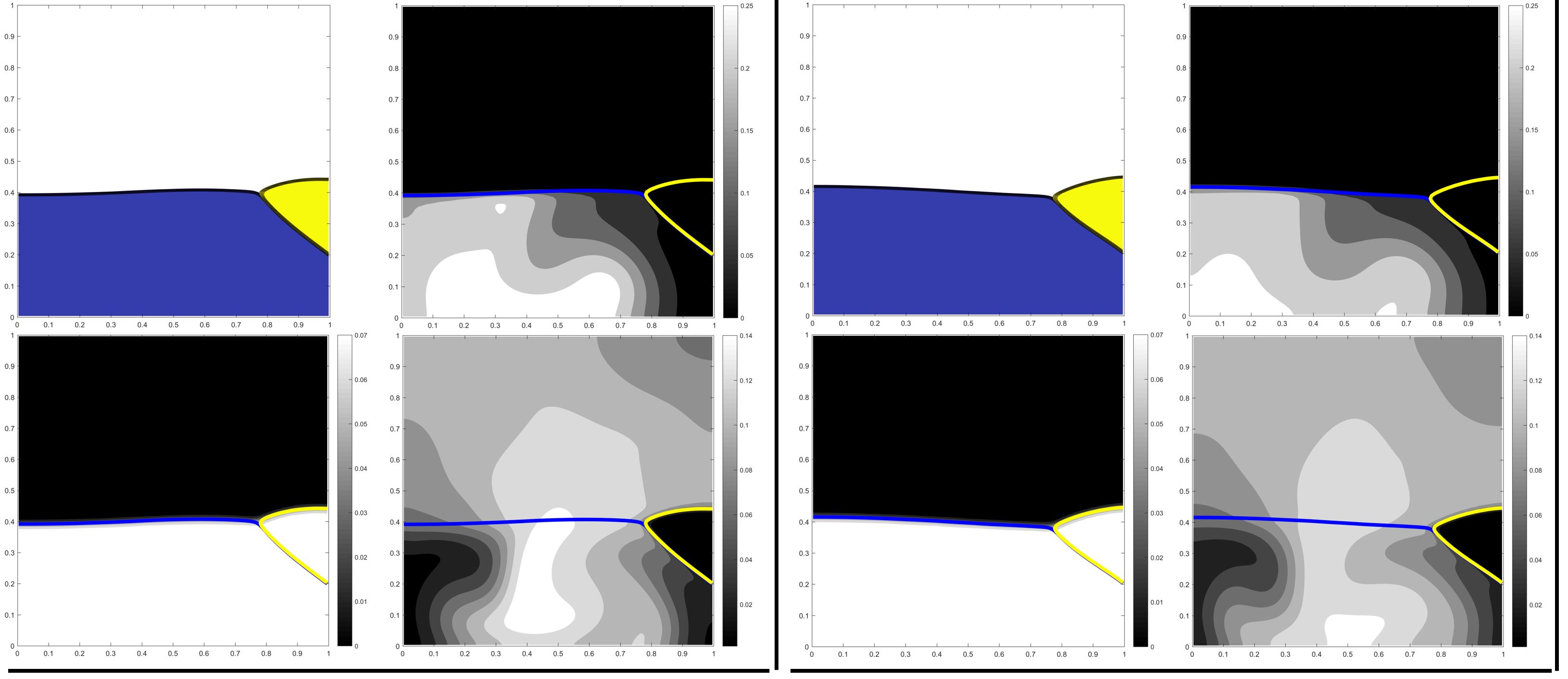}
	\includegraphics[scale=.18]{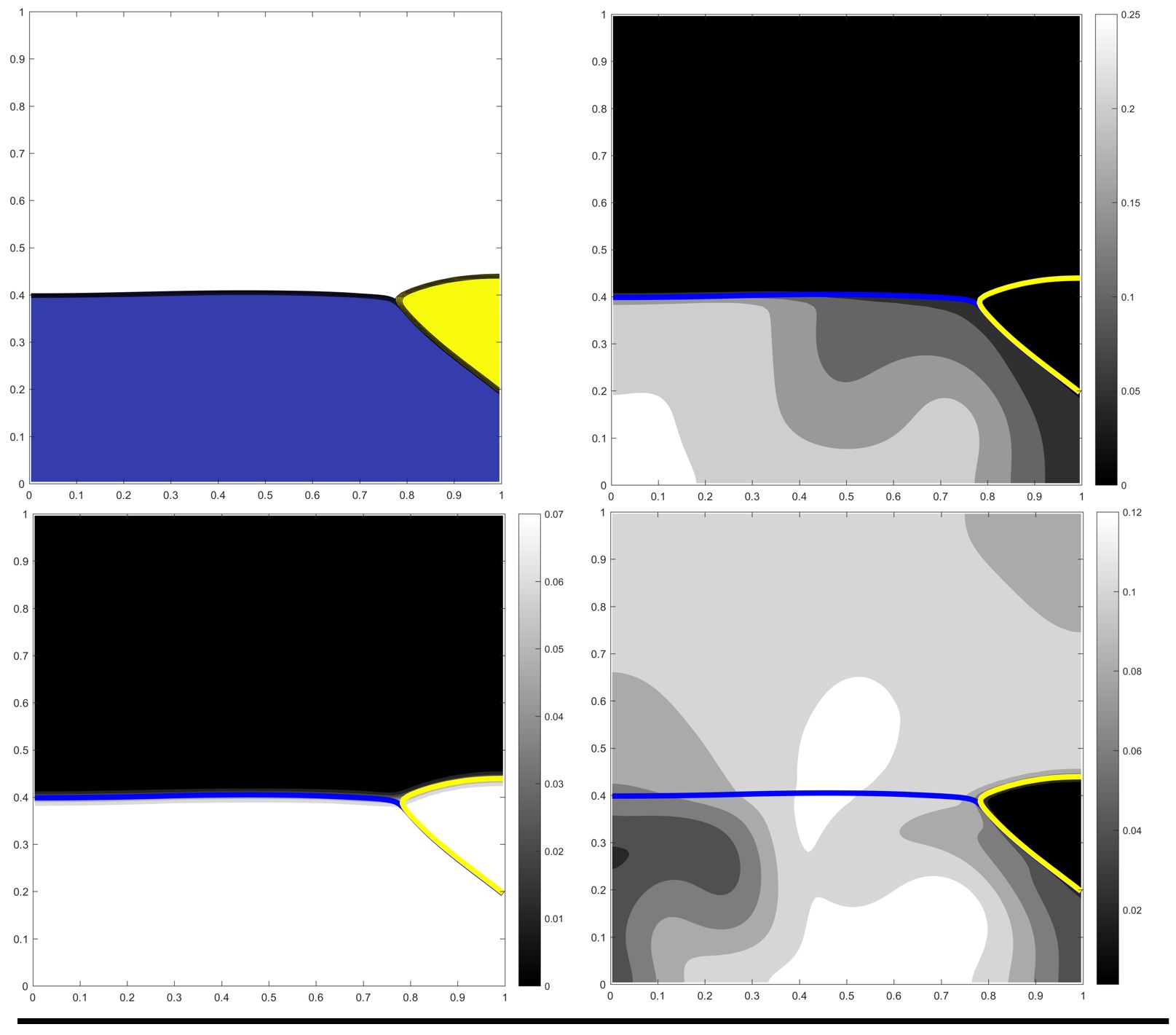}\\
	\includegraphics[scale=.18]{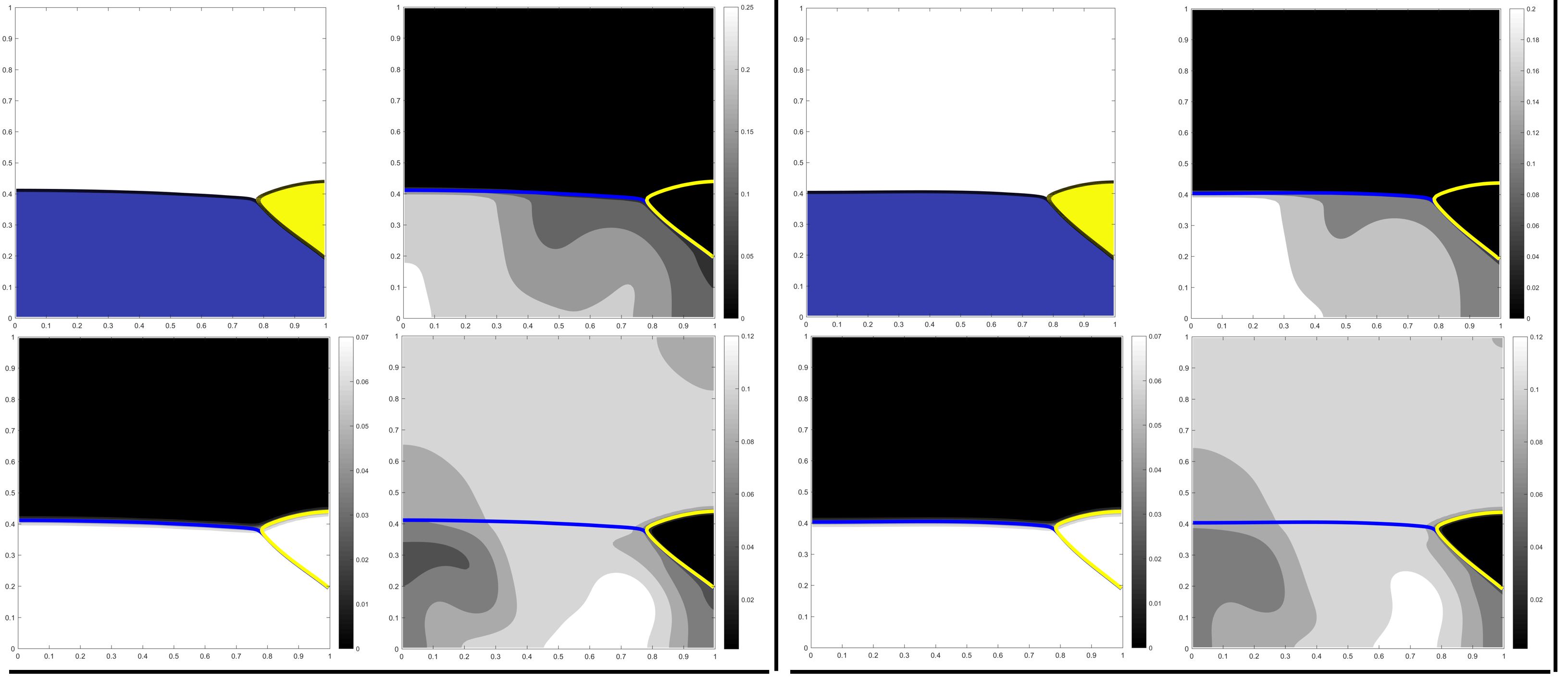}
	\includegraphics[scale=.18]{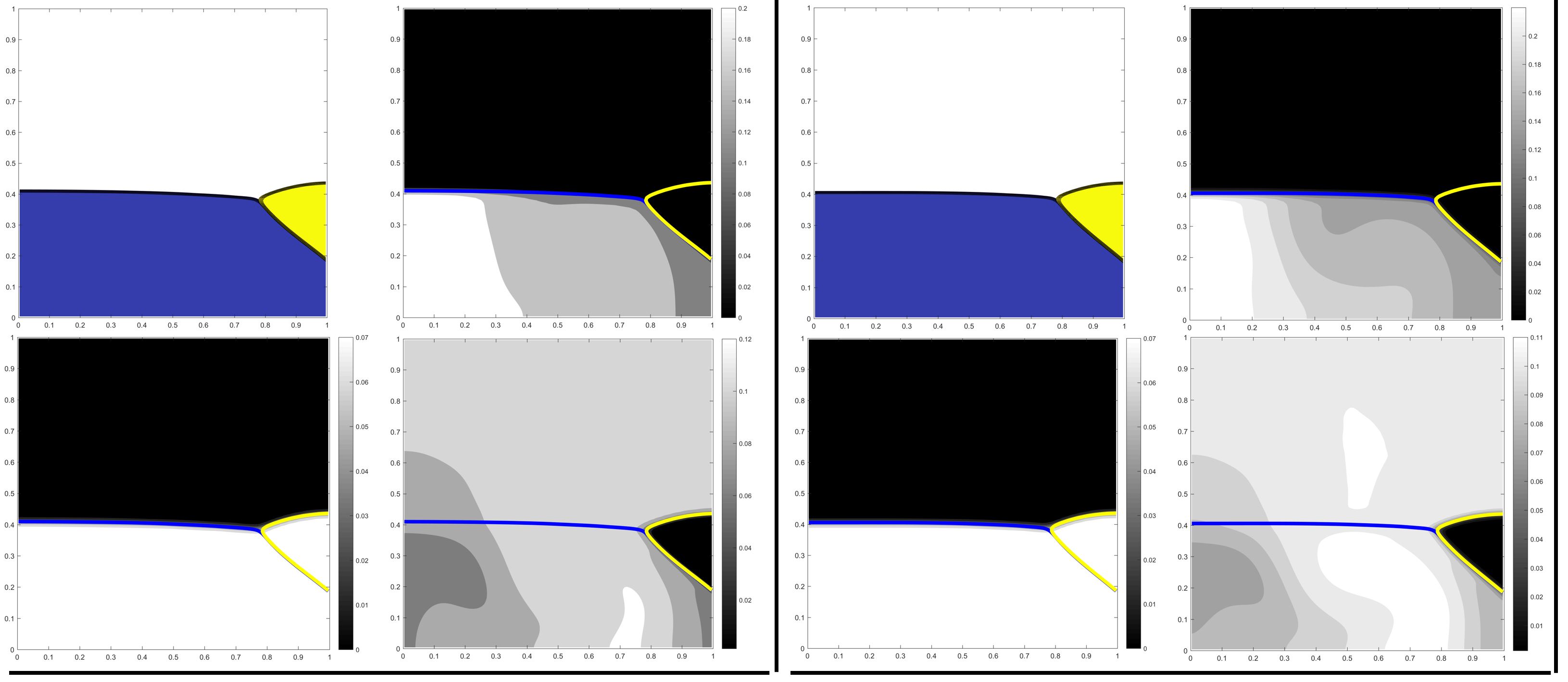}
	\includegraphics[scale=.18]{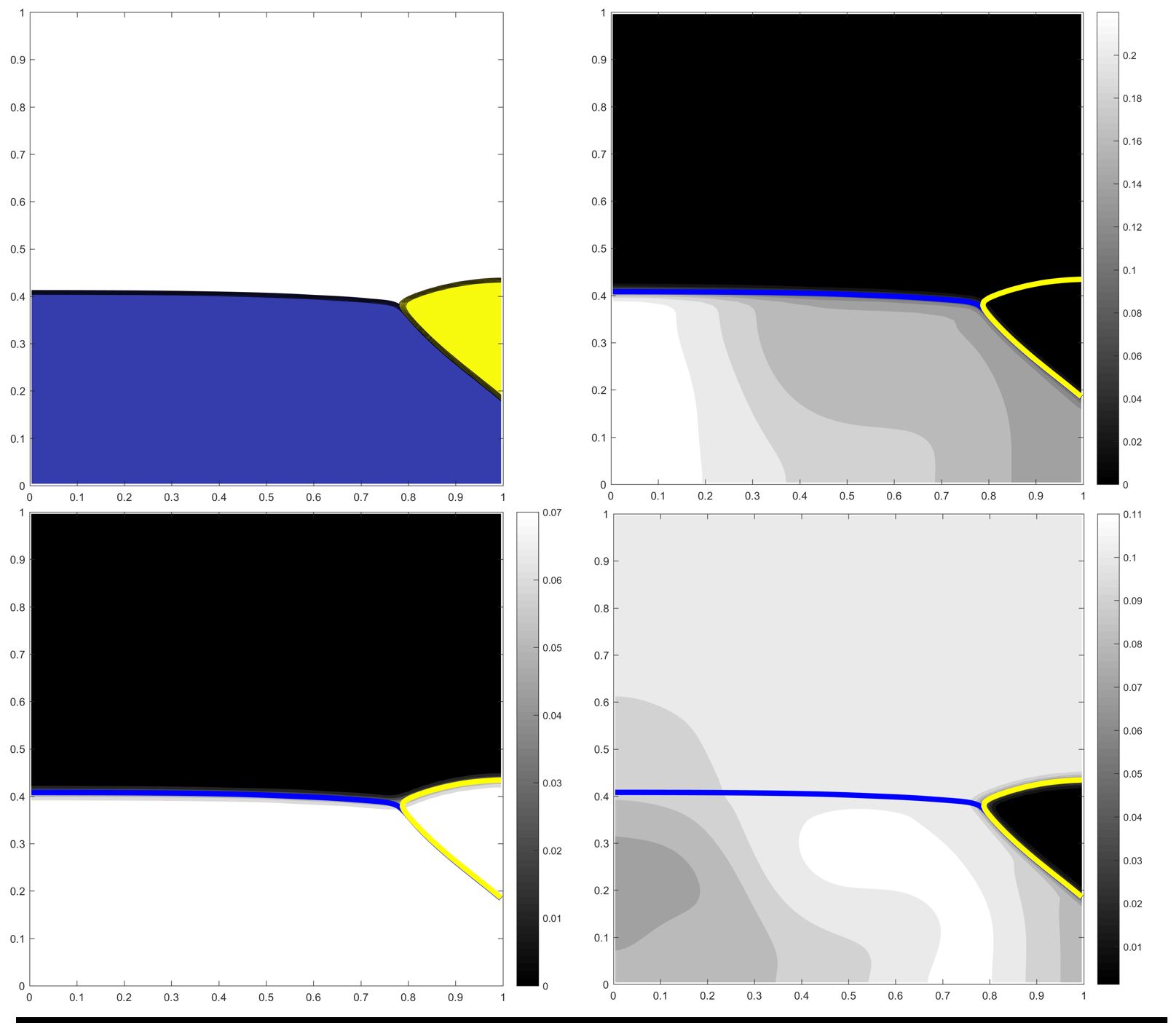}
	\caption{Results of the falling drops with moving contact lines, from left to right and top to bottom, $t=0.00$, $0.20$, $0.25$, $0.27$, $0.30$, $0.35$, $0.40$, $0.50$, $0.60$, $0.70$, $0.80$, $0.90$, $1.00$, $1.20$, $1.40$, $1.60$, $1.80$, $2.20$, $2.60$, $3.00$, $3.40$, $3.80$, $4.20$, $4.60$, $5.00$. Top-left of each panel: configuration of the phases, blue: Phase 1, yellow: Phase 2, white: Phase 3. Top-right of each panel: configuration of Component 1. Bottom-left: configuration of Component 2. Bottom-right: configuration of Component 3. In the top-right, bottom-left, and bottom right of each panel, blue solid lines: interfaces of Phase 1, yellow solid lines: interfaces of Phase 2. \label{Fig ContactLine} }
\end{figure}

\begin{figure}[!t]
	\centering
	\includegraphics[scale=.5]{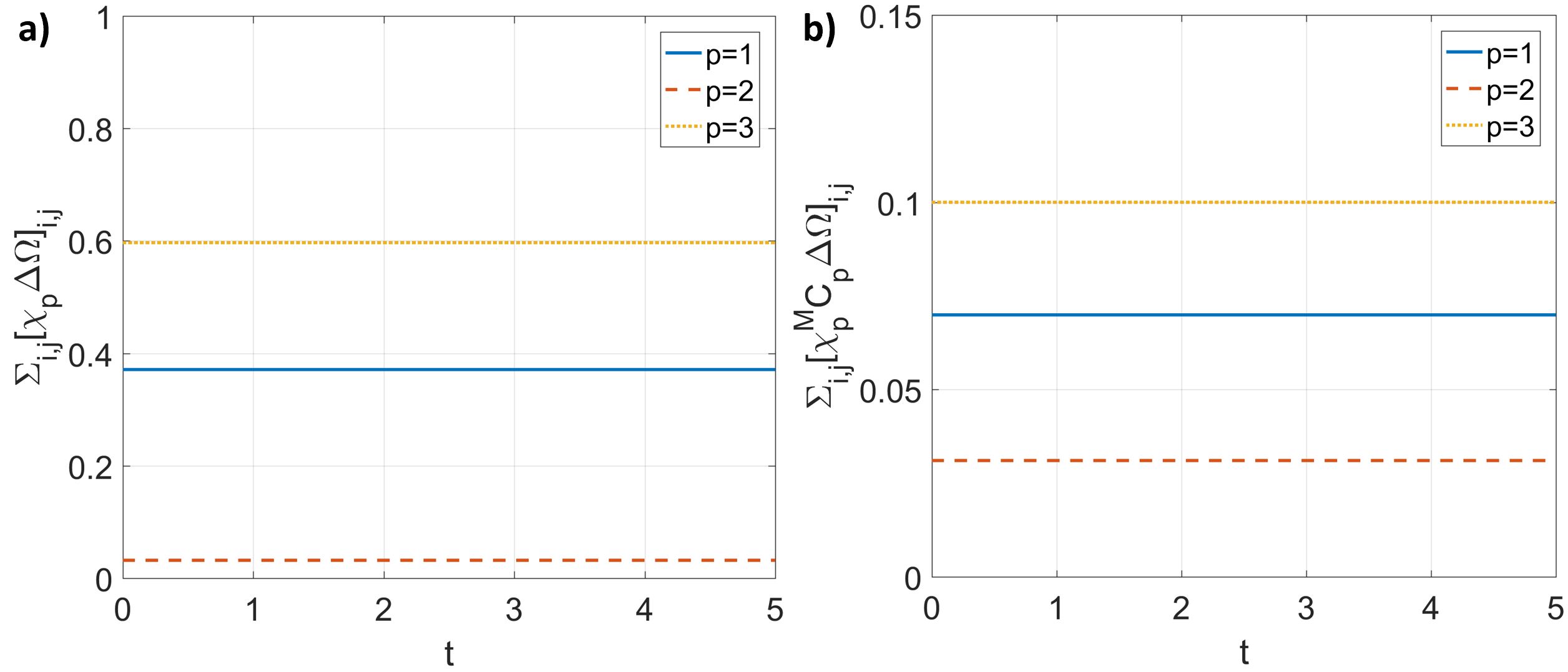}
	\caption{Time histories of the total volumes of individual phases and the total amounts of each component in its corresponding dissolvable region in the falling drops with moving contact lines. a) Time histories of the total volumes of individual phases. b) Time histories of the total amounts of each component in its corresponding dissolvable region. \label{Fig ContactLine-Mass} }
\end{figure}

The results are shown in Fig.\ref{Fig ContactLine}, and inside each panel, the top-left shows the configuration of the three phases, the top-right, bottom-left, and bottom-right show the concentrations of Components 1, 2, and 3, respectively. Phase 1 is filled by the blue color, Phase 2 is represented by the yellow color, and the white color is used to present Phase 3. The blue lines in the contours of the components represent the interface of Phase 1 and the yellow lines are the interface of Phase 2. The falling of the two drops is very similar to those in Section \ref{Sec Miscible falling drop}. Both drops deform little until they are close to the bottom tank of Phase 1, and the components inside the drops remain homogeneous. The diffusivity of Component 3 in Phase 3 is 4 times larger than that in Phase 1, and the flow in Phase 3 is more significant due to the falling of the drops. Component 3 is more distributed in Phase 3 and some amount of it has entered the drop of Phase 1, while it is still clustered inside the bottom tank of Phase 1. 
The drop of Phase 2 (the yellow one) first reaches the tank of Phase 1. It is floating on Phase 1 since Phase 2 is lighter than Phase 1. Component 2, carried by the drop of Phase 2, starts to diffuse into the bottom tank of Phase 1. The diffusivity of Component 2 in Phase 1 is 10 times larger than that in Phase 2, and it is the largest among all the diffusivities. As a result, Component 2 becomes homogeneous inside both Phases 1 and 2 in a very short period of time. After the drop of Phase 1 (the blue one) merges to the bottom tank, Component 1 is released to the tank. Since it is lighter than that in Section \ref{Sec Miscible falling drop}, it clusters close to the interface between Phases 1 and 3, follows the movement of the interface, and in the meantime keeps diffusing. The surface wave introduced by the merging of the drop of Phase 1 to the bottom tank pushes Phase 2 towards the right wall. After that, the drop of Phase 2 is lifted and then stretched as the surface wave moves up and down. The surface wave is gradually settled down by the viscous effect, and the $135^0$ contact angle between Phases 1 and 2 at the right wall is more clearly observable. As time goes on, the phases gradually reach their equilibrium configurations, and Components 1 and 3 keep homogenizing inside their corresponding dissolvable regions. Specifically, Component 1 only exists inside Phase 1 and none of it is observed in Phases 2 and 3 from our results. On the other hand, Component 2 is allowed in both Phases 1 and 3, and it is not observed in Phase 2 from the results.   
Again, we plot the time histories of the total volumes of individual phases and the total amounts of each component in its corresponding dissolvable region in Fig.\ref{Fig ContactLine-Mass}, and they are exactly conserved even though the problem considered is highly complicated and dynamical. 

\begin{figure}[!t]
	\centering
	\includegraphics[scale=.5]{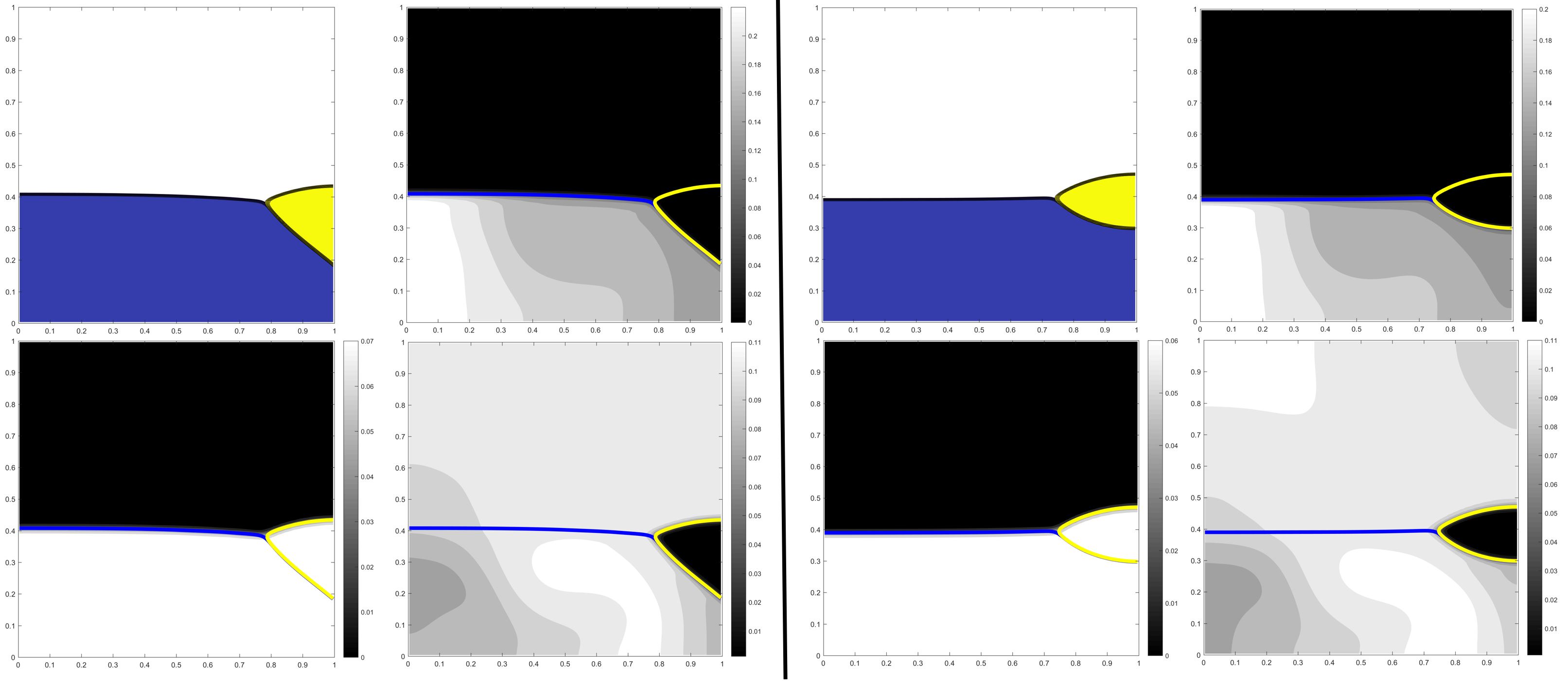}
	\caption{Comparison with different contact angle set-ups at $t=5$ in the falling drop with moving contact lines. Left: the contact angle between Phases 1 and 2 is $135^0$ at the right wall. Right: the contact angle between Phases 1 and 2 is $90^0$ at the right wall. Top-left of each panel: configuration of the phases, blue: Phase 1, yellow: Phase 2, white: Phase 3. Top-right of each panel: configuration of Component 1. Bottom-left: configuration of Component 2. Bottom-right: configuration of Component 3. In the top-right, bottom-left, and bottom-right of each panel, blue solid lines: interfaces of Phase 1, yellow solid lines: interfaces of Phase 2. \label{Fig ContactLine-Compare} }
\end{figure}

Finally, the effect of the contact angle is illustrated in Fig.\ref{Fig ContactLine-Compare} by comparing the present case to the case where all the contact angles are $90^0$. We observe that the effect of the contact angle doesn't change much of the dynamics of the problem while it becomes important for the equilibrium configuration, so we only show the results at $t=5$. It is clear that in the case with contact angles all being $90^0$, i.e., in Fig.\ref{Fig ContactLine-Compare} b), the interface of Phases 1 and 3 is finally horizontal, and the drop of Phase 2 looks like a section of an ellipse which is symmetric with respect to the interface of Phases 1 and 3. On the other hand, in the case with the $135^0$ contact angle between Phases 1 and 2 at the right wall, i.e., in Fig.\ref{Fig ContactLine-Compare} a), the interface between Phases 1 and 2 is close to an inclined straight line, and the interface between Phases 1 and 3 is not horizontal.

The problem considered is again very challenging. It includes multiple materials and their material properties are significantly different. The maximum density ratio is 1600, the maximum viscosity ratio is 5000, and the diffusivity of the components crosses two orders of magnitude. It also includes gravitation, interfacial tensions, topological change, large deformation of the interfaces, the effects of contact angles, and moving contact lines, which are the critical factors for multiphase and multicomponent flows. Nevertheless, our model and scheme are capable of overcoming those challenges and producing physically plausible results.

\section{Conclusions}\label{Sec Conclusion}
In the present work, we developed a consistent and conservative model and its corresponding scheme for multiphase and multicomponent, or $N$-phase-$M$-component, incompressible flows with $N\geqslant1$ and $M\geqslant0$. Phases are immiscible with each other while components are dissolvable in some specific phases. 
The model allows arbitrary numbers of phases and components appearing simultaneously. Each component can exist in different phases. Each phase is a ``solution'' of its pure phase (which is the background fluid of the phase) as the ``solvent'' and there can be multiple components as the ``solutes'' dissolved in that phase . The dissolvability matrix is defined to indicate these relations between phases and components.
Individual pure phases and components have their own densities and viscosities. Each pair of phases has a surface tension and each pair of a phase and a component has a diffusivity. At a wall boundary, every two phases have a contact angle. 

The model is developed based on the multiphase Phase-Field model including the contact angle boundary condition \citep{Dong2017,Dong2018,Huangetal2020N}, the diffuse domain approach \citep{Lietal2009}, and the consistency analysis \citep{Huangetal2020} on the consistency conditions proposed for multiphase and multicomponent flows. 
The locations of individual phases are represented by a set of order parameters, which is the volume fraction contrasts in the present model, and the sharp interfaces are replaced by interfacial regions inside which there are thermodynamical compression and diffusion to preserve the thickness of the region. The Phase-Field model also eliminates any generation of local void or overfilling, i.e., the summation of the volume fractions from the Phase-Field model is unity everywhere.
Each component has its own concentration governed by the convection-diffusion equation in the phases that it is dissolvable in, and its flux at the phase boundaries is either zero if it is not dissolvable in the neighboring phases or continuous otherwise. The concentrations represent the local amounts of the components and are generally interpreted as the ``molar concentrations'', while they can be considered as the ``volume fractions'' in some special problems. Since all the phases are evolving and deforming, it is challenging to solve the equation and assign the boundary condition of the components. We apply the diffuse domain approach where the original equation defined in a complex domain along with the boundary condition is replaced by an equivalent equation defined in the whole domain of interest, with the help of the indicator functions of the phases and phase boundaries. Then, the indicator functions are replaced by the smoothed ones, which makes the volume fractions from the Phase-Field model a directly available candidate. The component equation is finalized by incorporating the thermodynamical effect from the Phase-Field model so that the volume fraction equation derived from the component equation is consistent with the one from the Phase-Field model. We call such a consistency condition, the consistency of volume fraction conservation. This consistency condition is not realized in the original diffuse domain approach \citep{Lietal2009} and it plays an essential role in the energy law and Galilean invariance of the model.
The density and viscosity of the fluid mixture are computed using the volume fractions of the phases and concentrations of the components, and the motion of the fluid mixture is governed by the momentum equation. The consistency of mass conservation and the consistency of mass and momentum transport are applied to ensure the physical coupling between the mass conservation equation and the momentum equation. These consistency conditions are also essential for the energy law and Galilean invariance of the model. Finally, the analysis of the consistency of reduction is performed to ensure that the proposed model is not allowed to generate fictitious phases or components.
A physical energy law is satisfied with the proposed model. We show that, without external input, the total energy, which includes free energy, component energy, and kinetic energy, is not increasing as time goes on. Three factors contribute to the decay of the total energy. The first one is due to the thermodynamical non-equilibrium in the interfacial regions from the Phase-Field model. The second one is the inhomogeneity of the components in their dissolvable regions from the component equation. The last one is the viscous effect from the momentum equation.
It can also be shown that the proposed model satisfies the Galilean invariance. 

Consequently, the proposed model is consistent and conservative. It conserves the mass of individual pure phases, the amount of each component in its dissolvable region, and thus the mass of the fluid mixture, and the momentum of the multiphase and multicomponent flows. It ensures that the summation of the volume fractions from the Phase-Field model is unity everywhere so that there is no local void or overfilling. It satisfies a physical energy law and it is Galilean invariant. It satisfies all the proposed consistency conditions, which are the consistency of reduction, the consistency of volume fraction conservation, the consistency of mass conservation, and the consistency of mass and momentum transport. It should be noted that the consistency conditions play a critical role in avoiding fictitious phases and components, in deriving the energy law, and in proving the Galilean invariance of the model.
In addition to multiphase and multicomponent problems, the proposed model is also applicable to some multiphase problems where the miscibilities between each pair of phases are different. 
The proposed model is also flexible to different circumstances of cross-interface transport of a component which is dissolvable in both sides of the interface. Specifically, under different setup, a component, which is dissolvable in two neighboring phases, is either able to cross the phase interface with continuous flux at the interface or not allowed to cross the phase interface, i.e., with zero flux at the interface.
More importantly, the present study proposes a general framework that physically connects the dynamics of the phases, the components, and the fluid flows, using the proposed consistency conditions. Therefore, it suits to many models for the order parameters or components. Changing those models is equivalent to updating the definitions of the Phase-Field flux and component flux in the present framework and this doesn't affect the Galilean invariance of and the kinetic energy equation derived from the momentum equation.

A few assumptions have been made in the proposed model. 
First, we have assumed that the concentrations of the components are governed by the convection-diffusion equation. This is not a strong assumption. The framework we proposed is flexible to incorporate any other physical model that best describes the dynamics of the components.
Second, we only consider the Neumann-type boundary condition of components at phase interfaces since it suits many applications. The Dirichlet- or Robin-type boundary condition can also be incorporated inside the framework of the diffuse domain approach, see \citep{Lietal2009,Yuetal2020}. 
Third, we have not considered the effect of components on phase interfaces. For example, the appearance of the components at the phase interfaces can change the surface tensions as well as the contact angles, and therefore introduce the Marangoni effect. This will be an interesting extension of the present model in future work.
Last, the validity of the proposed model relies on the assumption of diluteness, although this assumption can be removed in a subset of problems where the components dissolved in a specific phase share an identical diffusion coefficient in that phase and cross-interface transports of components are not allowed. Complete removal of this assumption needs to incorporate the effect of components on changing phase volumes and will be another direction to improve the proposed model.

The corresponding numerical scheme is developed for the proposed model. 
The scheme is formally 2nd-order accurate in both time and space. 
The numerical solution of the model also converges to the sharp-interface one, which is validated by the convergence tests in the present work and in our previous work \citep{Huangetal2020,Huangetal2020CAC,Huangetal2020N,Huangetal2020B} for the multiphase incompressible flow model. 
The scheme preserves the properties of the model. It can be shown that the scheme conserves the mass of individual pure phases, the amount of each component in its dissolvable region, and therefore the mass of the fluid mixture. The momentum is exactly conserved by the scheme using the conservative method for the interfacial force, while it is essentially conserved with the balanced-force method. 
All the consistency conditions, i.e., the consistency of reduction, the consistency of volume fraction conservation, the consistency of mass conservation, and the consistency of mass and momentum transport, are shown to be satisfied by the scheme. This is very important for the problems including large density ratios, see the analysis in \citep{Huangetal2020}.
The scheme also ensures that the Phase-Field solutions are always in their physical interval, no fictitious phases or components are numerically generated, and the summation of volume fractions is always unity.
The numeral solution also preserves the Galilean invariance and the energy law, which is demonstrated by numerical experiments. 
In addition, the capability of the model for different circumstances of cross-interface transport of a component which is dissolvable in both sides of the interface is demonstrated. The effect of the newly proposed consistency of volume fraction conservation on eliminating unphysical fluctuations of the components around phase interfaces is illustrated.
Finally, the proposed model and scheme are successfully applied to two challenging problems, including multiple materials and various miscibility relations among the materials, and having many critical factors in multiphase and multicomponent flows, i.e., significant differences of the material properties, gravitation, interfacial tensions, topological change, large deformation of the interfaces, and the effects of contact angles and moving contact lines. In summary, the proposed model and scheme are general, effective, and robust for multiphase and multicomponent incompressible flows.

\section*{Acknowledgments}
A.M. Ardekani would like to acknowledge the financial support from the National Science Foundation (CBET-1705371). This work used the Extreme Science and Engineering Discovery Environment (XSEDE) \citep{Townsetal2014}, which is supported by the National Science Foundation grant number ACI-1548562 through allocation TG-CTS180066  and TG-CTS190041.
G. Lin would like to acknowledge the support from National Science Foundation (DMS-1555072, DMS-1736364, CMMI-1634832, and CMMI-1560834), and U.S. Department of Energy (DOE) Office of Science Advanced Scientific Computing Research program DE-SC0021142.

\bibliographystyle{plain}
\bibliography{refs.bib}

\end{document}